\documentclass[graybox, secnum]{svmult}

\usepackage{mathptmx}       
\usepackage{mathtools}
\usepackage{helvet}         
\usepackage{courier}        
\usepackage{type1cm}        
\usepackage{makeidx}         
\usepackage{graphicx}        
\usepackage{multicol}        
\usepackage[bottom]{footmisc}
\usepackage{hyperref}        
\usepackage{soul}            
\hypersetup{colorlinks=true,urlcolor=blue}
\usepackage[square,numbers]{natbib}
\makeindex             

\DeclareRobustCommand{\ion}[2]{%
\relax\ifmmode
\ifx\testbx\f@series
{\mathbf{#1\,\mathsc{#2}}}\else
{\mathrm{#1\,\mathsc{#2}}}\fi
\else\textup{#1\,{\mdseries\textsc{#2}}}%
\fi}

\begin{document}
\tableofcontents{}
\title*{Low Magnetic-Field Neutron Stars in X-ray Binaries}
\titlerunning{Low Magnetic-Field Neutron Stars} 
\author{Tiziana Di Salvo \thanks{corresponding author}, Alessandro Papitto, Alessio Marino, Rosario Iaria and Luciano Burderi}
\authorrunning{T. Di Salvo, A. Papitto, A. Marino, R. Iaria, L. Burderi}
\institute{Tiziana Di Salvo \at Dipartimento di Fisica e Chimica - Emilio Segr\'e, Universit\'a di Palermo, via Archirafi 36 - 90123 Palermo, Italy
\email{tiziana.disalvo@unipa.it}
\and Alessandro Papitto \at INAF—Osservatorio Astronomico di Roma, via Frascati 33, I-00076, Monteporzio Catone (RM), Italy
\email{alessandro.papitto@inaf.it}
\and Alessio Marino \at Institute of Space Sciences (ICE, CSIC), Campus UAB, Carrer de Can Magrans s/n, E-08193 Barcelona, Spain \\
Dipartimento di Fisica e Chimica - Emilio Segr\'e, Universit\'a di Palermo, via Archirafi 36 - 90123 Palermo, Italy 
\email{marino@ice.csic.es}
\and Rosario Iaria \at Dipartimento di Fisica e Chimica - Emilio Segr\'e, Universit\'a di Palermo, via Archirafi 36 - 90123 Palermo, Italy 
\email{rosario.iaria@unipa.it}
\and Luciano Burderi \at Dipartimento di Fisica, Universit\'a degli Studi di Cagliari, SP Monserrato-Sestu, KM 0.7, Monserrato, 09042 Italy
\email{burderi@dsf.unica.it}}
%
%
\maketitle
\abstract{
In this chapter we give an overview of the properties of X-ray binary systems containing a weakly magnetized neutron star. These are old (Giga-years life-time) semi-detached binary systems containing a neutron star with a relatively weak magnetic field (less than $\sim 10^{10}$ Gauss) and a low-mass (less than $1 M_\odot$) companion star orbiting around the common center of mass in a tight system, with orbital period usually less than 1 day. The companion star usually fills its Roche lobe and transfers mass to the neutron star through an accretion disk, where most of the initial potential energy of the in-falling matter is released, reaching temperatures of tens of million Kelvin degrees, and therefore emitting most of the energy in the X-ray band. Their emission is characterized by a fast-time variability, possibly related to the short timescales in the innermost part of the system. Because of the weak magnetic field, the accretion flow can approach the neutron star until it is accreted onto its surface sometimes producing spectacular explosions known as type-I X-ray bursts. In some sources, the weak magnetic field of the neutron star ($\sim 10^8-10^9$ Gauss) is strong enough to channel the accretion flow onto the polar caps, modulating the X-ray emission and revealing the fast rotation of the neutron star at millisecond periods. These systems are important for studies of fundamental physics, and in particular for test of Relativity and alternative theories of Gravity and for studies of the equation of state of ultra-dense matter, which are among the most important goals of modern physics and astrophysics. }

\section{Keywords} 
Neutron stars; Low magnetic field; X-ray Binaries; X-ray pulsars; Millisecond X-ray pulsars; Transitional pulsars; Fast variability; Type-I bursts; Burst oscillations; X-ray spectra; high-inclination sources; very faint X-ray sources.

\section{Introduction}
In this chapter we deal with weakly magnetized neutron stars (hereafter NSs) in Low Mass X-ray Binaries (hereafter LMXBs). In a LMXB the compact object interacts, through its gravitational field, with the other component, named the donor, which is usually less massive than the compact object. The donor is therefore a low mass (usually less than 1 solar mass, $M_\odot$) \textit{normal} main sequence star, or a degenerate dwarf (white dwarf), or an evolved star (red giant). These are therefore old systems with typical lifetime of the order of billions of years, long enough to dissipate most of the original NS magnetic field, resulting in fields of the order of $10^{8-10}$ Gauss. The donor star, which is typically not a strong wind emitter, fills its Roche lobe and hence transfers mass to the compact object through the inner Lagrangian point. The transferred matter has a high-specific angular momentum and forms an accretion disk before reaching the NS surface. More than two hundred of these systems have been detected in the Milky Way. Since these are old systems they are mostly found in the Galactic bulge and disk and $2-3$ dozen of these systems are found in Globular Clusters (see \cite{Liu_2007}, and the more recent catalogue by \cite{Avakyan_2023}). Few of them have also been discovered in nearby galaxies. 

LMXBs are among the brightest sources of the X-ray sky, suffice to say that the discovery in 1962 of the brightest of these sources, Scorpius X-1, using rocket-borne instruments, earned Riccardo Giacconi the Nobel prize in 2002 for the discovery of the first cosmic X-ray sources and marked the beginning of X-ray astronomy. A typical LMXB emits almost all of its radiation in X-rays, and typically less than one percent in visible light, so they are among the brightest objects in the X-ray sky, but relatively faint in visible light. The apparent magnitude is typically around 15 to 20. The brightest part of the system in this band is the accretion disk around the compact object, usually brighter than the donor star. Orbital periods of LMXBs range from ten minutes to days, with typical values below 1 day.

In NS LMXBs the matter transferred from the companion star eventually hits the NS surface generating electromagnetic radiation and making the accreting object a powerful source of energy \cite{Frank_1992}. The accretion disk also emits radiation produced via viscous dissipation, and hence the large majority of the emission from these systems is powered by gravitational potential energy release. Typically about $10\%$ of the rest-mass energy of the accreted matter is released and emitted mostly in the X-ray band (reaching temperatures from $0.1-1$ keV up to more than a hundred keV). The apparent luminosity of these systems depends on different parameters, such as the mass accretion rate and the geometry of the system, and can range from few $10^{31}$ erg/s up to few times the Eddington limit for a NS: 
\begin{equation}
    L_{Edd} = 4 \pi G M m_p c / \sigma_T = 1.26 \times 10^{38}\, M/M_\odot \; {\rm erg/s},
\end{equation}
where $M$ is the mass of the central object, $m_p$ the proton mass, $\sigma_T$ the Thomson cross section, and $M_\odot \simeq 2 \times 10^{33}$ g the mass of the Sun. 

Although the geometry of the innermost part of the system is still unclear, the low magnetic field of NSs in LMXBs in principle allows the accretion disk to approach the NS surface. The  region connecting the inner edge of the disk to the NS, where the azimuthal (Keplerian) velocity of matter in the disk smoothly approaches the velocity at the NS surface, is dubbed boundary layer and is an important source of radiation as well. The NS magnetosphere can affect the accretion process depending on the mass accretion rate and the magnetic field strength (see e.g.\ \cite{Ghosh_1977,Ghosh_1978}). The accreting matter transfers its angular momentum to the NS; when the inner disk is sufficiently close to the NS surface, the interaction of the NS magnetic field with the accreting matter can spin the star up, bringing the NS rotation to very short spin periods, close to the Keplerian frequency (given by $\Omega_K = \sqrt{G M_{NS}/r^3}$, where $r$ is the distance from the NS center) at the inner edge of the disk (spin equilibrium). This is consistent with the observed stellar spin frequencies, typically in the range $200-700$ Hz. 

Many LMXBs show type-I X-ray bursts, a spectacular eruption in X-rays every few hours to days. During these bursts, the observed X-ray intensity goes up sharply in $\sim 0.5-5$ s, and decays relatively slowly in $\sim 10-100$ s. The typical energy emitted in a few seconds during such a burst is $\sim 10^{39}$ ergs (the energy our Sun releases in more than a week).
These are interpreted as thermonuclear flashes on the NS surface, caused by material freshly accreted onto the surface that reaches densities and temperatures sufficient for nuclear ignition (see Sec.\ \ref{ss:bursts} for  details). The burning is unstable and propagates around the star, resulting in an X-ray burst typically characterized by a fast rise and an exponential decay (FRED). Type-I bursts provide a rare opportunity for the observer to measure the radius of a NS, that radiates like a spherical blackbody of luminosity $L_{burst} = 4 \pi R_{NS}^2 \sigma T^4 = f_X 4 \pi D^2$, where $D$ is the distance to the source, $f_X$ the flux during the burst, $T$ the temperature from the peak wavelength of the (blackbody) spectrum during the burst, and $\sigma$ the Stefan-Boltzmann constant. 

The fact that X-rays can be detected both from the NS and the inner part of the disk for many LMXBs provides an excellent opportunity to study the extreme physics of and around NSs, where special and general relativistic effects can play a role (see e.g.\ \cite{Bhattacharyya_2011} as a review).
Moreover, constraining mass and radius of NSs is important in order to infer constraints on the Equation of State (EoS) of matter at ultra-nuclear density (see \cite{Ozel_2016} as a review), which is one of the major goals of modern physics and astrophysics.  The EoS relates the pressure of a gas to its density, and also determines the equilibrium relation between the mass and radius of a NS.
However, the central X-ray emission from these sources cannot be spatially resolved with current X-ray facilities because of the large distances of these sources. Therefore, in order to investigate the properties of the accretion flow close to a NS we have to rely on the study of their spectral and timing properties. In the following we will describe some of these properties as well as the information they provide.     

\section{The zoo of low magnetic field Neutron Stars}

Low magnetic field NSs can be classified in several ways according to their phenomenology which in turn depends on the mass accretion rate, NS magnetic field strength, geometry of the inner disk region and its inclination with respect to our line of sight, orbital period, chemical composition and companion star type, and so on. In the following we will explore the zoo of X-ray emitting low magnetic-field NSs.

\subsection{Transient and persistent sources}\label{ss:transients}

One of these classifications is based on the long-term X-ray variability, according to which NS LMXBs can be classifed as transient or persistent sources. In persistent sources, the X-ray emission remains stable, although with variations of the luminosity up to one order of magnitude, whereas in transient sources (X-ray Transient, XRT) bright outburst phases (with luminosity up to the Eddington limit) are alternated with long-lasting quiescence phases (with luminosity down to a few $10^{31}$ erg/s). The large majority of NS LMXBs show a transient behavior whereas only a couple of dozen of these sources are persistent at different luminosity levels; from faint and very faint sources (typical luminosity below $\sim 10^{35}$ erg/s, see Section  \ref{ss:VFXT}) to the brightest sources (emitting at, or close to, the Eddington limit). X-ray outbursts often are accompanied by a brightening of the optical counterpart by $2-5$ magnitudes, usually interpreted as reprocessing of high-energy radiation from the accretion disk. 

The transient nature of these LMXBs is normally attributed to an accretion disk instability (see e.g.\ \cite{King_2001} as a review), that causes high accretion rate for certain periods of time and almost no accretion during other times. However, recurrence times and duration of X-ray outbursts are highly variable from source to source, even for very similar sources. Indeed some sources show regular outbursts every few years -- this is the case for instance of Aql X-1 or SAX J1808.4-3658. Other sources show very long (tens of years) periods of X-ray quiescence and rarely go into outburst. Many LMXBs showing coherent pulsations at millisecond periods, the so-called accreting millisecond pulsars (AMSPs, see Section \ref{ss:AMXPs}), have shown just one X-ray outburst in the last 25 years. Other sources show years-long periods of X-ray activity before starting similarly long periods of quiescence. This is the case of EXO 0748-676 which has been observed for more than 20 years as a persistent X-ray source before going into quiescence in 2008. Similarly, HETE J1900.1-2455 returned to quiescence in late 2015, after a prolonged accretion outburst of about 10 yr (e.g., \cite{DiSalvo_2021} and references therein). Some sources have also shown close-in-time outbursts (about a month or so apart), e.g. the 2008 outburst of IGR J00291+5934. This variegated phenomenology, together with other observational facts (mainly related to the observed orbital period evolution) led some authors to propose that other parameters may play a role in determining recurrence times and duration of X-ray outbursts, such as the magnetic field of a fast rotating NS (e.g.\ \cite{Burderi_2001}) or strong outflows and winds driving large mass loss from these systems \cite{Ziokowski_2018}.

During the X-ray outburst, usually characterized by a fast rising phase, a peak or {\it plateau} where the maximum luminosity is reached, and a slower decay of the X-ray flux, the source undergoes variation of its spectral properties. Typically the source transits from a hard/low-luminosity state to a soft/high-luminosity state and then back again to the hard/low-luminosity state. 
In the hard state most of the flux is emitted in the hard X-ray band, above 10 keV, and the X-ray spectrum is dominated by a Comptonized continuum due to soft photons Compton up-scattered off a hot, optically thin, electron cloud, named the {\it corona}. In the soft state, most of the flux is emitted in the soft X-ray band, below 10 keV, and the X-ray spectrum is dominated by low-temperature blackbody-like components. This spectral evolution is probably caused by the (Compton) cooling of the hot electron corona, that becomes denser and more compact around the NS, making the luminous inner accretion disk more visible. 
Hard-to-soft and soft-to-hard state transitions do not usually occur at the same X-ray luminosity but often an hysteresis is observed, such that the transition back to the hard state occurs at a luminosity several times lower than the hard-to-soft transition \citep{MunozDarias_2014}.  

Because of this spectral evolution towards soft states during the outburst, these sources are named Soft X-ray Transients (SXTs). However, some transients do not show a transition to the soft sate, their spectra remaining in the hard state even at the peak of the outburst. This is the case of the majority of AMSPs, that are therefore named Hard X-ray Transients (HXTs). It is unclear whether this feature may be related to the presence of a relevant magnetosphere in these sources which could prevent the accretion disk from significantly approaching the star, or to the mass accretion rate which remains relatively low (peak X-ray luminosity of $\sim 10^{36}-10^{37}$ erg/s) even during the outburst. 

These sources are often not detected during quiescent states due to their low luminosity, and for many scientific purposes it is preferable to observe them during their outbursts to maximize the statistics. Also, new transients are typically discovered during the outburst states. However, these outbursts are normally unpredictable, and therefore large field-of-view instruments 
(such as the ASM on board RXTE, MAXI on board the International Space Station, ISS, the ESA high-energy-observatory INTEGRAL, and BAT on board Swift) 
are required to continuously monitor the X-ray sky. 

However, there are scientific reasons that motivate observations of transient NS LMXBs in their quiescent states. For example, during the quiescence, the NS surface, heated up by previous outburst episodes, should be the primary X-ray emitting component, and this can be modelled to constrain the stellar radius. Also the study of the cooling history of the NS may give information on the micro-physics of the stellar interior and of the dense matter present in the crust
(e.g.\ \cite{Cackett_2008,Wijnands_2017}). In this case, high sensitivity imaging instruments (e.g.\ Chandra and XMM-Newton) are used to observe the sources in quiescence.
The observed X-ray emission from a transient NS LMXB during X-ray quiescence is expected to primarily originate from the NS surface/atmosphere (with sometimes some residual accretion). This atmosphere should be composed of pure hydrogen, and devoid of any X-ray spectral line; in fact, the composition of the atmosphere is likely to be affected by rapid gravitational settling, which may cause the accreted heavy elements to sink below the NS photosphere. Moreover, the relatively low magnetic field of NSs in LMXBs (unlike isolated NSs) makes the modelling simpler. Assuming a blackbody emission from the NS surface in a quiescent LMXB, 
the inferred radius at infinity, $R_\infty$, can be obtained from the relation: $R_\infty = (F_\infty/\sigma T_\infty^4)^{1/2} D$, where $F_\infty$ is the observed bolometric flux, and $T_\infty$ is the best-fit blackbody temperature.
Correcting for the hardening factor, $f$ (due to scattering of photons by the electrons and the frequency dependence of the opacity in the NS atmosphere), and for the effects of the gravitational redshift, we can write the actual radius as: $R_{BB} = R_\infty  f^2 / (1+z)$, where $1 + z = (1 - 2 G M / c^2 R)^{-1/2}$ is the surface gravitational redshift for a non-spinning NS. 

\subsection{Classical LMXBs: Z-sources and atolls}

Another widely used classification of LMXBs is based on their correlated spectral and timing behavior (see e.g.\ \cite{Bhattacharyya_2009} as a review). A convenient way to study how the X-ray spectrum of a LMXB changes with time and how it is correlated with changes in timing features is to compute a color-color diagram (hereafter CD). In this case, the available energy range  is divided into four (or three) contiguous energy bands, that we can indicate with numbers from 1 to 4 for increasing energy. The photon counts in each band in a given time interval (seconds or minutes or hours depending on the statistics) are used to define the soft and hard colors, with the ratio of the counts in the band 2 with respect the band 1 indicated as Soft Color (SC) and the ratio of the counts in the band 4 with respect the band 3 indicated as Hard Color (HC). In a CD the HC is plotted versus the SC for a given source at different times, and it is used to track changes of the spectral hardness of the source. The SC or HC can also be plotted versus the total source photon counts and in this way we get what is called a hardness-intensity diagram (HID). 

\begin{figure}
\vspace{-0.5cm}
\centering
\includegraphics[width=1.0\textwidth, angle=0]{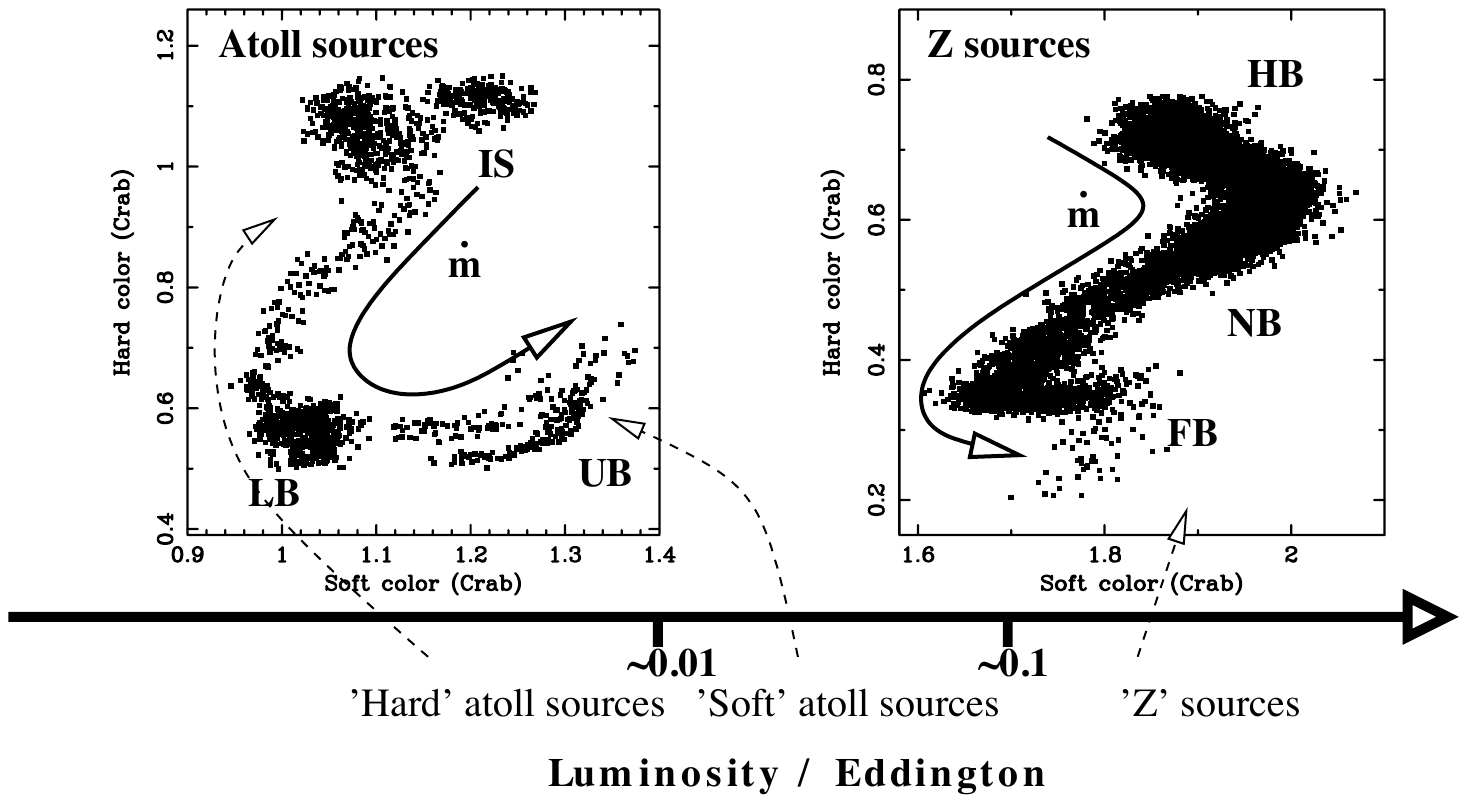}
\vspace{-1.0cm}
\caption{Typical CD of Atoll (left) and Z (right) sources. The (inferred) increasing mass accretion rate is indicated by the arrows. Two states are clearly defined for atoll sources, the island state and the banana state (LB, Lower Banana and UB, Upper Banana), corresponding to hard and soft states, respectively. As for Z sources, three branches are distinguishable: the horizontal branch (HB), the normal branch (NB), and the flaring branch (FB). From \cite{Migliari_2006} }
\label{fig:atoll_Z}
\end{figure}

NS LMXBs can be divided into two classes based on the shape of their CD; i) the so-called {\it Z sources}, that show a Z-like track in the CD (see Fig.~\ref{fig:atoll_Z}, right panel), which are sources persistently emitting at high luminosity, close to the Eddington limit for a NS, and ii) the so-called {\it atoll sources}, that show a C-like track in the CD (see Fig.~\ref{fig:atoll_Z}, left panel), which are persistent and transient sources that can become rather bright but typically show luminosities of few tenths of the Eddington limit. These two classes of sources, not only follow a different spectral evolution as traced by their CD, but also show spectral changes that are correlated with distinctly different timing properties (see \cite{vanderKlis_2006} as a review). 

The three branches of the Z track are called horizontal (HB), normal (NB) and flaring (FB) branch, respectively. The source moves along the Z track on timescale ranging from hours to a couple of days, performing a random walk and not jumping from a branch to another. The lower part of the CD of a typical atoll is called the banana state (BS, because of its shape) whilst the upper part is called the island state (IS). Like in Z sources, atolls move along the BS back and forth with no hysteresis on time scales of hours to a day or so, whilst the motion on the IS is much slower (days to weeks). Observational windowing can cause isolated patches to form, which is why this state is called island state. A group of four sources (that are GX 13+1, GX 3+1, GX 9+1, GX 9+9), called bright atolls or GX-atoll sources, are persistently bright sources and are nearly always in the BS (see \cite{vanderKlis_2006} and references therein). The Z and atoll tracks may show slow drifts (i.e.\ a \textit{secular motion} on timescales of months to years), usually more visible in the HID, that do not much affect the variability and its strong correlation with the position along the track, originating a phenomenon known as \textit{parallel tracks} (see Sec.\ \ref{ss:fastvar}). 

What drives the spectral evolution of these sources in the CD is not clear yet. Interestingly, in both atoll and Z sources, most of the X-ray spectral and timing properties depend only on the position of a source in the diagram. The main parameter that determines these properties, and thus the evolution on the diagram, is probably the mass accretion rate $\dot m$, which is linked to the accretion luminosity as $L_{acc} = \eta \dot m c^2$, where $\eta$ is the accretion efficiency ($\sim 10-15\%$ for a NS). Note, however, that the X-ray luminosity of the source is not clearly correlated with the position in the CD, especially in the NB of Z sources, where the X-ray luminosity decreases while the mass accretion rate should increase instead. Given the complexity of these paths, it is thought that more than one parameter must be involved. 

It is also not well understood what, apart from the luminosity, distinguishes between Z and atoll sources. It has been proposed that Z sources may have a higher accretion rate and a higher magnetic field strength ($B > 10^9$ Gauss, see e.g.\ \cite{Gierlinski_2002}). However, the discovery of the first transient Z source, XTE J1701--462 \cite{Homan_2007}, provided new important information on the nature of Z and atoll sources. This transient source started its outburst as a Z source, exhibiting both the typical Z-shaped track in the X-ray CD and the X-ray timing phenomenology
normally observed in Z-sources.
However, as its overall luminosity decreased, a transition from Z source to atoll source behavior was observed. This transition manifested itself both as changes in the shapes of the track in X-ray CD and HID, and as changes in the fast time variability,
X-ray burst behavior, and X-ray spectra \cite{Homan_2010}. This strongly suggests that the variety in behavior observed in NS LXMBs with different luminosity can be linked to changes in a single variable parameter, namely the mass accretion rate, without the need for additional differences in the NS parameters or viewing angle. Furthermore, the evolution of XTE J1701--462 along the outburst is at odd with the hypothesis that the position along the CD of Z sources is determined by the instantaneous mass accretion rate, since $\dot m$ variations along individual Z tracks appear to be quite small. On the other hand, variation in $\dot m$ seems to drive the evolution of the source in the atoll track. The motion along Z track may instead be the result of instabilities, whose presence does depend on the mass accretion rate, or may be determined by the instantaneous mass accretion rate through the disk normalized by its own long-term average, that has been also proposed to explain the parallel-tracks phenomenon. 

Thanks to the extensive monitoring provided by the instruments on board RXTE, we know that both atoll and Z-sources can show superorbital modulation, which are quasi-periodic variations on long timescales ($5-15$ years) with substantial modulation amplitudes for atoll sources and lower modulation amplitudes for Z-sources (see e.g.\ \cite{Charles_2011} and references therein). The different behavior between Z and atoll sources is attributed to variations in the accretion rate that are less pronounced in Z-sources since they are accreting close to the Eddington limit. The possible causes of this modulations are many; warping/tilting and/or precession of the accretion disc may lead to periodic/quasi-periodic superorbital modulation of the X-ray flux (see e.g.\ \cite{Kotze_2012}). Other possible mechanisms are long-term modulation of the mass transfer rate, as the one that induces spectral state transitions, or the presence in the system of a third body which can modulate the mass transfer rate, or magnetic-activity variations in the donor convection zone, similar to the solar cycle, leading to changes in the angular momentum of the donor, and hence to changes in the orbital period and the Roche-lobe size which, in turn, will modulate the mass transfer rate.

\subsection{Fast X-ray variability}
\label{ss:fastvar}

\begin{figure}
\centering
\includegraphics[width=1.0\textwidth, angle=0]{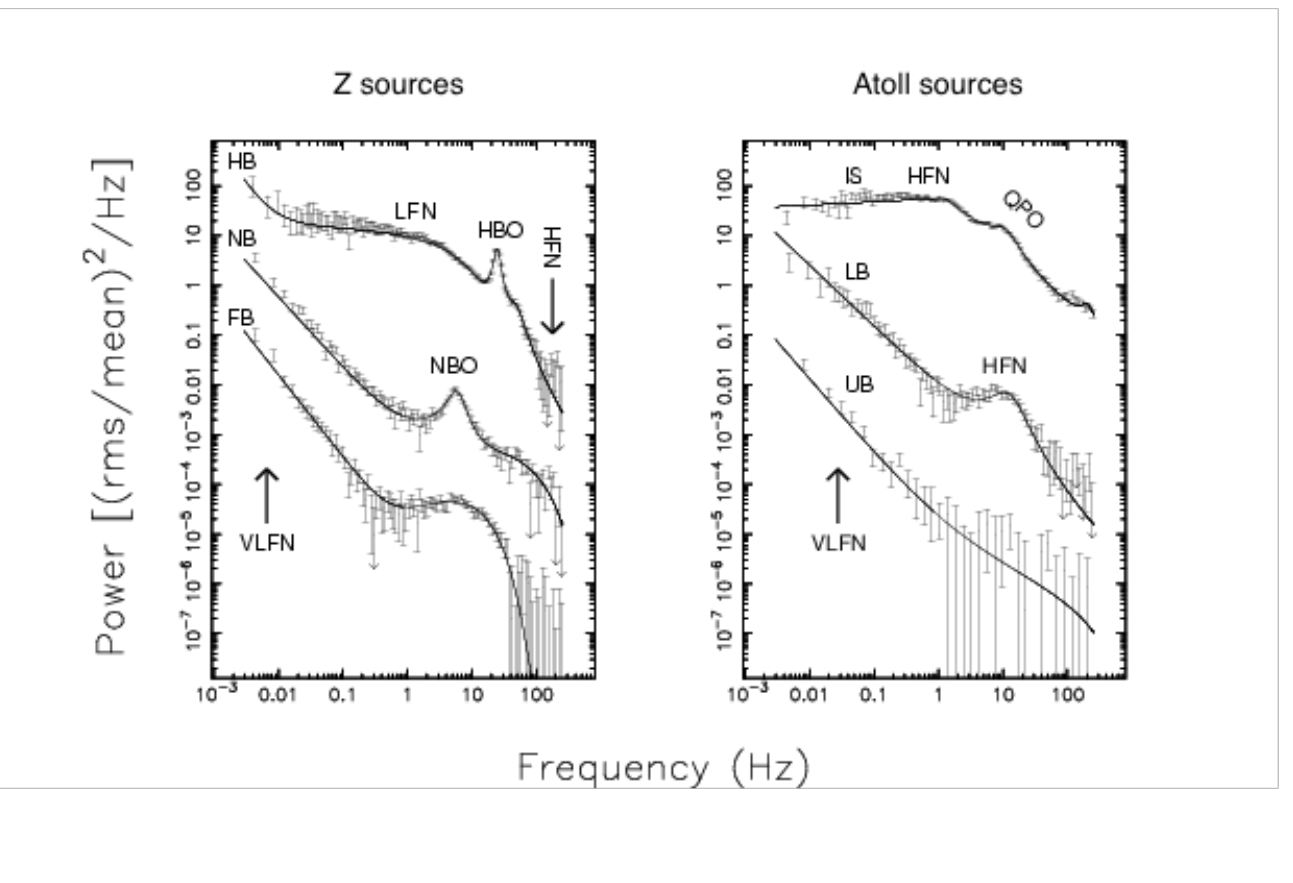}
\caption{Typical low-frequency ($<100$ Hz) power spectrum of NS LMXBs for Z sources (left) and atoll sources (right). Noise components and QPOs in the different branches of the Z and atoll tracks are indicated. Note that historically the BLN was called low frequency noise (LFN) in Z sources and high frequency noise (HFN) in atoll sources, respectively. Reprinted from \cite{Wijnands_2001} with permission from Elsevier}
\label{fig:LF_PSD}
\end{figure}

NS LMXBs show a variety of timing features in their power spectral density (PSD), many of which are not well understood yet. Continuum power, in excess of the mean random power, is often observed at frequencies less than $\sim 10-100$ Hz. The nature of this power, and the corresponding shape of the power spectral component, depends on the source state as determined by its position in the CD. For example, a power law-like continuum ($\propto \nu^{-\alpha}$, called very low frequency noise, VLFN) is usually observed below $\sim 1$ Hz in the softest states; for atoll sources in the end part of the BS (upper banana), and for Z sources on the FB. It has been variously ascribed to accretion-rate variations and unsteady nuclear burning. The detection of broad Lorentzian curves at $< 1$ Hz in the VLFN range has provided some support for models that produce the VLFN from a superposition of finite events such as \textit{nuclear fires} (e.g.\ \cite{Bildsten 1995}).
At lower luminosity (ISs in atolls and HB and NB in Z sources, respectively) the PDS is characterized by a band limited noise (BLN), approximately constant at low frequencies and steepening at higher frequency, usually described by a cutoff or broken power-law  or a zero-centered Lorentz function (see Fig.\ \ref{fig:LF_PSD} for details). Since a Lorentzian is the Fourier transform of a (oscillating) function whose amplitude drops exponentially with time, such as a shot, these features have tentatively interpreted in terms of a combination of shot noise components, with different periods, amplitudes and decay times (the decay time of a shot in the light curve is inversely proportional to the FWHM of the Lorentzian in the Fourier PSD), that add up together to produce the observed light curve (see, however, \cite{Mendez_2021} and references therein for a wider discussion).  

At higher frequencies with respect to the noise components there are Lorentzian-shaped features interesting a narrow frequency range called low-frequency quasi-periodic oscillations (LF-QPOs). These features are present, although with different parameters and centroid frequencies (from few Hz up to $10-20$ Hz), in almost all the branches of the Z and atoll tracks (see Fig.\ \ref{fig:LF_PSD}). The LF-QPOs observed along the Z track (sometimes observed together their harmonic components) show different characteristics from a branch to another and are called HBO, NBO, FBO in the HB, NB, and FB, respectively. The characteristic frequencies of both the noise components and LF-QPOs are correlated with the position in the CD, and usually increase monotonically in each branch of the CD in the sense of increasing mass accretion rate.

\begin{figure}
\vspace{-1.0cm}
\centering
\includegraphics[width=0.4\textwidth, angle=0]{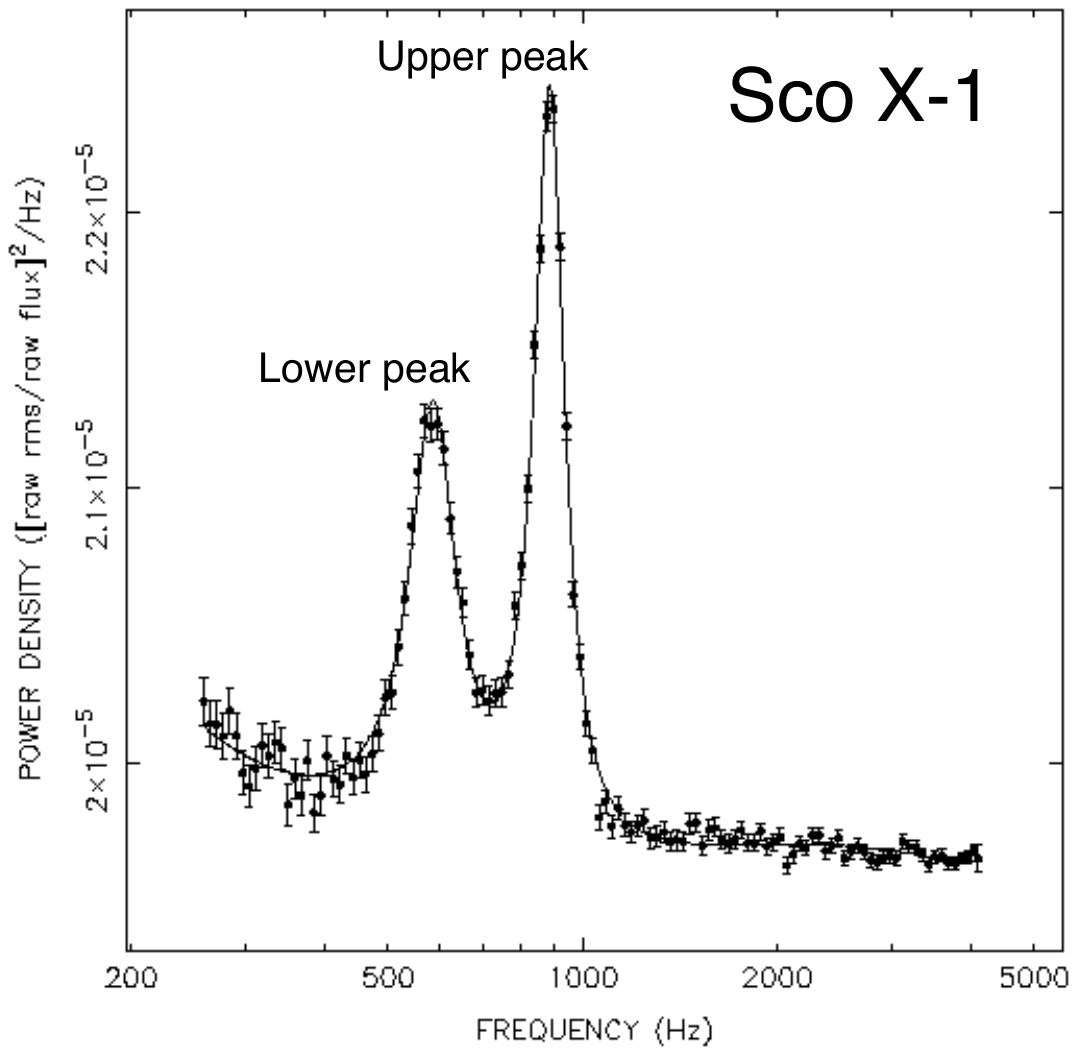}
\includegraphics[width=0.44\textwidth, angle=0]{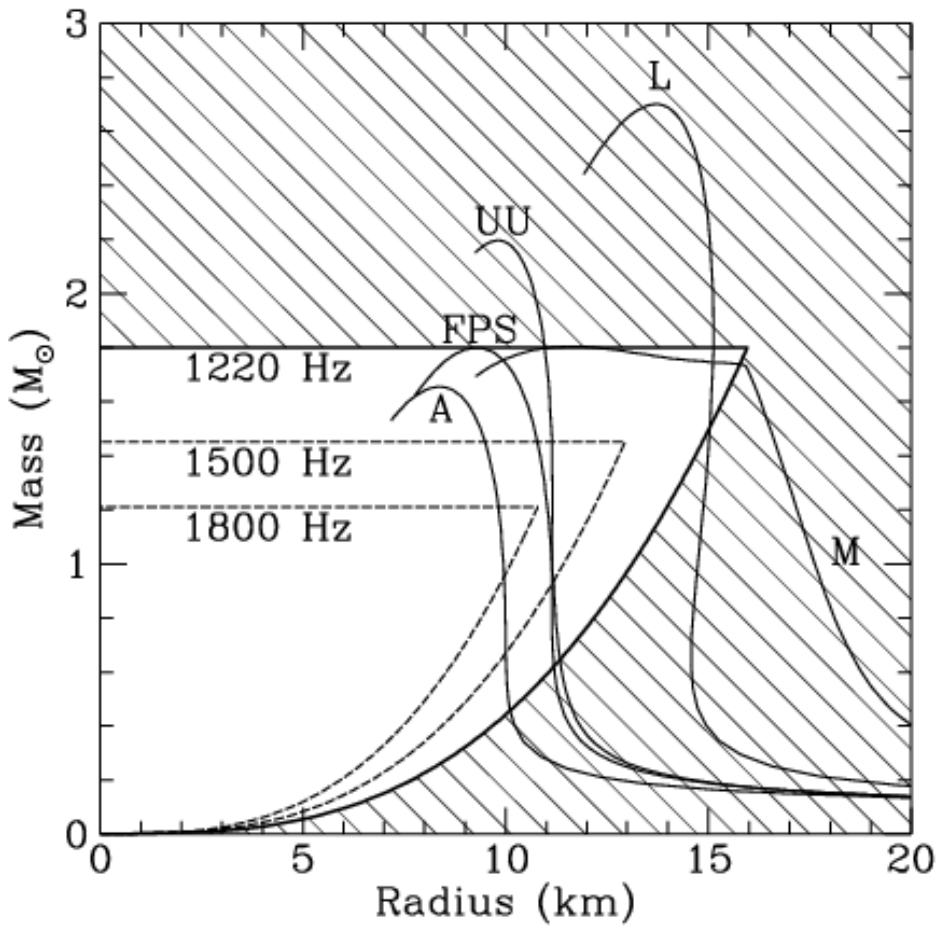}
\vspace{-1.0cm}
\caption{\textit{Left:} Power spectral density of the LMXB and Z source Sco X--1 showing a pair (lower and upper peak) of kHz QPOs. Adapted from \cite{vanderKlis_2006}.
\textit{Right:} Constraints on NS mass and radius from 
the observed maximum kHz QPO frequency (identified as the orbital, Keplerian frequency of the matter in the disk) as indicated; the hatched area is excluded if the maximum orbital frequency is 1220 Hz. Credit: \cite{Miller_1998} \textcopyright AAS. Reproduced with permission}
\label{fig:twin_kHz}
\end{figure}

At the end of 1995 the launch of RXTE, with its large collecting area ($\sim 6500$ cm$^2$) and optimal time resolution (up to $1\, \mu s$), allowed to extend in frequency the study of the fast-time variability in these sources. For NS systems, new quasi-periodic peaks at frequencies of hundreds of Hertz, called kilohertz QPOs (kHz QPOs), were discovered first in the brightest source, Sco X–1 \cite{vanderKlis_1996}, and then in many other NS LMXBs (to date one or two kHz QPO peaks have been observed in about 30 NS LMXBs, see \cite{Mendez_2021} for a recent and comprehensive review). These QPOs often occur in a pair (see Fig.\ \ref{fig:twin_kHz}, left panel), and these {\it twin peaks} usually move together in the frequency range from $\sim 200$ up to $\sim 1300$ Hz, in correlation with the source state. These represent the fastest variability ever observed from an astronomical object, and the minuscule time scale (less than a millisecond) of this variability indicates that kHz QPOs are produced in the accretion flow very close to the NS surface, giving the opportunity to study the dynamics of the accretion flow in these extreme conditions.

For a given source the lower and upper peak of the twin kHz QPOs are seen to move in frequency together, without changing their frequency separation too much. This frequency separation is normally found within $20\%$ of the NS spin frequency (for slow rotators, with spin frequency below $\sim 400$ Hz) or half of that (for fast rotators, with spin frequency above $\sim 400$ Hz). However, recent works have suggested that this frequency separation, $\Delta \nu$, is consistent with being independent of spin
\cite{Mendez_2007}. Given that the frequency of the upper kHz QPO resembles the Keplerian orbital frequency of a test particle at radii close to the radius of a $\sim 1.4\, M_\odot$ NS, the kHz QPOs were immediately associated to motion of matter at the innermost region of the accretion disc. If this were the correct interpretation, the observation of the highest frequency kHz QPO can reach might provide a strong constraint on the inner disk radius and hence on the NS mass and radius, i.e. on the EoS of ultra-dense matter. 
The highest observed kHz QPO frequency of about 1220 Hz might already provide interesting constraints (see Fig.\ \ref{fig:twin_kHz}, right panel). As one can see, 
the two stiffest EoS lie in the excluded regions. A kHz QPO of 1500 Hz would exclude many neutron star EoS \cite{Miller_1998}. 

\begin{figure}
\centering
\vspace{-1.8cm}
\includegraphics[width=0.7\textwidth, angle=270]{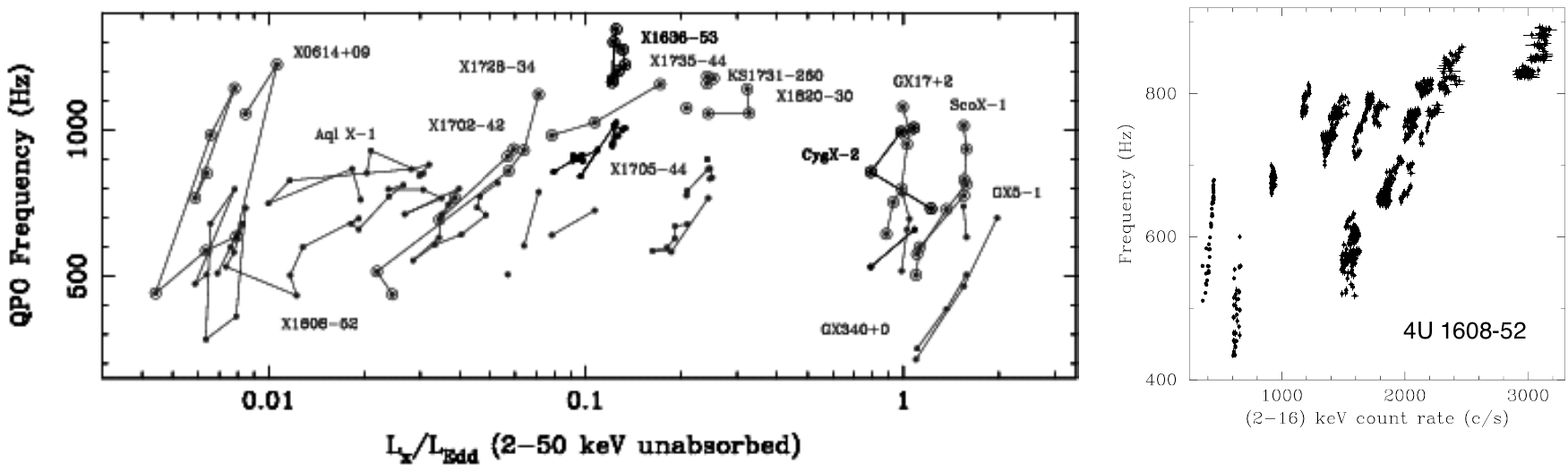}
\vspace{-1.6cm}
\caption{The parallel tracks phenomenon cross sources (left panel; upper and lower kHz peaks are indicated with different symbols), and in time for the atoll source 4U 1608–52 (right panel), where the lower peak frequency is plotted vs.\ the source count rate. Adapted from \cite{vanderKlis_2006} and references therein}
\label{fig:parlines}
\end{figure}

Since the X-ray luminosity is expected to increase with the mass accretion rate, $\dot m$, while the inner inner radius of the disk is expected to decrease, i.e.\ to approach the NS, it was expected that the QPO frequency would increase with X-ray intensity. While this was the case over short time intervals (a day or less), on longer time scales, and across sources, this correlation was more complex, with the QPO frequency tracing several, more or less parallel, tracks in a plot of QPO frequency vs.\ X-ray intensity. This kind of plots were then, indeed, called parallel tracks (see Fig.\ \ref{fig:parlines}). This may suggest that the frequency of the kHz QPOs depend not only on the X-ray luminosity but also on some other parameter, such as the magnetic field strength of the source or a particular component of the X-ray luminosity (excluding for instance the contribution given by nuclear burning on the NS surface or some radial inflow). Interestingly, the parallel tracks collapse into a single track (one for each kHz QPO) when the frequencies of the QPOs are plotted against the HC or the position along the CD, meaning that these frequencies depend on the source spectral state more than on its instantaneous X-ray luminosity. Furthermore, kHz QPO frequencies appear correlated with the characteristic frequencies of many narrow and broad timing features, including the LF-QPOs, whose frequencies also depend on the source spectral state. 

Other properties of the kHz QPOs regard their fractional rms amplitude and quality factor, defined as the ratio of the centroid frequency and the full width at half maximum of the QPO, $Q = \nu_0/\Delta$, which is a measure of the coherence of the QPO (features with $Q<2$ are considered peaked noise instead of QPOs). The rms amplitude increases with increasing the photon energy and, in similar energy bands, decreases with increasing the source luminosity. The coherence of the lower peak has a maximum at intermediate QPO frequencies whilst the quality factor of the upper peak increases with the frequency all the way to the highest detectable frequencies (\cite{vanderKlis_2006} and references therein). These properties make the kHz QPO more evident in intermediate states than at the extremes of the Z or atoll track. 

Other two features that may be present in the power spectra of these sources and that where discovered in the RXTE era are the so-called Hecto-Hertz (hHz) QPO and the milli-Hertz (mHz) QPO. The first feature is detected in atoll sources at about $\sim 100-200$ Hz and its frequency remains approximately constant when the source moves along the CD. It is probably linked to similar high-frequency QPOs detected in black-hole (BH) binary systems (see \cite{vanderKlis_2006} and references therein). The mHz QPO is observed in the frequency range $(5-15) \times 10^{-3}$ Hz only in a very particular X-ray luminosity range ($L_{2-20\, keV} \simeq (0.5-1.5) \times 10^{37}$ erg/s) and at soft energies ($<5$ keV), and is interpreted as possible variations in nuclear burning on the NS surface. In fact, in the atoll source 4U 1608–-52 the frequency of the kHz QPO was anti-correlated with the $2–5$ keV X-ray count rate variations during the mHz oscillation. This anti-correlation suggests that the inner edge of the accretion disc increases due to the additional flux produced in each mHz QPO cycle on the NS surface. These features are also linked to type-I bursts, since these usually disappear before the onset of a type-I burst. In the NS transient IGR J17480-–2446, these mHz QPOs appear as the result of a smooth evolution of the thermonuclear bursts properties; these observations have strengthened the unstable nuclear burning interpretation (see, however, \cite{Ferrigno_2017} for a different interpretation).

\begin{figure}
\centering
\includegraphics[width=0.7\textwidth, angle=90]{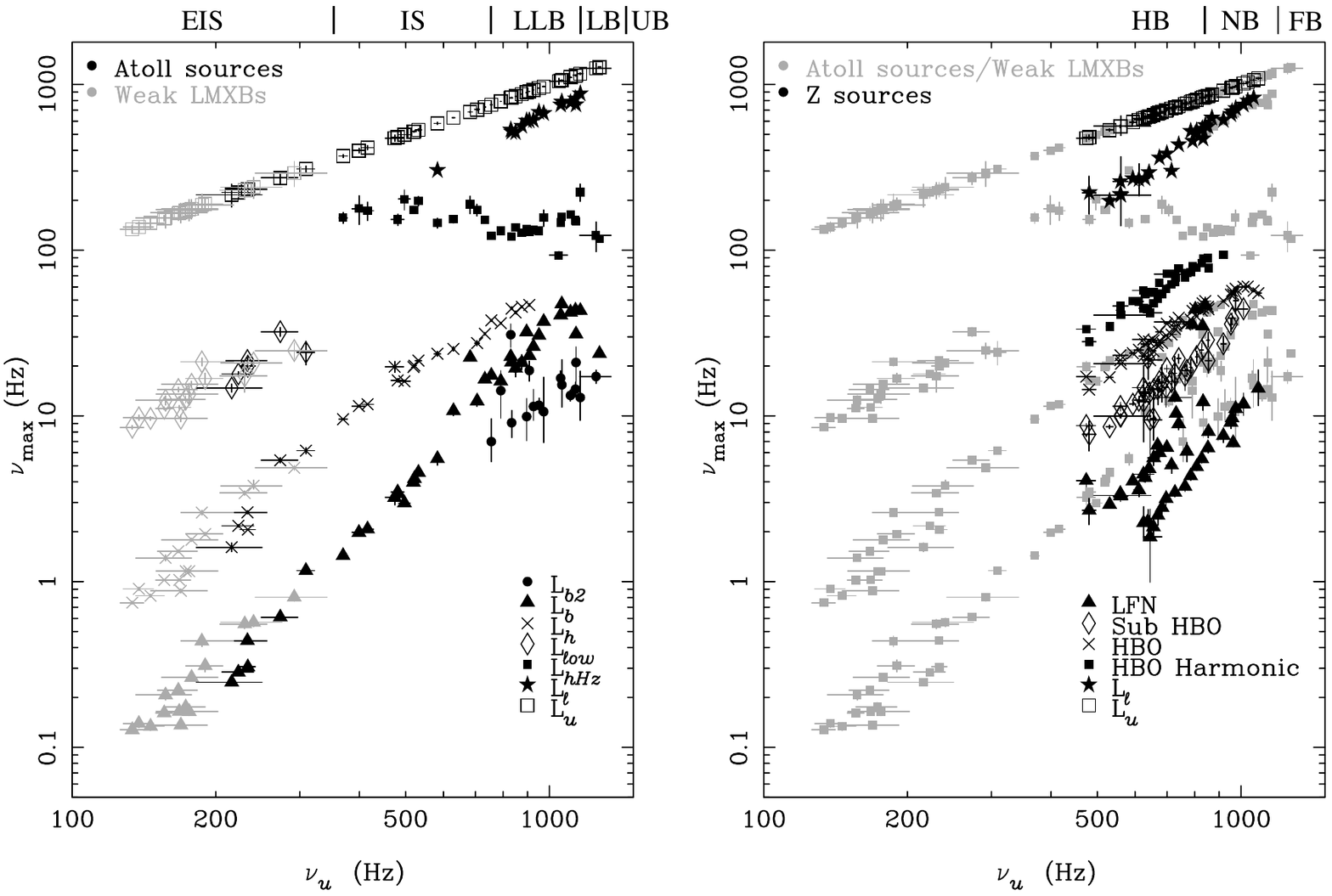}
\caption{Frequency of the most important PSD components plotted vs.\ the frequency of the upper kHz QPO for: \textit{Left:} Atoll sources and weak LMXBs, and \textit{Right:} Z sources compared with these objects. The characteristic frequencies $\nu_{max}$ of the components are plotted as indicated; approximate source state ranges are indicated at the top. Adapted from \cite{vanderKlis_2006} and references therein}
\label{fig:freq_corr}
\end{figure}

Both the low-frequency features and the kHz QPOs change their properties in a correlated way as the source moves along the CD. In particular, their frequency increases along the CD in the sense of increasing mass accretion rate, that is from the top to the bottom of the CD. In particular, the frequency of these features increases with increasing the frequency of the upper kHz QPO, as it is shown in Figure \ref{fig:freq_corr} for different source types. Note that sources covering an order of magnitude in luminosity when in the same state, yet display very similar power spectra and essentially the same frequency correlations. The only feature that does not show any correlation is the hHz QPO, whose frequency is quite stable in atoll sources, whilst this feature is absent in Z sources. The dependence of the frequency of the BLN and LF-QPOs on the frequency of the upper peak, $\nu_u$, is approximately quadratic in both Z and atoll sources, suggesting that Lense-Thirring precession of an orbit with frequency $\nu_u$ might be causing the LF-QPO (see below).

Although the phenomenology of kHz QPOs is very regular, repeatable and similar across different sources, there is no a universal consensus about its interpretation and modeling. 
The observation that the difference in frequency between the two kHz peaks was similar to the spin frequency, or half of it, led to propose a \textit{beat-frequency model}, according to which a blob of matter orbiting with a Keplerian frequency, identified with the upper-peak frequency, beating with the spin frequency of the NS, reproduces the frequency of the lower peak. However, this model had to be abandoned since it could not explain new data about the two kHz QPOs' frequency difference, that may not be related to the spin frequency of the NS.  

\begin{figure}
\centering
\includegraphics[width=1.0\textwidth, angle=0]{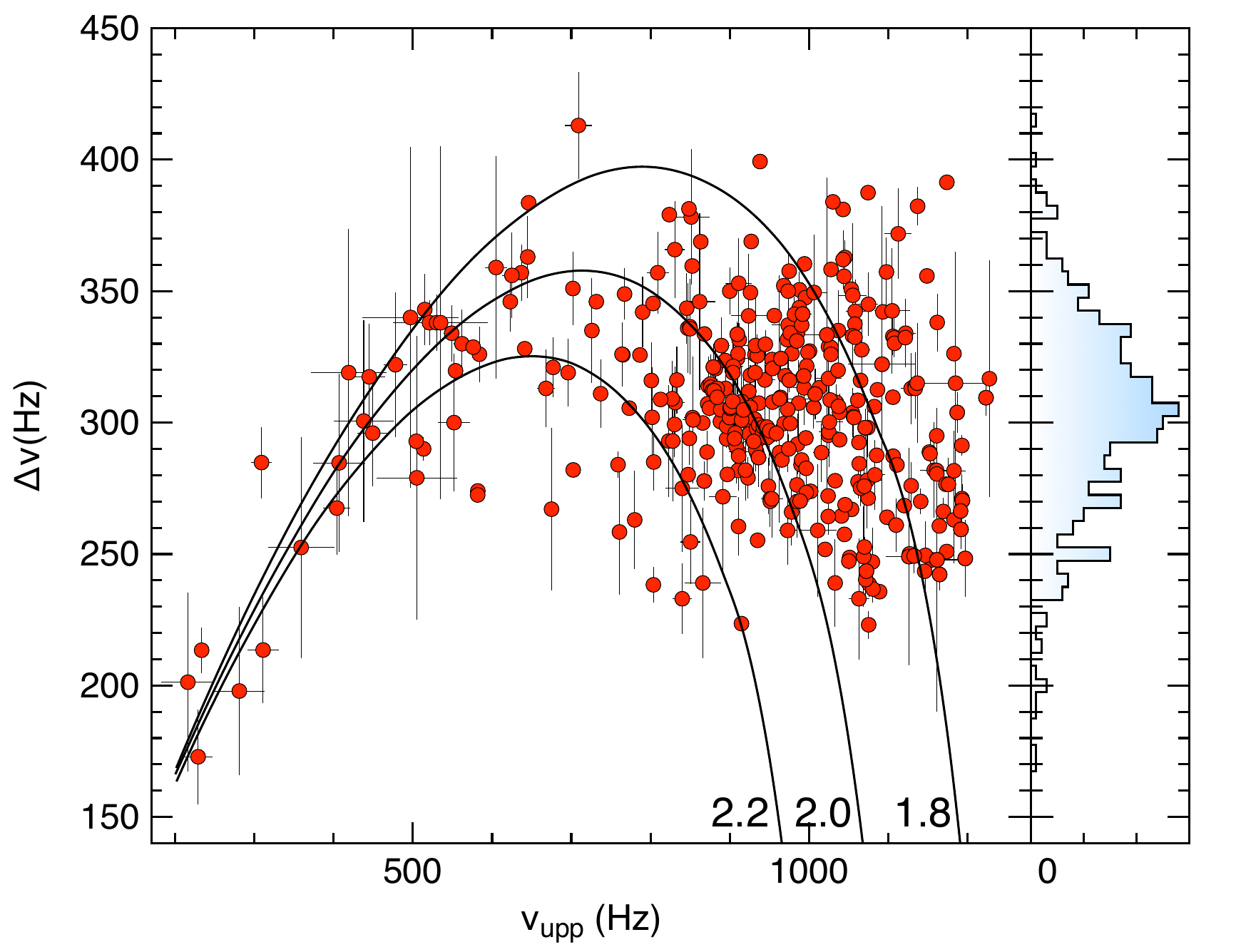}
\caption{$\Delta \nu$ vs.\ $\nu_{u}$ for all values published in the literature. The three curves correspond to the prediction of the relativistic-precession model for a mass of 1.8, 2.0 and 2.2 solar masses, respectively. On the right, the distribution of all $\Delta \nu$ values on
the Y axis of the plot on the left is shown. The distribution has a Gaussian-like shape with centroid $\sim 305$ Hz. Courtesy of T. Belloni; see \cite{Mendez_2021}, and references therein, for further details.}
\label{fig:stella_new}
\end{figure}

Many other models try to interpret the frequencies of the QPOs using general relativistic frequencies in the disk at some preferred radius and the spin frequency (or the beating between any two of them).
As an example, in the so-called \textit{relativistic precession model},
considering the periastron precession of the innermost parts of the accretion disc, the upper kHz QPO was identified as the orbital (Keplerian) frequency of a test particle at the inner edge of the disc, $\nu_u$, under the influence of the general relativistic potential of the NS. In this model, the lower kHz QPO would be the difference between this azimuthal and the epicyclic radial frequency, the so-called periastron precession frequency, at the same region in the disc. The difference between the azimuthal and the periastron precession frequency in the model changes generally in the same way as in the observations (Fig.\ \ref{fig:stella_new}; see, however, a discussion in \cite{Mendez_2021}), with no relation between the frequencies of the kHz QPOs and the spin of the NS. Moreover, the LF-QPO was identified with the nodal, or Lense-Thirring, precession of the orbit, predicting that the frequency of the first would be proportional to the square of $\nu_u$, as indeed is observed.
The relativistic-precession model is, in fact, an extension of the Lense-Thirring model that was proposed a year earlier to explain the correlation between the frequency of the upper kHz QPO and the LF-QPO in NS LMXBs. The Lense-Thirring and the relativistic-precession models became one consistent model for both the low- and the high-frequency variability \cite{stella_1998, stella_1999}.

Apart from the observed frequencies of kHz QPOs, the modulation and decoherence mechanisms should also be explained. Some of the possible explanations have been discussed in \cite{vanderKlis_2006} (and references therein). 
Note that, although the kHz QPO frequencies are plausibly determined by the characteristic disk frequencies, the modulation mechanism is likely associated to the high energy spectral component (e.g., the corona, or boundary layer between the disk and the NS). This is because the disk alone cannot explain the large observed amplitudes, especially at hard X-rays where the contribution of the disk is negligible \cite{Mendez_2006}. 

\subsection{X-ray spectral properties}

The spectroscopic study of the persistent (i.e.\ excluding type-I bursts, dips and eclipses which will be described in the next sections) X-ray spectrum of NS LMXBs is useful to understand the physical properties of the accretion flow close to the compact object, in the strong gravity regime, and its geometry. There are several components in a NS LMXB that can emit in the X-ray band or affect the propagation of photons, such as the accretion disk, an accretion disk corona or a cloud of energetic (thermal and/or non-thermal) electrons (at various locations in the system) acting as a Comptonization medium, the boundary layer, 
the NS surface itself, and possibly a jet. Because of the weak magnetic field of the NS, X-ray spectra of NS LMXBs are often very similar to those of BH X-ray binaries, with, however, an additional component given by the fact that the NS, contrary to a BH, has a solid surface. If the NS has an appreciable magnetic field, as in AMSPs, X-rays may be generated in the accretion column on top of the magnetic polar caps; in this case the accretion disk may be truncated by the NS magnetosphere and may not reach the NS surface. In general, based on the similarity of the spectra of BH and NS binaries, and on similarities in their timing properties, there has been a tendency to assume a common mechanism for X-ray emission, related to the properties of the accretion flow more than to the presence (or absence) of a solid surface for the compact object. 

The spectral properties of the X-ray emission of all the components mentioned above depend on the mass accretion rate. Indeed high and low rates of accretion, usually associated with softer and harder spectra, respectively, also lead to different source geometry and/or different dominant emission mechanisms. As an example, an accretion disk is always present, but likely extends close to the NS at high accretion rates, perhaps up to the NS surface or to the boundary layer. 
In this case the spectrum is expected to be dominated by soft thermal components originated in the accretion disk and the NS surface and/or boundary layer. During periods of low accretion, the inner disk may not emit enough to cool down the temperature of the hot corona via inverse Compton scattering ({\it Compton cooling}) resulting in harder spectra, with temperature up to a hundred keV, with a weak (if any) soft thermal emission.

\subsubsection{The continuum spectrum: an historical overview}
\label{ss:history}
In the mid-1980s, the X-ray spectra of NS LMXBs were generally described as the sum of a soft and a hard component and were interpreted in the framework of two models, dubbed the \textit{Eastern model} (also called the TENMA model), and the \textit{Western model} (or EXOSAT model, see e.g.\ \cite{Barret_2001} and references therein). The soft component of the Eastern model is a multi-color disk component approximating the emission from an optically thick, geometrically thin accretion disk. Given the local blackbody emissivity of the multi-temperature accretion disk, this spectrum consist of a power law with photon index $\Gamma \simeq -2/3$ up to an exponential cutoff corresponding to the temperature of the innermost (and hottest) disk region. The hard component is a weakly Comptonized blackbody component; Comptonization of seed photons emitted at the NS surface/boundary layer would take place in the inner parts of the accretion disk. In the Western model, the soft component is a single temperature blackbody, attributed to an optically thick boundary layer, whereas the hard one represents Comptonization taking place in a hot region at the inner part of the accretion disk.
Both the Western and the Eastern models were generally capable of providing good fits to the spectra of LMXBs.

In the 1990s, there was increasing evidence that a (saturated or unsaturated) Comptonization component was most of the time required to fit the observed persistent emission continuum spectrum, and an additional blackbody-like component was needed for more luminous and spectrally softer sources. A new model, the so-called \textit{Birmingham model}, closely related to the Western model, was proposed, in which the two continuum components were a simple blackbody from the NS plus Comptonized emission from an extended accretion disc corona (ADC) above the accretion disc \cite{Church_1995}, as suggested by observations of high-inclination sources during dips and eclipses (see e.g.\ \cite{Church_2004} for a discussion; see also Sec.\ \ref{ss:high-incl}).  
Also, the Eastern model could not adequately fit the spectra of some of the lower luminosity atoll sources, where the spectrum extended without any evidence for a break up to the highest measured energies ($\sim 100$ keV). Therefore the Eastern model was modified in order to include the effects of the Comptonization of the $\sim 1$ keV blackbody coming from the NS. In fact, the unsaturated Comptonized component, also present in Western/Birmingham model, can be approximated by a power-law with a high-energy cutoff; the power-law slope and cutoff energy are essentially determined by the Thomson depth and electron temperature of Comptonizing region. These parameters can vary over a wide enough range to fit the $1 - 20$ keV spectra of the hard state of atoll sources as well. In some low-luminosity atoll sources, the additional blackbody component is not required (see e.g.\ \cite{DiSalvo_2002} as a review). 

This (improved) spectral decomposition proved adequate for fitting the $1-20$ keV spectra of Z-sources; typical temperatures for the blackbody component are in the $1 - 2.5$ keV range both in the Western and Eastern model. The fraction of the $1 - 20$ keV luminosity in the blackbody component is $\leq 20\%$ in most cases. This value, however, is substantially lower than that expected from boundary layer emission ($\sim 50\%$, unless the NS rotates very close to mass shedding limit). 
Theoreticians have, in general, considered it natural that the Comptonizing region is located close to the NS, where most of the gravitational energy of the accreting matter is released, and so have supported the Eastern model.
Moreover, an extended ADC cannot be optically thick, as it is required in soft spectra dominated by saturated Comptonization, since it would hamper the detection of accretion disk features, that are often evident in the spectra of these sources (see below).

Luminosity related spectral changes were observed already by TENMA from the LMXB 4U 1608--52. With decreasing luminosity, the degree of Comptonization increased, and the X-ray spectrum approached a power-law shape. Deviations from a simple power law, appearing as a broad absorption edge between 8 and 10 keV, were first observed by GINGA in the spectrum of 4U 1608--52. These deviations could be interpreted as due to the reflection of the incident power law by a relatively cold medium (such as the accretion disk). Line emission centered between 6.4 (neutral Fe) and $6.7-6.9$ (Fe~XXV and Fe~XXVI, respectively) keV, with equivalent width of $70-130$ eV and Full Width Half Maximum (FWHM) of $\sim 1$ keV was also reported.

In the 2000s, significant advances have been possible thanks to the broad band spectral capabilities of BeppoSAX ($0.1-200$ keV) and RXTE ($3 - 100$ keV), followed by Chandra ($\sim 0.1-10$ keV) and XMM-Newton ($\sim 0.3-10$ keV) with enhanced spectroscopic capabilities in the classical X-ray range. Low-energy coverage (below $\sim 2$ keV) enabled to resolve the soft ($\leq 1$ keV) component of the spectra, simultaneous hard X-ray coverage could constrain the physical parameters of the hard Comptonization component. In addition, a good overlap region between the soft and hard X-ray bands was useful to reveal the Compton reflection component. Good spectral resolution below $\sim 10$ keV proved useful to resolve the  emission line and absorption edge features associated to neutral or ionized Iron, that produces the strongest discrete features in the classical X-ray range being one of the most abundant element with a large fluorescence yield.  

\subsubsection{Soft and hard spectral states}

As first noted by \cite{Paradijs_1994}, the spectral hardness over the $10-80$ keV energy range of NS LMBRBs (and accreting X-ray sources in general) appears to be higher at lower X-ray luminosity. Therefore the mass accretion rate appears to be the main parameter driving the spectral hardness. 
Hard and soft spectral states are usually observed both in persistent and in transient NS LMXBs. Transient sources usually behave as atolls during their X-ray outbursts, sharing similar spectral and timing properties. Only two sources to date have shown transitions between atoll and Z-state: XTE J1701--462 and the Terzan 5 transient IGR J17480--2446. 

\begin{figure}
\centering
\includegraphics[width=0.32\textwidth, angle=0]{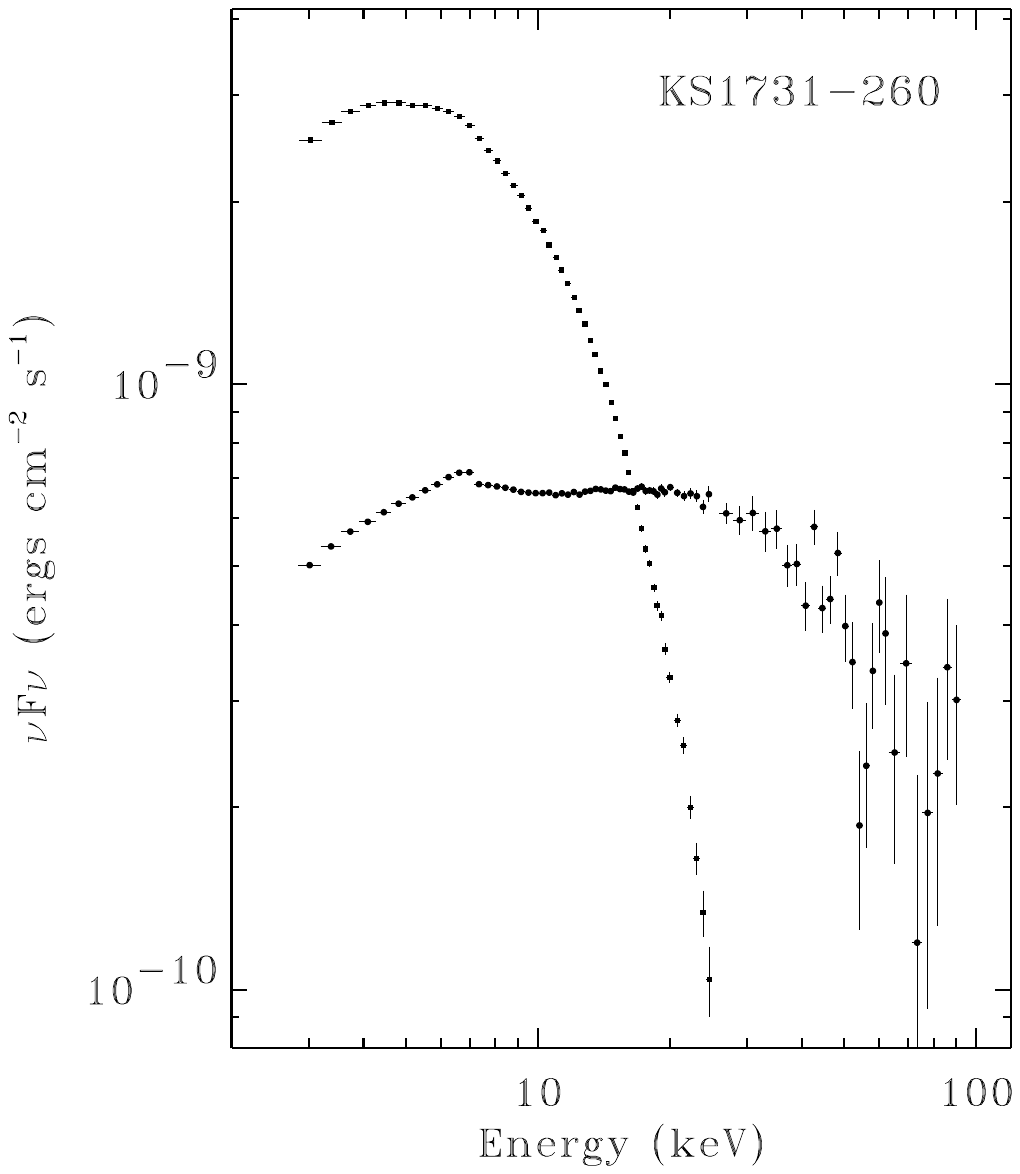}
\includegraphics[width=0.32\textwidth, angle=0]{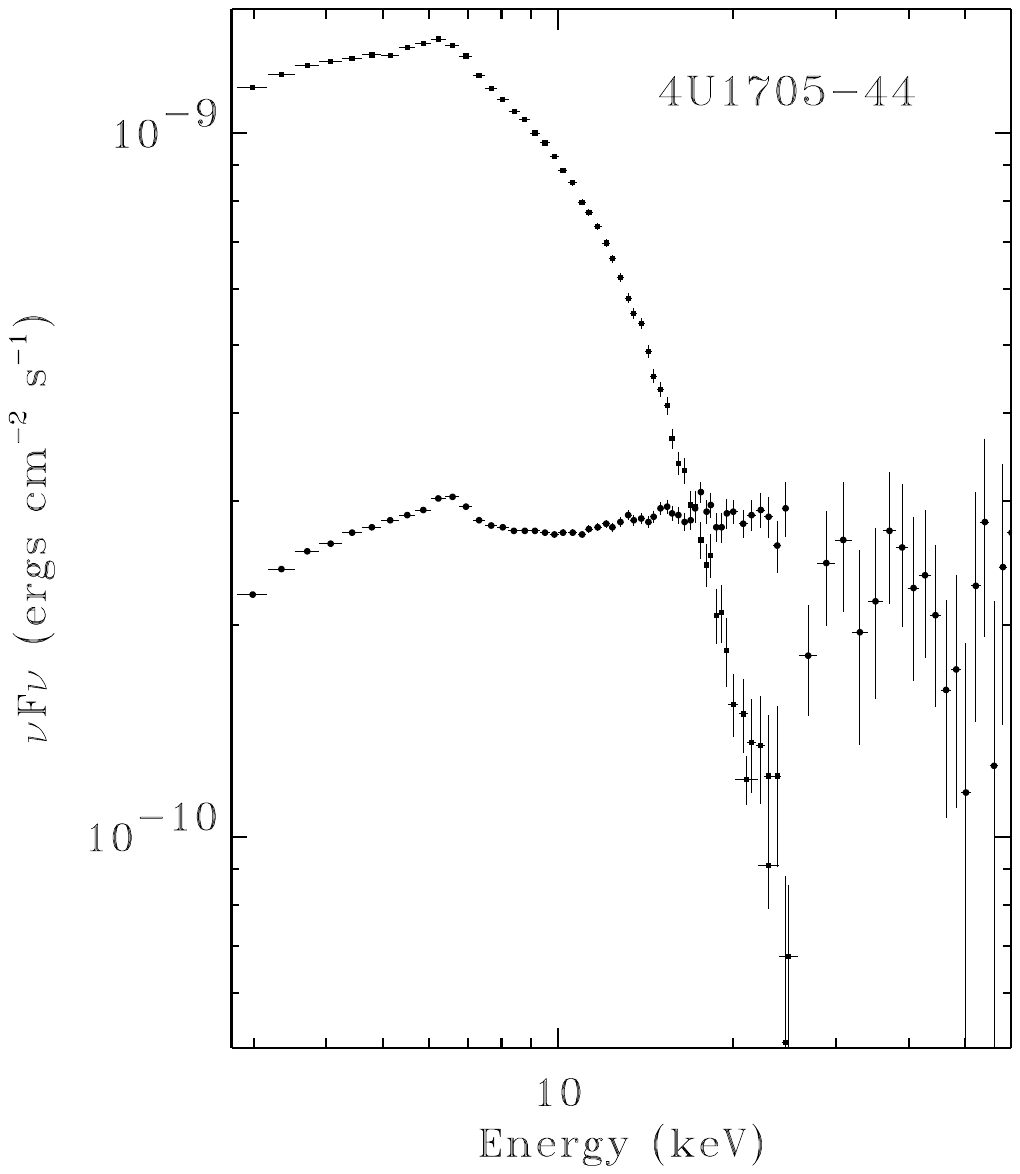}
\includegraphics[width=0.32\textwidth, angle=0]{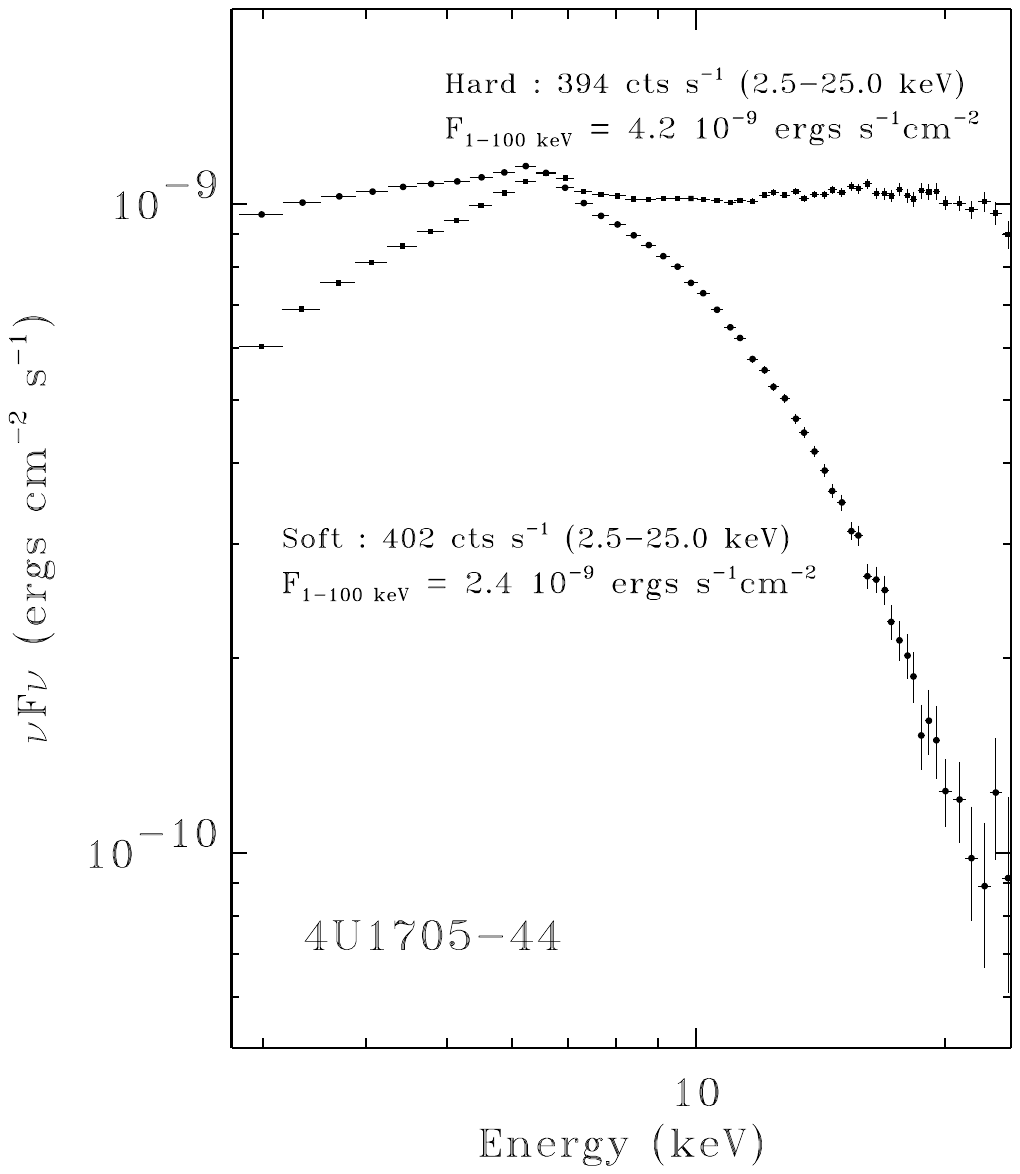}
\caption{Soft and hard spectral states from the LMXBs KS 1731--260 (left) and 4U 1705--44 (middle) as observed by RXTE (PCA and HEXTE spectra are combined). PCA spectra taken by RXTE from 4U 1705--44 during a spectral transition that occurred in February 1999 (right). The broad-band luminosity in the hard spectrum is about twice the one associated with the soft spectrum, because of the presence of a strong hard X-ray component. This illustrates that the X-ray count rate alone is not always a good indicator of the spectral state. Reprinted from \cite{Barret_2001} with permission from Elsevier}
\label{fig:states}
\end{figure}

Yet there is evidence that at least on occasions an additional parameter controls the soft/hard spectral transitions. This is especially apparent from a study of 4U 1705-–44, in which the source underwent a soft to hard state transition while the $0.1-200$ keV bolometric luminosity of the source decreased by a factor of $\sim 3$ from the soft to the hard state and increased by only a factor of $\sim 1.2$ in the opposite transition from the hard to the soft state (Fig.\ \ref{fig:states}, right panel). On another occasion the same source displayed hard and soft states which were found to differ in luminosity by a much larger factor, up to one order of magnitude (Fig.\ \ref{fig:states}, left and middle panels). It has been suggested that the second parameter regulating the spectral state transitions might be the truncation radius of the optically thick disc. However, what determines the radius at which the disc is truncated is unclear: this could be the mass accretion rate through the disk normalized by its own long-term average (as proposed to explain the \textit{parallel tracks} observed in the kHz QPO frequencies vs.\ X-ray flux diagram), but also magnetic fields or the formation of jets could play a role.

\subsubsection{Soft spectral states}

\begin{figure}
\centering
\includegraphics[width=0.4\textwidth, angle=0]{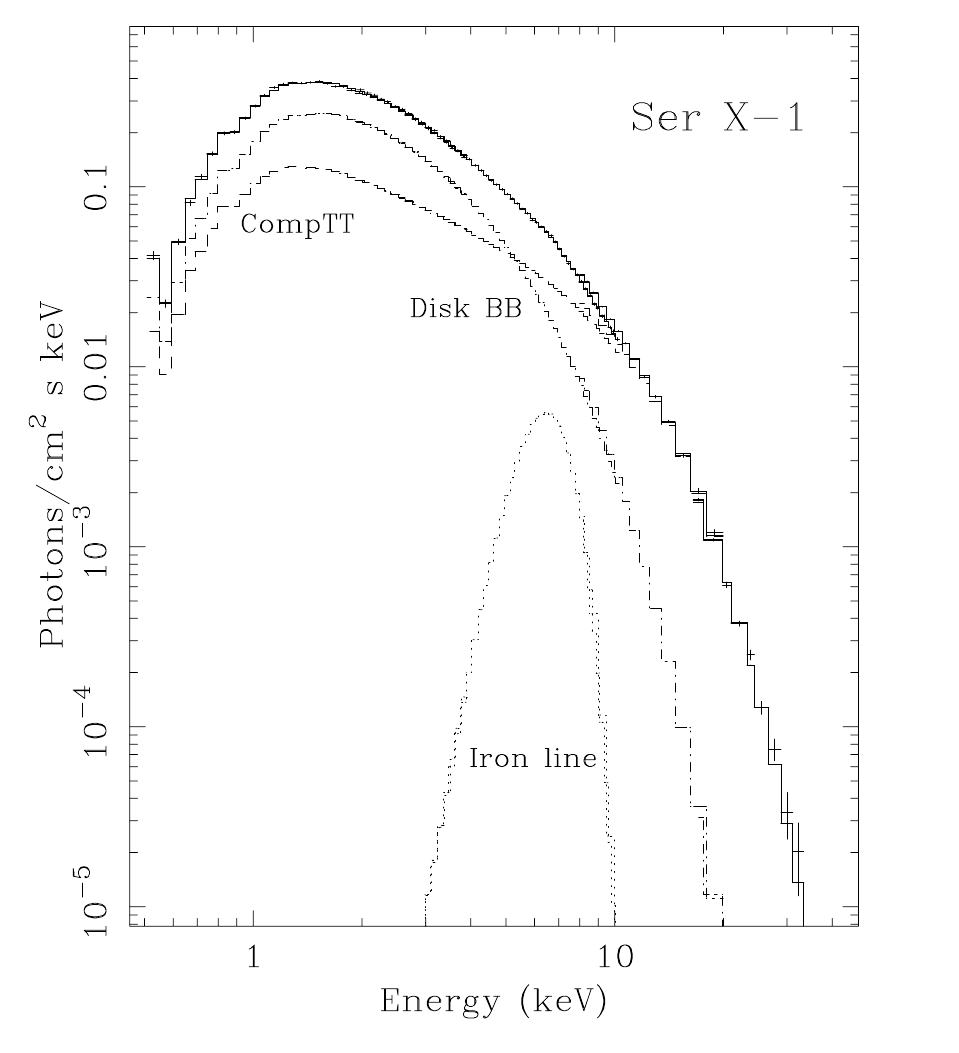}
\hspace{1cm}
\includegraphics[width=0.4\textwidth, angle=0]{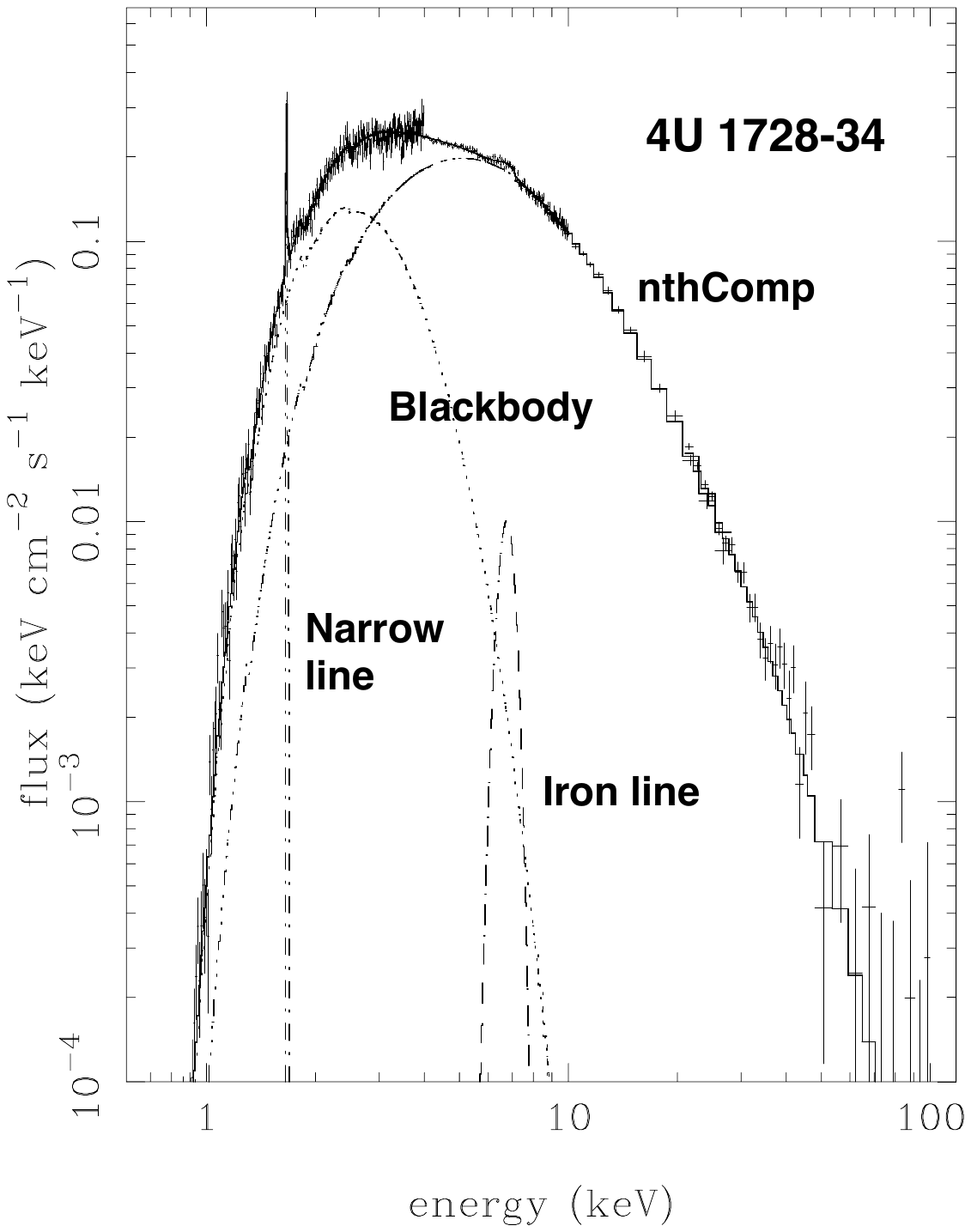}
\caption{A soft spectrum of Ser X--1 (left panel) and 4U 1728--34 (right panel) as measured by BeppoSAX. In both cases, the spectrum is the sum of a soft component (disk blackbody or blackbody) and a harder Comptonized component, plus a broad emission line at 6.4 keV and 6.7 keV for Ser X--1 and 4U 1728--34, respectively. An additional line at 1.7 keV is detected in 4U 1728--34. 
Reprinted from \cite{Barret_2001} with permission from Elsevier}
\label{fig:continuum}
\end{figure}

Soft spectra have been observed from both Z and atoll sources over a range of luminosity going from $\sim 10^{37}$ ergs/s to $\sim 10^{38}$ ergs/s. In the modern modelling these spectra are usually decomposed as the sum of a soft component and a saturated Comptonization component (see e.g.\ \cite{Barret_2001} as a review; see Fig.\ \ref{fig:continuum} showing a couple of examples). Despite the good sensitivity and spectral resolution at low energies, which resolves nicely the soft component, usually it is not possible to distinguish between a single temperature blackbody and a multicolor disk blackbody model. Blackbody temperatures of less than 2 keV are generally observed. For the multicolor disk model the typical range for the color temperature is $\sim 0.5$ to 1.5 keV. For the latter model, sometimes very small values of the projected inner disk radius, $R_{in} \cos i$ where $i$ is the inclination angle of the disk, are derived, typically a few kilometers. However, the inner disk radius measured in this way may underestimate the true inner disk radius \cite{Merloni_2000}.
The observed value has to be corrected by a spectral hardening factor, which varies with the accretion rate and the fraction of energy dissipated outside the disk. When corrected for an invariant spectral hardening factor of $\sim 1.7$ a plausible value of a few tens of km for the effective inner disk radius is usually obtained.  

For the Comptonization component, a temperature of a few keV and a relatively large optical depth of $\sim 5-15$ are observed. The Comptonization model also provides the characteristic temperature of the seed photons for the Comptonization. Seed photon temperatures range from 0.3 to 1.5 keV, i.e.\ the same range of temperatures measured for the thermal components. The ratio between the fluxes of the soft (blackbody or multicolor disk blackbody) and Comptonized components varies typically between 0.1 and 0.5, thus indicating that the soft component does not dominate the source luminosity. This has generally led to the interpretation that the soft component originates from an optically thick accretion disk, whereas the harder Comptonized component arises from a hot inner flow and/or a hot boundary layer with the seed photons coming from both the inner accretion disk and the NS surface. Clearly, this is a revisited version of the aforementioned Eastern model.

The (soft) photons, often observed in the X-ray spectrum as a thermal blackbody-like component, may provide information on the inner accretion flow. In fact, it is possible to infer the equivalent spherical radius, $R_W$, of the emission area of the seed-photon Wien spectrum from the bolometric source flux, $f_{bol}$, once the distance to the source, $D$, is known, and correcting for the energy gained by the photons in the inverse Compton scattering. The relative gain is given by the Comptonization parameter $y = (4 kT_e / m_e c^2) \max(\tau, \tau^2)$, then the radius is given by:
\begin{equation}
    R_W = 3 \times 10^4 D [f_{bol}/(1+y)]^{1/2} / (k T_W)^2 \; {\rm km},
\end{equation}
where $D$ is in kpc, $f_{bol}$ in erg cm$^{-2}$ s$^{-1}$, and $k T_W$ is the seed photon temperature in keV  \cite{intZand_1999}.

\begin{figure}
\centering
\includegraphics[width=0.5\textwidth, angle=0]{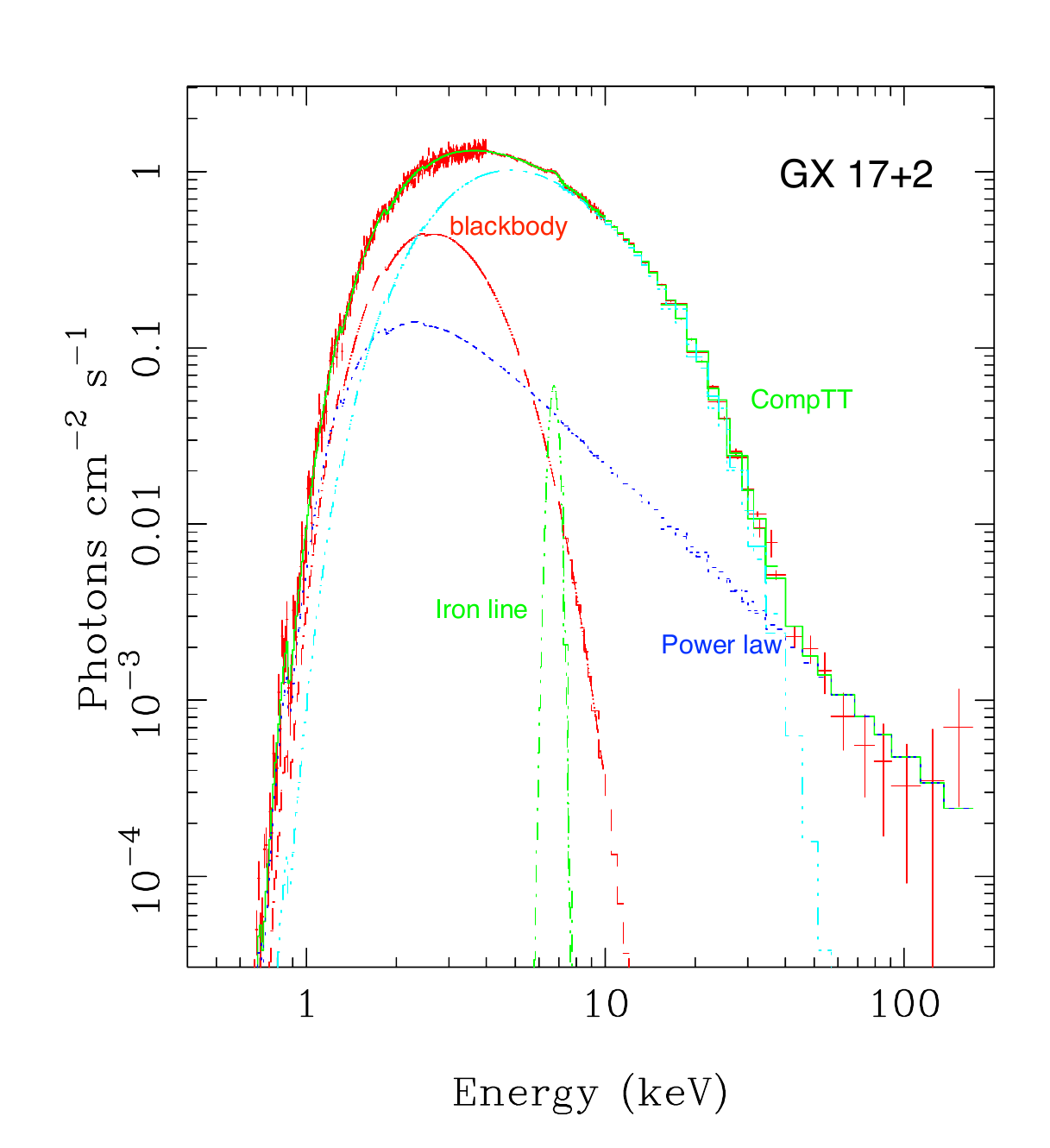}
\includegraphics[width=0.42\textwidth, angle=0]{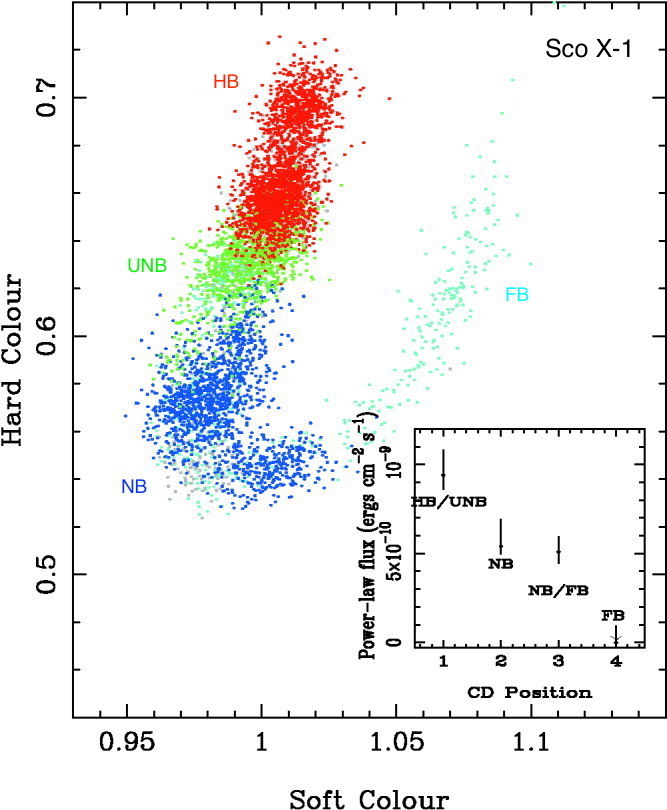}
\caption{Left: BeppoSAX photon spectrum of GX 17+2 in the HB. The solid line represents the best fit model; individual model components are also indicated. 
The hard power law, dominating the spectrum above 30 keV, is plotted as a dotted line. 
Right: Color‐color diagram of Sco X‐1 from PCA data during the simultaneous INTEGRAL/RXTE observation taken in 2003. 
The four colors indicate the four regions in which the CD has been divided, from which INTEGRAL spectra were extracted. Inset: Measured power‐law flux in the $20–200$ keV range plotted for each of the four CD‐resolved spectra.  
See \cite{DiSalvo_2002, DiSalvo_2006} and references therein}
\label{fig:GX17p2}
\end{figure}

Broad-band spectral studies have shown that many LMXBs in soft states display variable hard, power-law shaped components, dominating their spectra above $\sim 20-30$ keV. Power-law hard tails in Z-source spectra were occasionally detected in the past (e.g.\ \cite{DiSalvo_2002} and references therein), but large progress in the study of these hard tails
has been achieved through observations with BeppoSAX, RXTE and INTEGRAL. 
This hard component can be fitted by a power law, with photon index in the range $1.9-3.3$, contributing up to $\sim 10\%$ of the source bolometric luminosity, although this contribution may be different in different sources. 
In most of the cases the hard component becomes weaker when the source moves in the CD towards increasing the (inferred) mass accretion rates.
This was observed for the first time in a BeppoSAX 
observation of GX 17+2 (see Fig. \ref{fig:GX17p2}), where the hard tail was observed to vary systematically with the position of the source in the CD. In particular, the hard component (a power-law with photon index of $\sim 2.7$) showed the strongest intensity in the HB of the Z-track (see the corresponding spectrum in Fig. \ref{fig:GX17p2} left panel), and a factor of $\sim 20$ decrease was observed when the source moved to the NB. 
%

Observations of Sco X-1 performed with INTEGRAL and RXTE have shown a highly-significant (power-law) hard tail, with photon index $\sim 3$ and without any clear exponential cutoff up to $\sim 200$ keV, whose intensity appeared correlated with the position of the source in the CD, showing a phenomenology similar to GX 17+2 (Fig.\ \ref{fig:GX17p2}, right panel; see e.g.\ \cite{DiSalvo_2006} and references therein). These results were subsequently confirmed by a systematic study of all the RXTE archival data of Sco X-1 from 1997 to 2003, where the broad-band ($3-200$ keV) CD-resolved spectra of the source were fitted to a hybrid thermal/non-thermal Comptonization model, where the non-thermal fraction of the power injected in electron heating counts up to half of the total injected power \cite{Dai_2007}. 
These hard power-law tails are also often observed in the soft (banana) state of atoll sources (see e.g.\ \cite{Pintore_2016} and references therein). 

The origin of these hard tails is currently debated. A thermal origin of these components is unlikely; since a high energy cutoff is not observed up to energies of $\sim 100$ keV or higher, this would imply extremely high electron temperatures. In the case of bright LMXBs these high temperatures would be difficult to explain considering that the most of the source emission is very soft. These components may be related to the non-thermal power-laws observed in soft spectral states of BH binaries. The analogy between the hard X-ray spectral components of NS and BH binaries suggests a similar emission mechanism and geometry in all these systems. 
However, in bright NSs, the luminosity in the tail remains a small fraction of the source luminosity, whilst this is not always the case for BHs. This could again be related to the additional Compton cooling from the NS surface/boundary layer emission. 

Some scenarios have been proposed in order to explain these features; they can be either produced by non-thermal Comptonization off relativistic electrons in a local outflow or in a hybrid thermal/non-thermal corona (i.e.\ a corona in which a fraction of the energy can be injected in form of electrons with a non-thermal velocity distribution), or by bulk motion of accreting material close to the NS. However, it should be noted that at high accretion rate radiation pressure due to emission from NS surface can slow down the bulk motion of matter causing quenching of bulk Comptonization. Another scenario suggests an origin from synchrotron emission from relativistic electrons in a jet escaping from the system and, in this case, one would expect to observe a correlation with radio emission from these systems, that has been indeed observed in a couple of cases (see \cite{Migliari_2007} and references therein). More in general, radio emission from bright LMXBs, which probably arises in jets, is anticorrelated with the inferred mass accretion rate. This seems to be a fairly general behaviour, holding for different kinds of accreting compact objects, such as BHs, Z-sources, as well as atoll sources (although with weaker radio emission). Future simultaneous multiwavelength observations will be very useful to further investigate the nature of such components.

\subsubsection{Hard spectral states}

While Z sources are always bright, with X-ray luminosity normally above $\sim 10^{37}$ erg/s, and show soft spectra, atoll sources experience much larger variation of their X-ray luminosity, up to more than one order of magnitude when the source moves along its CD. A dramatic spectral transition occurs at a luminosity of $(1-3) \times 10^{37}$ erg/s: above this luminosity the high-energy cutoff of the Comptonized emission in all sources is at a few keV. The NS and the Corona are thought o be in thermal equilibrium at $\sim 1-2$ keV. 
Below this luminosity, the high-energy cutoff gradually increases towards 100 keV or more causing the hardness of the IS. Thermal equilibrium is lost, the corona becomes much hotter than the NS via an additional coronal heating mechanism; these are the hard spectral states of atoll sources (see some examples in Table~1 of \cite{DiSalvo_2002}).
In this state, the spectrum is dominated by a hard (unsaturated) thermal Comptonization component, often approximated with a power law with photon index in the range $\Gamma \sim 1.5 - 2$, and a high-energy cutoff. The inferred optical depth of the Comptonizing cloud is now a few ($\sim 2-3$ for a spherical geometry). The optical depth of the Comptonization component is related to the photon index and electron temperature via the following equation (e.g.\ \cite{Zdziarski_1996} and references therein):
\begin{equation}
    \Gamma = \left[\frac{9}{4} + \frac{1}{(kT_e/m_e c^2) \tau (1+\tau/3)}\right]^{1/2} -  \frac{1}{2}
\end{equation}

There are sources that appear to spend most of the time in this state (e.g.\ 4U 0614+091). In others a gradual transition from the soft to the hard state has been observed in response to a decrease of the source X-ray luminosity and/or the source drifting from the banana to the islands (see few examples of hard and soft spectra in Fig.\ \ref{fig:states}). 
The electron temperature of the corona in NS LMXBs seems to be lower than the typical values observed from BH systems. This has been interpreted as the signature of the NS surface acting as a thermostat for the Comptonization region. However, in a few cases so far, no clear high-energy cutoffs were observed in the hard X-ray spectrum (e.g.\ Aql X-1 or 4U 0614+09). These non attenuated power laws have photon index in the range $2.0-2.5$.

At low energies, the weak thermal component, contributing modestly to the source luminosity (about $10-20\%$ of the total) is consistent with the accretion disk being cooler than in the soft state and possibly truncated farther from the compact object. Evaporation of the inner parts of the disk, or some advection-dominated accretion flow, could explain this geometry but it remains a hypothesis to date. Sometimes the soft component is not detected, possibly because it is absent, or too faint, or because the bulk of its emission is radiated below the observed energy range.
For NS binaries, in addition to the disk, thermalized emission from the NS surface can also supply seed photons. Note that both components (i.e.\ their unscattered fraction) may be directly visible in the X-ray spectrum. 
In the hard spectral state, the disk does not contribute much to the X-ray spectrum, and the seed photons are predominantly provided by the NS surface (limited to a hot spot in the case of accreting millisecond pulsars in which the inner flow is channeled by the NS magnetic field, as in the case of SAX J1808.4-3658). A fraction of the NS (or hot spot) emission may be visible through the scattering cloud which usually has a moderate optical depth.

On the other hand, in the soft spectral state, seed photons may be still provided by the NS surface but the contribution from the accretion disk is no longer negligible. Only the unscattered disk emission is normally directly observed, as the contribution from the NS surface is hidden by the Comptonizing cloud of increased optical depth. The changing optical thickness of the corona/boundary layer, the relative strength of its unscattered emission and the contribution from the disk may explain the complex spectral evolution of NSs, as tracked by their X-ray color-color diagrams (\cite{Barret_2004} and references therein). 
In both soft and hard spectral states of NSs, the Comptonized component carries out most of the energy and dominates over the disk contribution. The site of Comptonization must therefore involve the boundary layer, in which the kinetic energy of the accreted material must eventually be released. In general relativity, it is known that the luminosity of the boundary layer can exceed that of the disk by about a factor of 2. In the hard state, the observations suggest that the boundary layer is optically thin and probably smoothly merges with the hot inner flow.

Three classes of models are currently competing to explain hard state spectra (see \cite{Barret_2004} as a review and references therein), of which the first one is probably the most popular today. The first one, is the so-called disk-spheroid model, in which a hot inner flow is surrounded, with a partial overlap, by a cool accretion disk acting as a source of soft seed photons (together with the NS surface). The hot inner flow may be a two-temperature plasma, with the protons much hotter than the electrons. The two-temperature plasma cools either by Compton scattering or by advection. In the second class of models, electrons are stored in magnetic flares above a non truncated accretion disk. These flares could be generated by the magneto-hydrodynamic dynamo responsible for the disk viscosity. The third one takes advantage from the fact that hard states are usually associated with compact and quasi-steady radio jets. It has been suggested that the X-rays could arise directly from (synchrotron or inverse Compton) emission at the base of the jet.

\subsubsection{The reflection component}
\label{ss:reflection}

Below the continuum, a reflection component, in the characteristic shape of a broad bump (the so-called Compton hump) centered around energies of a few tens of keV, is usually detected, and comprises broad and skewed features at the energy of the most abundant elements, and in particular of the Iron K-shell fluorescence line and edge. The broadness of these features is caused by the  relativistic smearing introduced by the fast orbital motion of the matter in the disk which is responsible of the reflection component.
This reflection component offers the possibility to constrain the physical properties and geometry of the accretion flow in the immediate vicinity of the NS, since it depends on the inclination of the disk with respect to our line of sight, the inner and outer radii of the reflecting region of the disk, its ionization and element abundances, its emissivity law, expressed as a power-law radial dependence, $r^{-\alpha}$, where $\alpha$ is normally between $\sim 2$ (for an irradiation-dominated disk) and $\sim 3$ (close to the intrinsic disk emissivity law, see \cite{Fabian_1989}). The strength of the reflection component also depends on the so-called reflection amplitude (in most models defined as the solid angle, $\Omega/2 \pi$, subtended by the reflector as seen from the source of the illuminating continuum); the typical value for a spherical inner corona and an outer accretion disk is $\Omega/2 \pi \sim 0.3$, whilst a plane-parallel geometry of the corona over the disk would correspond to $\Omega/2 \pi \sim 1$. 

The parameters of the reflection component and associated features can be constrained from spectral fitting, but requires a precise knowledge of the continuum spectrum (hence high-quality broad-band spectra).
This reflection component has now been observed in the spectra of many NS systems, both in soft and hard spectral states. However, it is much easier to detect the reflection continuum in hard state spectra, whilst in soft state spectra the Compton hump is overwhelmed by the soft thermal Comptonization continuum, the reflection component being instead dominated by a strong Iron line, and sometimes broad low-energy discrete features are also visible. In the hard state, it is generally found that the reflector is mostly neutral, subtends a small solid angle, much less than $2 \pi$ (thus excluding a static plane-parallel geometry). Also the amount of relativistic smearing is much less than what would be expected if the reflector extended down to the last stable orbit or the NS surface. 

On the other hand, in soft states, the properties of the reflection component (e.g.\ the increased smearing) suggest that the reflector extends closer to the central object. Note that a correlation between the strength of the reflection component and the photon index of the illuminating Comptonization continuum has been reported, meaning that softer spectra are associated with more reflection. A similar correlation has been found in Seyferts AGNs, Galactic BH systems and the X-ray bursters GS 1826--238 and 4U 1608--522, thus suggesting that a similar accretion geometry occur in a wide range of systems, independently of the nature and mass of the compact object \cite{Zdziarski_1999}. This correlation can be interpreted with the reflecting medium (i.e.\ the accretion disk) playing a dominant role as a source of seed photons for the Comptonization in the irradiating source; when the disk moves in, the solid angle it subtends to the hard source increases, leading to more reflection and more cooling, hence softer spectra (see e.g.\ \cite{Barret_2004} and references therein for a detailed discussion).

The most evident feature of the reflection component is a broad and asymmetric emission line usually found at 6.4 (Fe~I) or 6.7 (Fe~XXV) keV. These features are identified with K-shell radiative transitions of Iron at different ionization stages. Sometimes a smeared absorption edge at $7-8.8$ keV (Fe~I -- Fe~XXV, respectively) has also been detected. High-resolution Chandra spectra of these sources have shown that these features are intrinsically broad and cannot be the result of a blending of lines (see e.g. \cite{DiSalvo_2006} and references therein). The large width of these lines is therefore ascribed to the combination of relativistic Doppler effects, caused by the high orbital velocity of the matter in the disk, and gravitational effects due to the strong field in the vicinity of the NS surface, that smears the features produced by reflection in the accretion disk. As the matter in the inner accretion disk rotates fast close to the compact object, Doppler effects broadens the line and makes it double-peaked, special relativistic beaming makes it asymmetric, and the gravitational redshift shifts it towards lower energies.

\begin{figure}
\centering
\includegraphics[width=1.0\textwidth, angle=0]{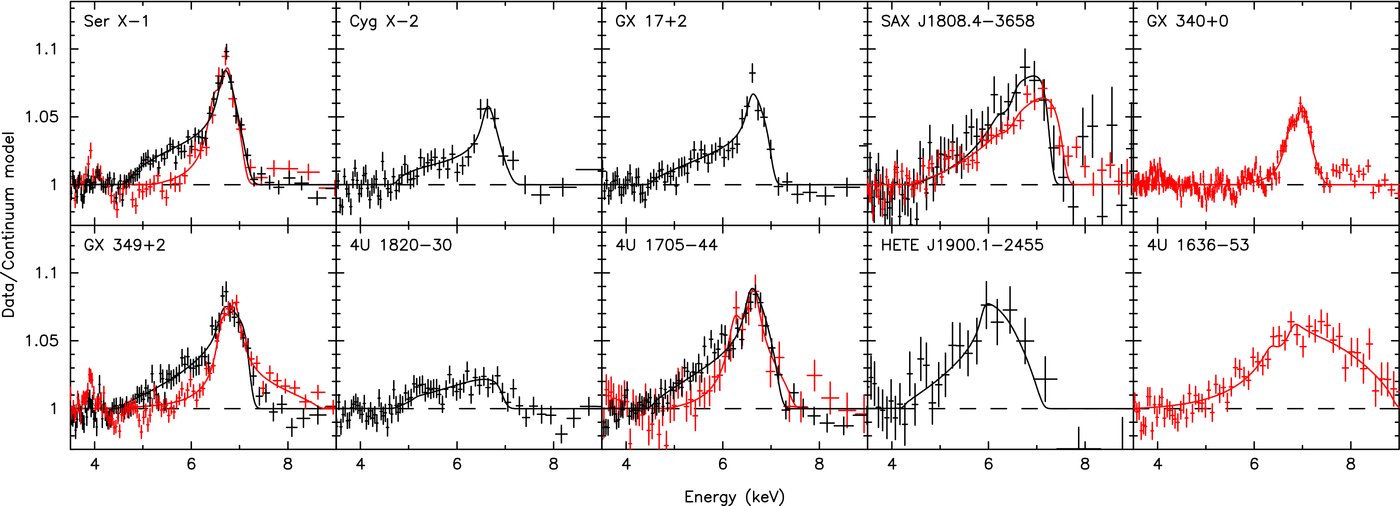}
\caption{Examples of the Fe K emission lines in NS LMXBs. Plotted is the ratio of the data to the continuum model. The solid line shows the best-fitting {\it diskline} model. Data from Suzaku are shown in black (combined front-illuminated detectors), data from XMM-Newton are shown in red (PN camera). Credit: \cite{Cackett_2010} \textcopyright AAS. Reproduced with permission}
\label{fig:Felines}
\end{figure}

As of now, the inner disk origin of broad iron line has been confirmed for most of the NS LMXBs (see e.g.\ \cite{Cackett_2010} and references therein). A few examples of Iron line profiles observed with XMM-Newton and Suzaku are shown in Fig.\ \ref{fig:Felines}. More recently, smeared Iron lines in several NS LMXBs have also been detected or confirmed with broad-band spectra obtained with NuSTAR (see e.g.\ \cite{Ludlam_2017} and references therein). 
Indeed only a few sources have shown no sign of reflection features to date, as for example the Z source GX 5--1 and the bright atoll source GX 9+9. Possible explanations discussed in literature are a large ionization of the disk, that increases the amount of Compton broadening of the line making it hard to detect, or a relatively large inclination angle of the system coupled with a relatively small reflection amplitude and/or strong Comptonization of the reflection features, that may produce a weak reflection component overwhelmed by the strong illuminating continuum. In the latter case, higher statistics spectra, that can be obtained with large-area observatories such as eXTP, would be able to reveal weak reflection components if present.

\begin{figure}
\vspace{-1.5cm}
\centering
\includegraphics[width=0.8\textwidth, angle=90]{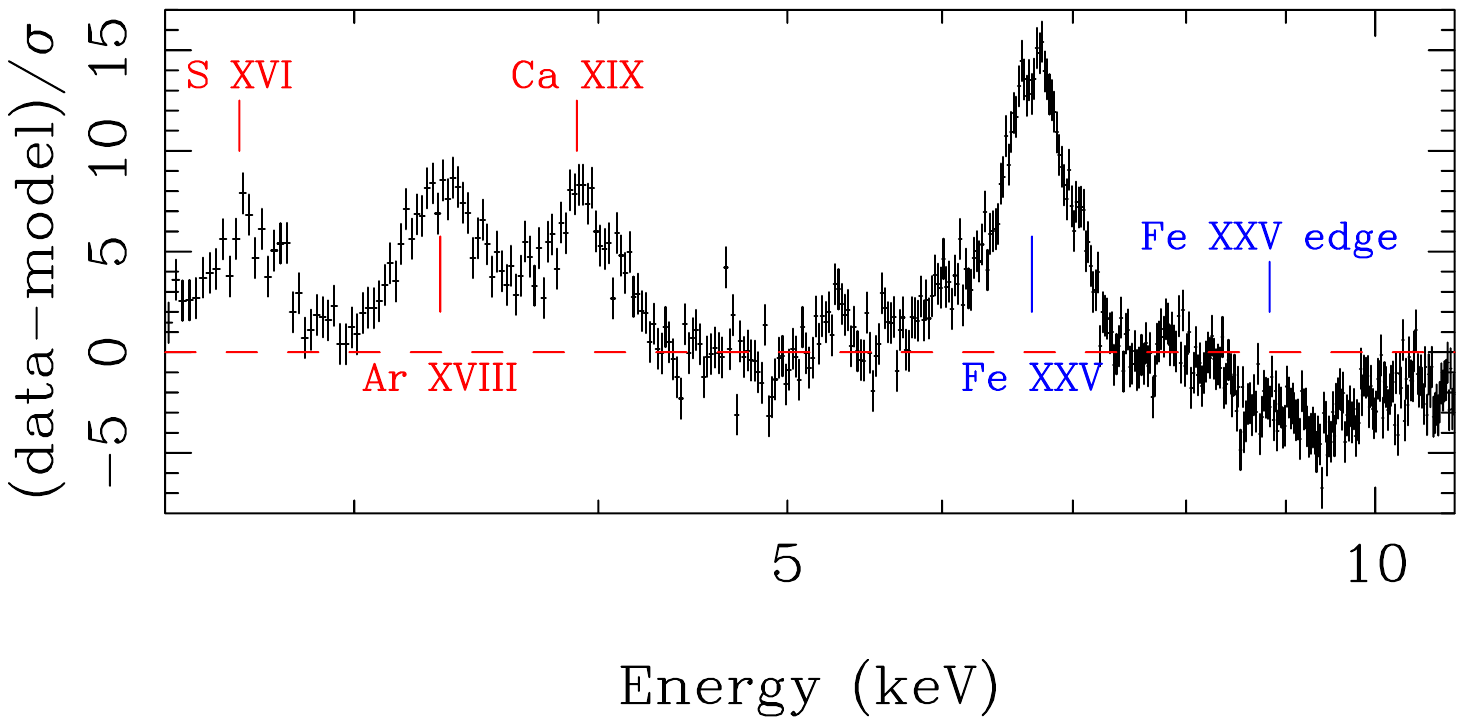}
\vspace{-1.5cm}
\caption{Residuals in units of $\sigma$ of the $2–10$ keV XMM-Newton/pn spectrum of 4U 1705--44 with respect to the continuum model. The iron line profile, which appears to be broad and double peaked, is clearly visible in the residuals, at more than $10\, \sigma$ above the continuum model. Other discrete features are also visible in the residuals, indicated by vertical lines that are placed at 2.62 ($K\alpha$ transition of S XVI), 3.32 ($K\alpha$ transition of Ar XVIII), 3.90 ($K\alpha$ transition of Ca XIX), 6.67 ($K\alpha$ transition of Fe XXV) and 8.828 (Fe XXV smeared edge) keV, respectively. All these discrete feature appear intrinsically broad and their profiles are compatible with being produced by reflection in the same disk region where the Iron line is produced. From \cite{DiSalvo_2009}}
\label{fig:4U1705}
\end{figure}

An example of a clearly asymmetric Iron line and other broad discrete features is given by the XMM-Newton spectrum of 4U 1705--44, that is shown in Figure \ref{fig:4U1705} (see \cite{DiSalvo_2009}, see also \cite{DiSalvo_2006} and references therein reporting on the Chandra high-resolution spectrum of 4U 1705--44). The whole reflection component in 4U 1705--44 can be well described by a self-consistent reflection model allowing for a slight overabundance of Ar, Ca and Fe by a factor $1.5-2.5$ with respect to their solar abundance \cite{Egron_2013}. 
Fitting the Iron line profile to a relativistic model can therefore give valuable information on the inner accretion disk and can constrain the inner disk radius and inclination to the line of sight. We have mentioned that kHz QPOs' frequencies are likely associated with the Keplerian frequency at the inner accretion disk. If this is true, then the relativistic Iron line and the kHz QPOs are possibly physically related, and their simultaneous observation and joint analysis may, not only shed light on the physics of these features, but also be useful to constrain the NS parameters. For example, if the upper kHz QPO frequency ($\nu_u$) can be identified with the orbital frequency at the inner edge of the accretion disk, and thus from a radius close to the radius $R_{in}$ at which the Iron line is produced, then the NS mass can be inferred from the relation: $\nu_u = \sqrt{G M / R_{in}^3}$, where $R_{in}$ is inferred from the line profile fitting and is given in units of the gravitational radius, $Rg = G M / c^2$.

An alternative location of the line emitting region may be the corona itself. In this case, the large width of the line may be attributed to Compton broadening, i.e.\ thermal Comptonization of the line photons in the corona. This produces a genuinely broad Gaussian distribution of line photons with $\sigma > E_{Fe} (k T_e / m_e c^2)^{1/2}$, where $E_{Fe}$ is the centroid energy of the line, and $k T_e$ is the electron temperature in the corona (see \cite{DiSalvo_2006} and references therein). This mechanism can explain the observed width of the Iron line for temperatures of the emitting region of few keV (assuming a moderate optical depth of a few). Note that this mechanism cannot work for the hard state of LMXBs, where the temperature in the corona is tens of keV. Note also that in most self-consistent reflection models available today, the effects of Compton broadening are taken into account, nevertheless relativistic smearing is usually required.

\subsection{Bursting sources}\label{ss:bursts}

NS LMXBs often display sudden episodes of intense activity, during which the X-ray luminosity increases by several orders of magnitude in very short time-scales, i.e. typically a few seconds or less, and then decays in a quasi-exponential fashion: these phenomena are called type-I X-ray bursts. The name distinguishes them from the rather less common type-II X-ray bursts, which are likely due to spasmodic episodes of accretion and are not covered in this review so that in the following we will simply refer to type-I X-ray bursts as bursts. They occur when the accreted material piled up onto the NS surface undergoes unstable ignition and triggers a thermonuclear flash. The observation of bursts is considered an unambiguous proof of the NS nature of the primary star in a X-ray binary. Indeed, accreting BHs can not produce bursts, since they do not have a solid surface. Furthermore, bursts spectral emission can be quite satisfactorily described by blackbody spectrum coming from a spherical region of typically $1-10$ km radius, i.e.\ consistent with being radiated by the NS surface or a portion of it. 

Since the observation of the first burst in 1969 in the NS LMXB Cen X-4, thousands of bursts have been detected from over a hundred of NS LMXBs (see e.g.\ \cite{Galloway_2020} for recent reviews). They appear in all classes of NS LMXBs, from atolls to Z sources, from pulsating to non-pulsating sources, from slowly rotating NS systems to AMSPs. Furthermore, bursts have been observed both in hard and soft states. Despite their ubiquity, bursts physical properties, such as duration, intensity, recurrence time and spectral energy distribution, are far from being homogeneous, as they are influenced by, e.g., the chemical composition of the transferred material, the mass-accretion rate and the spectral state they occur in. Subsequently, studying the X-ray emission from these phenomena has provided, for instance, information on the nature of the donor in these systems, estimates of the mass, radius and spin period of the accreting NS, constraints on the distance to the source.

\subsubsection{Observational properties of bursts}

Bursts light curves are typically described with a Fast Rise Exponential Decay (FRED) profile, since they show an initial, very rapid phase (time scale of seconds) where the flux increases linearly followed by a slower, quasi-exponential relaxation lasting tens or hundreds of seconds. Shorter, i.e. flashes of the total duration of tens of seconds, and longer bursts have been observed, even if the latter ones are far less common (about 1\% of the total \cite{Galloway_2017}). In particular, we distinguish between intermediate bursts, lasting from minutes to tens of minutes, and the very peculiar superbursts, which take hours to reach completion. 

The duration of these bursts is associated to the chemical composition of the ignited material. Standard FRED bursts are indeed fueled by accretion of mixed H/He material. In case of H-poor and He-rich material, the duration tends to be sensibly reduced since hydrogen burning requires beta decays, which in turn imply longer time-scales. Intermediate bursts arise from the ignition of a particularly thick, and therefore time-consuming, layer of He formed from accumulation of H-poor material at low mass-accretion rate. Finally, unstable burning of C in the depths of the NS envelope is thought to power the longest and most energetic bursts, i.e. the superbursts \citep[][]{Cumming_2001}. 

A burst typically radiates an enormous amount of energy, with energy outputs ranging from 10$^{39}$ erg up to 10$^{42}$ erg in the superbursts case, and it therefore takes consistent time for the fuel to be restored and a subsequent burst to occur. Recurrence times between bursts are usually irregular and go from hours to days for "normal" bursts to months or years in the case of intermediate and superbursts. A particular case is represented by systems that show periodic, regular and therefore predictable bursts, typically due to low metallicity matter accreted at high mass-accretion rate. The most notable case is the "Clocked Burster" GS 1826--24 \citep{Ubertini_1999}, but such behaviour has been also observed in at least two other systems; during its 2010 outburst, IGR J17480--2446 displayed regular bursts, but with a recurrence time which progressively decreased as the persistent flux increased, going from 24 min to about 3 min (see, e.g.\ \cite{Motta_2011}). More recently, another transient NS LMXB, i.e. 1RXS J180408.9--342058, was discovered as a {\it part-time} clocked burster, as it exhibited regular bursts only in the intermediate spectral state (see, e.g.\ \cite{Marino_2019b}). 
Duration and recurrence times for bursts vary, even for the same source, depending on the spectral state: bursts occurring in hard states, which are typically characterized by lower mass-accretion rates, last longer and rarely reach the Eddington limit, while during soft states, bursts tend to be shorter, brighter and (on average) less frequent. Photospheric Radius Expansion (PRE), i.e. a signature of the burst reaching the Eddington limit (see Sec.\ \ref{ss:PRE}) and burst oscillations (see Sec.\ \ref{oscillations}) are also characteristic of soft states' bursts. Interestingly, bursts are also observed at very low accretion regimes in Very Faint X-ray Transients (see Sec.\ \ref{ss:VFXT}). It is the case of the so-called \textit{burst-only} systems, intriguing NS LMXBs which are normally too X-ray faint to be detected by all-sky monitors, like e.g. MAXI onboard the ISS or the ASM onboard RXTE, but become visible within the duration of the burst (see, e.g.\ \cite{Cornelisse_2002}). 

As mentioned above, during thermonuclear bursts, the X-ray spectrum emitted by the system changes drastically and becomes dominated by the radiation liberated during the thermonuclear explosion. Burst spectra are typically described with a blackbody spectrum of a $\sim 2-3$ keV temperature at the peak. Deviations from this simple spectral shape are observed due to the interaction between the photons emitted during the burst and the electrons and ions in the NS photosphere. Compton scattering is expected to shift the observed temperature, so that the "color" temperature obtained by fitting the X-ray spectra may be significantly different from the effective temperature, i.e. by a factor $f_{\rm c} \approx 1.5-2 $ . In principle, radiation can also be absorbed by (non-completely) ionized species present in the photosphere or created by nuclear burning during bursts, thereby producing absorption lines and edges superimposed on the burst spectrum. The shape and energy of these lines is expected to be affected by Doppler broadening, connected to the NS spin, and gravitational redshift, proportional to its compactness. Measuring the latter effect could therefore provide constraints on the EoS of ultra-dense matter, at least in principle. In fact, despite several uncertain claims (see, for a few examples, \cite{Galloway_2017}), finding absorption lines from the NS surface has so far proven to be particularly challenging. 

Bursts are also observed at lower wavelengths, in particular in the optical/UV band and in the infrared band (see, e.g.\ \cite{Vincentelli_2020}), with delays of seconds with respect to the X-ray radiation. In these cases, the emission is interpreted as due to reprocessing of the X-ray photons radiated during the burst by the accretion disk, that emit in the optical/UV band, or by the secondary star, responsible for the infrared emission. In the latter case, studying the delay between the radiation at different wavelengths could therefore provide estimates on the orbital period of the system, as done for 4U 1728--34, for which a $\sim$1.1 h orbital period was inferred \citep[][]{Vincentelli_2020}. 


\subsubsection{Photospheric Radius Expansion bursts}
\label{ss:PRE}

\begin{figure}[t]
\centering
\includegraphics[width=0.99\textwidth, angle=0]{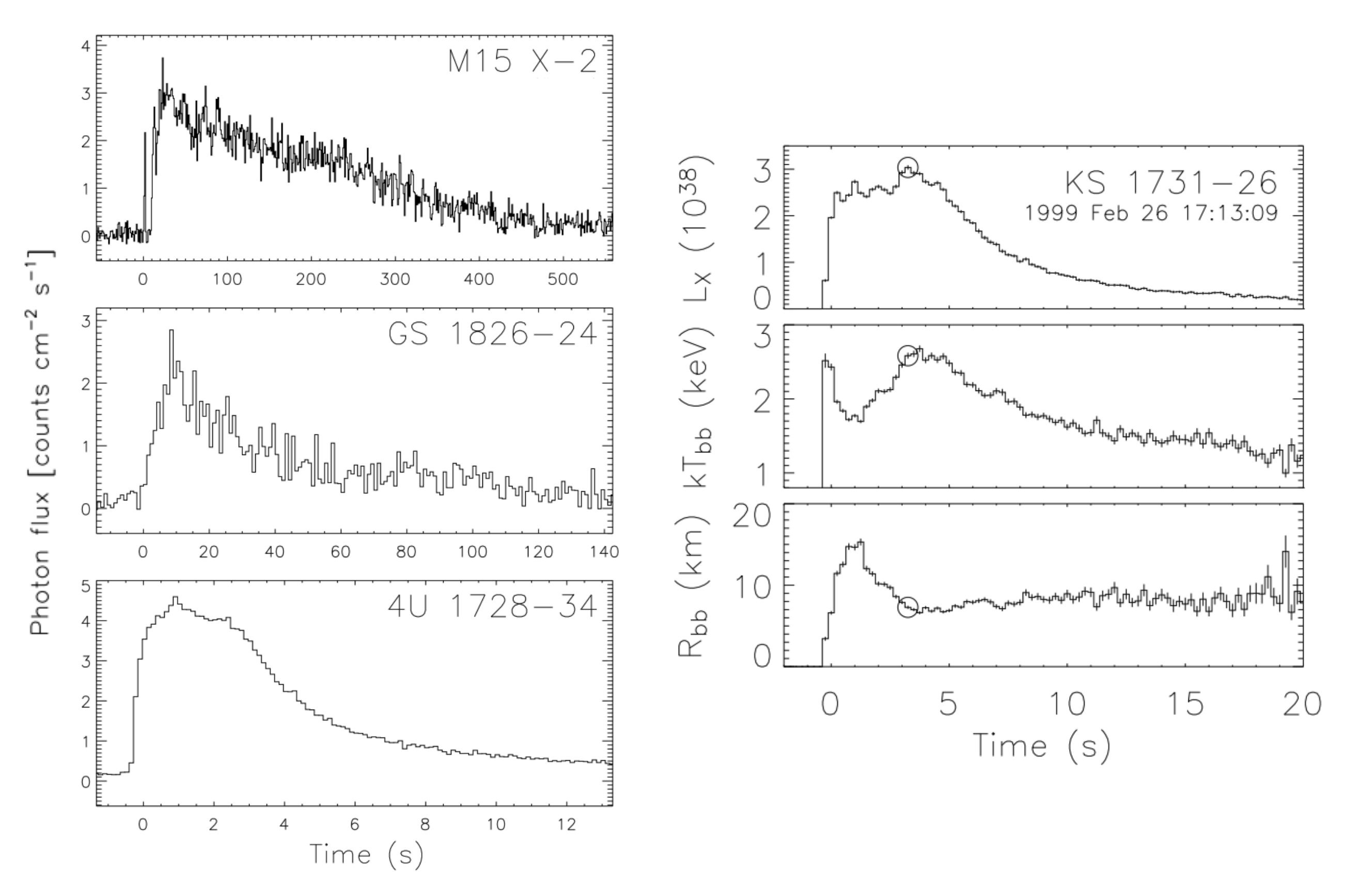}
\caption{({\it Left}) Examples of {\it RXTE}/PCA light curves of bursts with different duration. Credit: \cite{Galloway_2020}. ({\it Right}) PCA light curve of a PRE burst (top panel) and evolution of the blackbody temperature (middle panel) and blackbody radius (bottom panel). The simultaneous increase in temperature and radius at the peak is visible at $\sim$2 s. Credit: \citep[][]{Galloway_2008}. \textcopyright AAS. Reproduced with permission}
\label{fig:bursts_profile}
\end{figure}

During the burst, the X-ray luminosity increases by several orders of magnitude and sometimes can even reach the Eddington limit. When the limit is reached, the strong radiation pressure lifts up the NS photosphere and forces it to expand: this phase is called Photospheric Radius Expansion, PRE hereafter. Such inflation lasts only for time-scales of seconds, during which, as the photospheric radius $R_{\rm ph}$ expands, the blackbody temperature $T_{\rm bb}$ has to cool down, in order to keep the luminosity - already at its limit - constant. The photosphere will later gradually return to its original radius, at the so-called "touch-down" phase. The delicate interplay between $R_{\rm ph}$ and $T_{\rm bb}$ is witnessed by the typical double peaked profile in PRE bursts light curves: during the expansion phase, the cooling of the blackbody emission causes a sensible breakdown in the flux in a limited energy band\footnote{Throughout the PRE, the system has reached the Eddington limit and its luminosity is unchanged, therefore the observed dip in the flux is only artificial and is a consequence of the blackbody spectrum moving away from the limited energy band of the X-ray detector.}, while its contraction is accompanied by an increase of the flux and thereby by the appearance of the second peak. This phenomenon can also be observed by directly monitoring the evolution of $T_{\rm bb}$ and $R_{\rm bb}$ as obtained from time-resolved spectroscopy of the burst (see fig. \ref{fig:bursts_profile}, {\it right}). Indeed, if we divide the duration of the burst in short, subsequent time intervals and collect a spectrum in each of them, modelling these time-resolved spectra with a blackbody allows us to follow the evolution of $T_{\rm bb}$-$R_{\rm bb}$ over time. PRE bursts should therefore display at the peak a decrease in $T_{\rm bb}$ and a simultaneous increase in $R_{\rm bb}$. 

Knowing that the system reached the Eddington limit, it is possible to compare their luminosity with the measured flux to estimate the distance \citep{Kuulkers_2003}. The values of the distances obtained through this method have to be considered with an appropriate level of caution, due to several systematic biases affecting them. Indeed, without details on the composition of the NS atmosphere and therefore of its opacity, the value of the Eddington luminosity for a particular source is known only approximately. Moreover, even for the same source, the peak fluxes attained by PRE bursts seem to scatter around a mean value, with variations within 15\% \citep[][]{Kuulkers_2003}. Keeping these caveats in mind, using PRE bursts to estimate distances has proven to be the best, and sometimes only, method to obtain even loose constraints on the distance of the system and it has been therefore applied to a large number of NS LMXB \citep{Galloway_2008}. 

Furthermore, PRE bursts have the potential to deduce mass and radius of the NS and put constraints on the EoS of ultra-dense matter (see \citep[][]{Degenaar_2018_eos} for a review). Putting together the X-ray flux measured at the peak and the (approximately) known luminosity and distance of the system, spectral modelling of the blackbody emission from the burst can be used to obtain a measure of the radius of the emitting region. Unfortunately, the expansion and subsequent contraction of the photosphere when the Eddington limit is reached, make the apparent radius of the emitted blackbody spectrum and the NS not coincident, apart from the {\it touchdown} phase, when the photosphere settles back onto the NS surface. It is therefore possible to use the normalization of the blackbody at touchdown to obtain a measurement of the NS radius and, in turn, solving for the mass as well \citep[][]{Ozel_2016}, a method often referred to as the {\it touchdown method}. However, only spectral models which take into account the spectral distortion due to the NS atmosphere, embodied in the color correction factor $f_c$ (see above), are adequate to obtain reliable estimates of mass and radius. At current, the method has been applied to 5 NS LMXBs and yield a combined range for the radius of $9.8-11$ km, for a fixed mass of 1.4 M$_\odot$ (see \citep[][]{Ozel_2016}, and references therein). 

One of the main drawbacks of the touchdown method is its dependence on prior knowledge of the distance and the chemical composition of the atmosphere. Recently, an alternative method to obtain constraints on the NS M/R which does not require this knowledge has been developed, the so-called {\it cooling tail} method. In the quasi-exponential tail of the burst, the photospheric radius is not expected to change, so that the normalization of the blackbody emission $K_{\rm bb}$ depends only on $f_c$. Atmosphere models predict $f_{c}(t)$ to be sensitive to different values of mass and radius, so that the observed $K_{\rm bb}$ values in the late phases of the burst can be fitted to measure the NS parameters. In real-life spectra of NS LMXBs, however, the burst emission at its final stages is hard to disentangle from the persistent emission from the accretion flow, especially in bright sources. The cooling tail method has indeed provided the most reliable constraints in bursters at low mass-accretion rate \citep[][]{Nattila_2016}. Interestingly, a systematic discrepancy has been reported between the radii obtained with these two methods, sometimes even when applied to the same source, with the cooling-tail radii being somewhat larger than the touchdown radii. As both techniques suffer from systematic biases, the origin of such discrepancy and whether one method is more reliable than the other are currently not established (see \citep[][]{Degenaar_2018_eos} for a discussion).

\subsubsection{Burst oscillations}
\label{oscillations}

\begin{figure}[t]
\centering
\includegraphics[width=0.9\textwidth, angle=0]{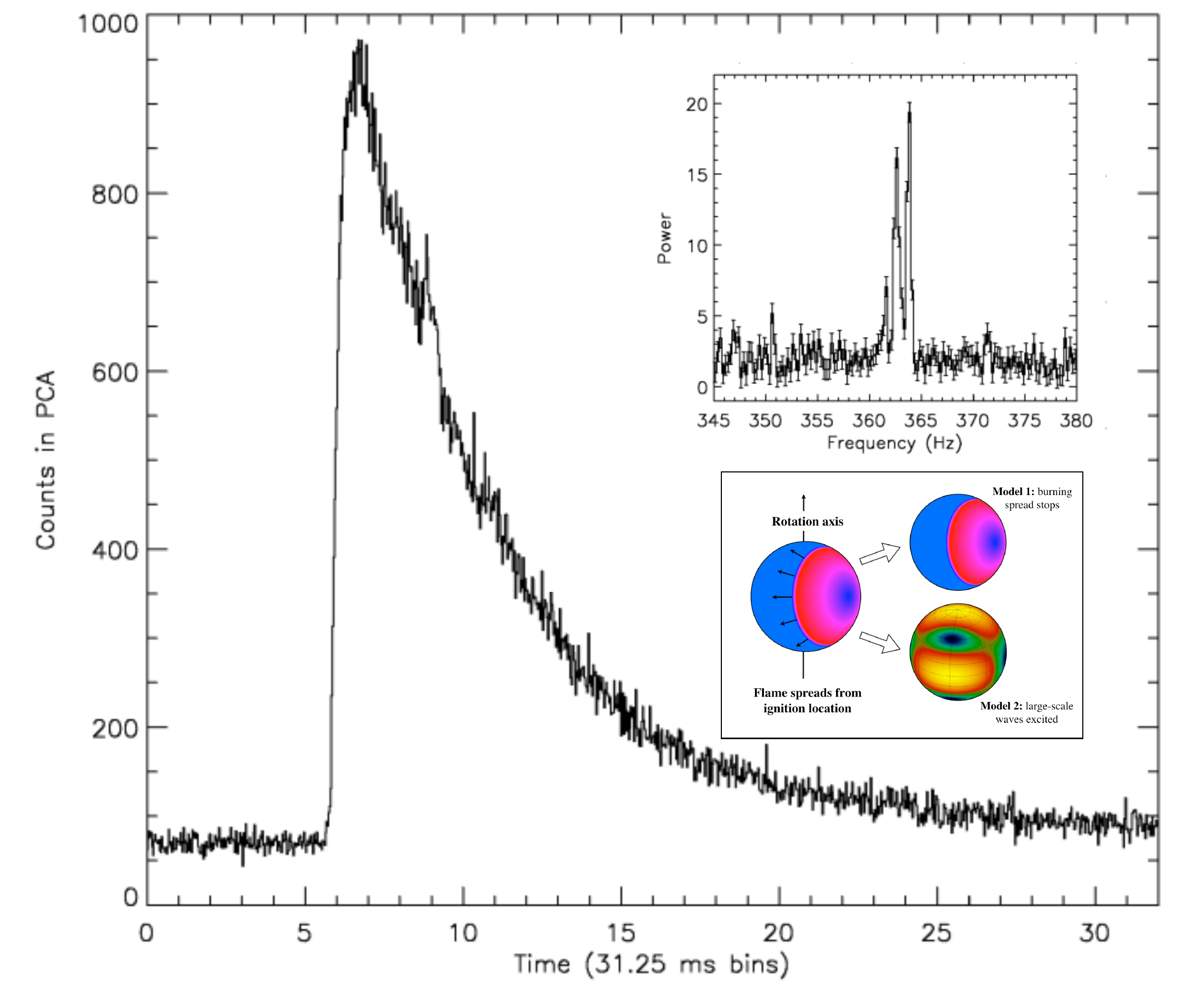}
\caption{In the main panel, we show the {\it RXTE}/PCA light curve of a burst from 4U 1728-34 occurred in 1996. The small top inset displays a portion of the corresponding power spectrum, with apparent peak at the frequency of the burst oscillations. Credit: \cite{Strohmayer_1996} \textcopyright AAS. Reproduced with permission. The bottom inset panel displays sketches showing two proposed explanations for burst oscillations (see text), adapted from \cite{Watts_2012}}
\label{fig:bursts_osc}
\end{figure}

During type-I X-ray bursts, LMXBs sometimes show coherent, periodic variations of the X-ray intensity (Fig.\ \ref{fig:bursts_osc}, {\it left}), dubbed burst oscillations (see \citep[][]{Watts_2012} for a review).  So far, they have been observed in a subset of 17 bursters, 7 of them being AMSPs. Their characteristic frequency is usually consistent, within a few Hz in the worst case, with the spin frequency of the NS when known. Broadly speaking, they are thought to arise from some sort of asymmetry in the brightness of the NS surface, e.g. hot spots, which is then modulated by the rotation of the compact object. However, the exact mechanism underlying burst oscillations is still to be determined.  

As mentioned above, burst oscillations are not always observed during bursts. Even for the fraction of systems showing burst oscillations, their appearance is typically irregular and not systematic, with the exception of four AMSPs. These systems are tireless X-ray pulsators, displaying pulsations in their persistent emission and in any burst exhibited, irrespective of the burst properties or the accretion state. In the other sources, a correlation between the accretion rate and the presence of oscillations seems to exist. Indeed, burst oscillations tend to appear when the system is in soft state, i.e. usually at high accretion rate. Furthermore, oscillations are typically detected in "normal", H/He bursts, while they have only been observed once during superbursts \citep[][]{Strohmayer_2002} and never in intermediate bursts. Even throughout the whole burst duration, oscillations are not always detected but rather show up during the rise and the tail of the burst \citep[][]{Galloway_2008}. The scattered nature of burst oscillations detections may be due to a selection bias, i.e. they may be always present but their amplitudes may be too faint to be detected with state-of-the-art technology. 

As mentioned above, in all cases oscillations frequency appears to be associated to the NS spin frequency, i.e. it is either coincident with it or deviates from it by only a few Hz \citep[][]{Watts_2012}. This has been first confirmed in the AMSP SAX J1808.4-3658 \citep[][]{Chakrabarty_2003}. In several cases, oscillations were found in NSs with previously unknown spin periods, thereby enabling the measure of their spin for the first time. It is the case of the fastest accreting NS known to date, i.e. 4U 1608-52 spinning at $\sim$620 Hz frequency \citep[][]{Galloway_2008}, which is a burster but not a pulsar. Moreover, the frequency of the oscillations does not appear to be fixed but exhibits the tendency to drift to upper frequencies, with an increase of typically a few Hz. This drift could be connected to the photosphere expansion and it is indeed less pronounced in AMSPs and non-PRE bursts \citep[][]{Watts_2012}. 

The most natural explanation for burst oscillations, in analogy with pulsars,
invokes the development of hot spots on the NS surface during burst ignition (see Fig. \ref{fig:bursts_osc}, {\it right}). According to this model, ignition starts from a point in the NS surface and then the liberated flame spreads in all directions. However, some mechanism, able to confine the heat within a limited area and prevent it to make the surface brightness homogeneous, has to be at play. This model is particularly suited to explain oscillations at the burst rise. Alternatively, burst oscillations could arise from large scale waves in deeper layers of the NS excited from the flame fronts expanding on the surface. The temperature disuniformities created by these waves could in turn explain the brightness asymmetry originating the oscillations (see e.g.\ \citep[][]{Heyl_2004}). This mechanism could be active for oscillations detected at the decay of the burst, when the heat has already spread over a large portion of the NS surface.

Burst oscillations could in principle play an important role in constraining mass and radius of the NS. Giving us the opportunity to measure the spin of non-pulsating NSs, burst oscillations could indeed permit the discovery of very short, e.g. sub-millisecond, spin periods, which will give important information on the NS EoS. In fact, at some spin rate, i.e. the mass-shedding frequency, the NS is expected to break apart as the gravitational attraction holding it together is overcome by centrifugal forces. The exact value of such spin frequency is a function of the composition of the NS and the EoS (see \citep[][]{Lattimer_2016}, and references therein). Furthermore, the compactness of the NS is also expected to alter the shape of the oscillation profile. Indeed, the more compact the NS, the stronger the relativistic effects on the photons emitted by the hot spots and originating the oscillations. Pulse profile modelling, i.e. the technique to identify the compactness of the NS from the study of the profile of a pulsating signal, is a very promising method already applied to millisecond pulsars (e.g.\ \citep[][]{Bogdanov_2019}) taking advantage from the large effective area at $\sim 1$ keV of NICER, which could also be used for burst oscillations in the future.



\subsubsection{Probing the surrounding accretion environment}

All burst properties discussed so far - duration, recurrence, peak flux, presence of burst oscillations just to mention a few - are critically influenced by the accretion flow surrounding the NS. Indeed, as already mentioned, bursts look and evolve differently at different mass accretion rates, spectral states and for different composition of the material in the accretion flow. Reversing the point of view, one could wonder whether the accretion flow is in turn somehow influenced when hit by the huge energy output liberated by the burst. Historically, bursts and accretion environments were conceived as detached; according to this assumption, burst spectroscopy has been usually performed by simply subtracting the flux measured before the onset of the burst. However, in the last few decades, the impact that bursts have on the surrounding accretion flow has gained growing attention (see \citep[][]{Degenaar_2018_burst} for a review). It is now clear that, for instance, thermonuclear bursts can alter the structures, i.e. disk and corona, of the accretion flow, drive outflows such as radiatively driven winds, originate reflection features emitted by the disk. 

First of all, the persistent flux emitted by the accretion flow during the burst is typically enhanced by an amplification factor $f_a$. The values attained by $f_a$ range from $\approx$1 up to $\approx$80, as obtained by \cite[][]{Worpel_2015}, with the highest enhancement factors found for PRE bursts. Flux amplification could be interpreted as due to the reprocessing by the disk of the radiative output emitted during the burst \citep[][]{Intzand_2013}, as confirmed by the occasional detection of discrete reflection features (see below), or to a temporary increase mass-accretion rate likely arising from Poynting-Robertson drag on the disk material. Amplification factors of order unity were found for Z sources, likely because they have already near-Eddington mass-accretion rates. However, in very few cases of super-expansion bursts, the estimated values for $f_{a}$ were found negative, possibly due to shielding from an optically thick shell of plasma ejected by the burst or simply to a consistent part of the spectrum moving out of the observed energy band. Interestingly, not only the flux level is influenced by the burst, but also the spectral shape of the continuum. A striking 80\% reduction in the hard X-ray spectrum, i.e. $30-75$ keV band, has been observed in GS 1826-238 in correspondence of the bursts, along with other observational evidences of a softening in the persistent emission at hard X-rays subsequent to the blast (for a list, see e.g.\ \citep[][]{Speicher_2020}). This hard deficit has been explained with an extra-cooling of the hot corona caused by the injection of soft photons radiated during the burst. Furthermore, several studies showed that such hard deficit is state-dependent, i.e.\ it is stronger during the hard state, probably due to the different geometry of the accretion flow in this state \citep[][]{Degenaar_2018_burst}. These evidences highlight that attempting to model the spectrum of the persistent emission along with the burst emission may be a more adequate and physically reliable strategy than simply subtracting the pre-burst flux. 

As the burst energizes the accretion flow around the NS, discrete absorption and emission features may evolve or appear only in association with the burst. Fe K- and L-shell lines have been detected during several bursts, typically intermediate and/or superburst since the short duration of normal bursts makes obtaining detailed time-resolved spectra particularly challenging. From the spectral shape of the iron line, the radius of the inner disk can be constrained (see Section \ref{ss:reflection}), so that detecting these features during thermonuclear bursts could provide information on how the optically thick disk responds to the burst, i.e. whether its inner regions are evacuated or material in the disk penetrates even further in. For instance, in 4U 1820-30, the disk was found to recede during a superburst \citep[][]{Ballantyne_2004}. Recent simulations of bursts interacting with thin disks showed clearly that, in response of the blast, disks become hotter, geometrically thicker and optically thinner. Moreover, their inner edges tend to recede in the rise and peak of the burst, due to a local increase in mass-accretion rate resulting from Poynting-Robertson drag, and then move inwards again during the decay \citep[][]{Fragile_2020}. During two superbursts from 4U 1820-30 and 4U 1636-536, a significant increase in the equivalent hydrogen column density during the burst was associated to material pushed away from the photosphere or from the disk in the form of a wind. Furthermore, the combination of blueshifted absorption features between $8-9$ keV and irregularities in the tail of the burst profile brought a strong evidence for winds caused by the burst in IGR J17062–6143 \citep[][]{Degenaar_2013}. It is worth mentioning that reflection features were instead notably absent in several other normal bursts analyzed with high resolution spectrometers such as the Reflection Grating Spectrometer (RGS) on board {\it XMM-Newton}\citep[][]{Degenaar_2018_burst}. Whether this lack of evidences is due to detectability issues, to the accreted material being highly ionized or to geometrical effects is unclear.

\subsection{High-Inclinations Sources}
\label{ss:high-incl}

A small subset of NS LMXBs are viewed at a high-inclination angle; these are dipping and eclipsing sources. The angle $i$ between the perpendicular to the plane containing the binary system and the line of sight is called the inclination angle. When the line of sight is along the perpendicular to the orbital plane, the  angle $i$ is null and the source is  observed \textit{face-on}, whereas when $i$ is 90$^\circ$, the line of sight lies in the orbital plane and the source is observed \textit{edge-on}. These almost edge-on sources offer the unique opportunity to study the properties of the matter above the disk and possible outer-disk structures. 
In fact, the radiation emitted by the system passes through this matter and absorption lines, and edges can be observed in the X-ray spectrum. These give us information on the chemical composition, ionization state, and velocity field. Moreover, a variable height of the disk at its outer edge, and/or the \textit{bulge} produced by the stream of matter transferred from the companion star at the impact point with the outer disk, can modulate the X-ray flux from the central source and produce dips, i.e., a drop in the X-ray flux caused by periodic obscuration of the central X-ray source, and eclipses, caused by the companion star along the line of sight, in the light curve. These latter features will appear periodically at every orbital cycle, giving us the opportunity to measure the otherwise elusive orbital period of the system and its evolution with time.

The X-ray flux emitted by an LMXB typically shows periodic modulation or features if $i$ is larger than 60$^\circ$, and the orbital period is less than one day. In particular,  the light curve of the source shows dips for $i$ in the $60^\circ$-$75^\circ$ range (pure dipper sources), dips and total eclipses for $i$ in the 75$^\circ$-$80^\circ$ range, and partial eclipses for $i$ larger than $\sim 80^\circ$ (the so-called ADC sources) \citep{frank_02}.
The shape, duration, and spectral properties of the dips change from source to source and from (orbital) cycle to cycle. To date, 13 LMXBs hosting a NS and 6 LMXBs hosting a BH candidate have shown clear dips in their light curves \citep{dai_14}.
The physical mechanism  explaining the occurrence of  dips is  still debated in the literature.  \citep{white_82}
proposed that a variable, azimuthal-dependent height of the outer rim of the accretion disk could obscure or partially absorb the X-ray emission from the central source and/or an extended ADC. 
However, some dipping sources (hereafter dippers) have experienced the temporary disappearance of the dips, which points to a strong variability of the occulting regions. An alternative mechanism was proposed by \cite{frank_87} who suggested that  if matter from the companion star flows across the thickness of the outer accretion disk, part of the stream may hit the disk at a much closer radius. At the outer disk, the stream matter is quickly dynamically and thermally virialized, but a fraction of it gets energy from the impact shock forming a two-zone medium: blobs of cold condensed gas, surrounded by a lower density and hotter plasma at large scale heights above the disk. 

The first  observational constraints  on the properties and geometry of these systems were obtained studying the continuum emission from the dippers. At first, the spectra were fitted using the so-called Birmingham model (see Sec.\ \ref{ss:history}).
Seed photons for the Comptonized spectrum come from the accretion disk, and Comptonization is thought to occur at large disk radii in an extended corona, whose radius is $\gg 10^9$ cm. Analyzing the spectra obtained during 
deep dips (where emission is totally blocked at the dip bottom), it has been shown that the corona emission is gradually covered (progressive covering model), implying it is extended, with a disk-like geometry, while the blackbody emission is point-like and attributed to the NS emission \cite{church_04}. The main assumption in deriving the estimates for the ADC radius is that the dip is caused by the bulge located at the outer accretion disk, 
whose main effect is a progressive photoelectric absorption of the primary incident source flux.

However, this progressive covering approach, implying an increase of the covering fraction and column density of a neutral absorber, cannot explain all the features observed in the spectra of these sources. In fact,  
in the past two decades, the detection of  absorption lines associated with  highly ionized elements in the X-ray spectrum  of high-inclination systems harboring a BH has provided new methods to investigate the atmosphere of the accretion disk.  These features often appear to be blue-shifted pointing to a fast outflow or wind along the disk or more in general to an outflowing photoionized plasma, similar to that observed in Seyfert galaxies.   The ionization state of the optically thick absorbing plasma is variable on the  kilosecond  time scale,  and the estimated  wind velocity is thousands of km s$^{-1}$ (e.g., \cite{ueda_04} and references therein).
In all cases, the most clearly resolved lines are from H-like and He-like transitions of iron (\ion{Fe}{xxv} and \ion{Fe}{xxvi}), which implies that the ionization parameter $\xi$ of the warm absorber is  larger than one hundred. 
\begin{figure}
\centering
\includegraphics[width=1.0\textwidth, angle=0]{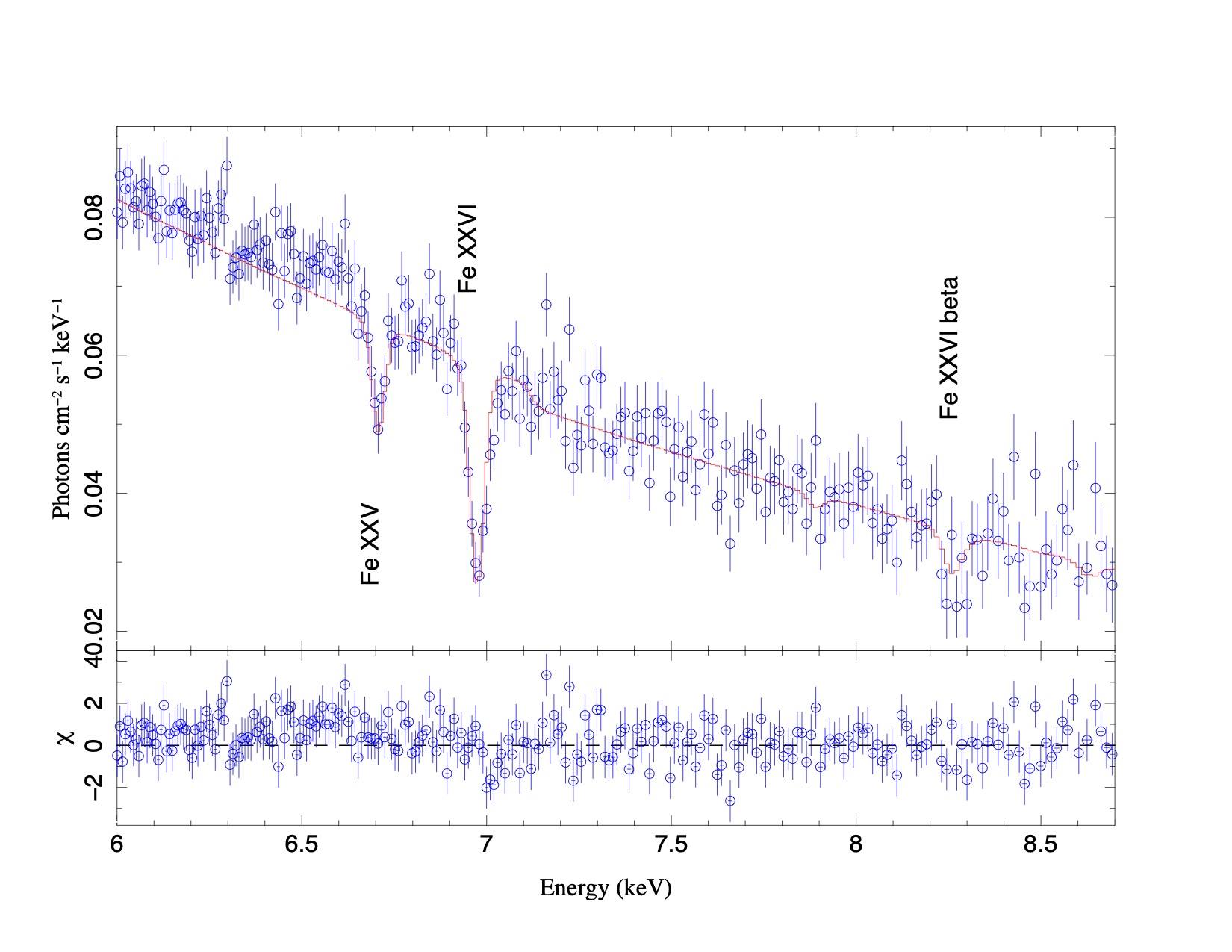}
\caption{Data, best-fit model, and residuals for the \ion{Fe}{xxv} and \ion{Fe}{xxvi}  absorption lines observed in  the Chandra  (HEG+MEG data) non-dip average spectrum of GX 13+1. The adopted model consists of an absorbed accretion disk plus a  thermal Comptonized component. The lines are fitted with the XSTAR warm-absorber model. (Credit: \cite{dai_14}, reproduced with permission \textcopyright ESO)}
\label{fig:feline}
\end{figure}

The key role played by the presence of photoionized plasma  in dippers harboring a NS  was highlighted  by 
\citep[][and references therein]{diaz_06}, who, analyzing a sample of dippers,  modeled the changes in the \ion{Fe}{xxv} and \ion{Fe}{xxvi}  absorption lines and the continuum during the dips  taking into account an increase in the column density and a decrease in the amount of ionization of a photoionized absorbing plasma (implying an increase of a neutral absorber column density). Outside of the dips, the properties of the absorber do not vary strongly with orbital phase, which suggests that the ionized plasma has a cylindrical geometry with a maximum column density close to the plane of the accretion disc and is not confined to the locus of the bulge. 
In this scenario, there is no more need for a partial covering of an extended corona because the soft excess observed during dips is naturally accounted for by a combination of a strong increase in the column density of the warm medium and a decrease of its ionization parameter. Since dipping sources are   LMXBs viewed from close to the orbital plane, this implies that photoionized plasma at the outer edge of the disk is a common feature of LMXBs. 

All the dippers show Fe absorption lines at the rest energy except for the bright atoll GX 13+1, which shows a clear blue-shift of the ionized Iron lines, indicating a wind velocity close to 400 km s$^{-1}$ \cite{ueda_04,dai_14} (see Fig. \ref{fig:feline}). \cite{diaz_12} suggested that most of the dippers 
have a  moderate X-ray luminosity (usually $10^{36} - 10^{37}$ erg s$^{-1}$), unable to produce strong winds.
In this case  the ionized corona above the accretion disk surface is in a stable equilibrium under the radiation  and gravitational pressure exerted by the central region of the system, and the absorption lines  are produced in the hot and static atmosphere. 
On the other hand, in bright sources such as GX 13+1 radiation pressure can launch strong winds along the disk producing blueshifted absorption lines.

\begin{figure}
\centering
\includegraphics[height=0.22\textwidth, angle=0]{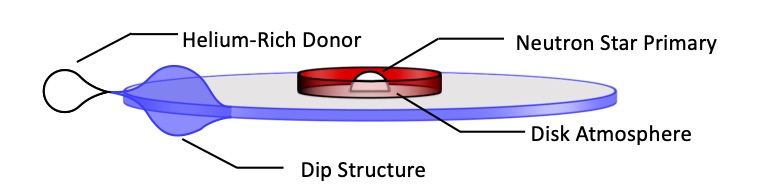}
\includegraphics[height=0.20\textwidth, angle=0]{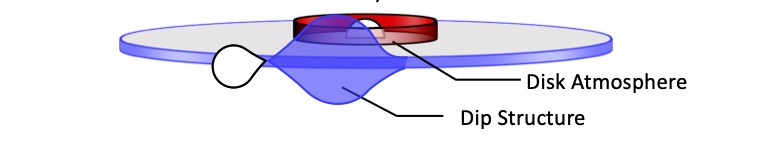}
\caption{Upper panel: Sketch of X1916-053 out of the dip. The atmosphere close to the NS is directly observed (red region). The lines formed in this region are gravitationally red-shifted.\\
Lower panel: Sketch of X1916-053 during  the dip. The central regions are occulted by the bulge at the outer rim of the accretion disk. The observed lines are not red-shifted. Adapted from \cite{trueba_20}}
\label{fig:1916}
\end{figure}

Particularly interesting are the recent studies of the dipper XB 1916-053, belonging to the class of the ultra-compact X-ray binary systems (hereafter UCXB) having an orbital period close to 50 min. \cite{trueba_20} 
have found evidences in long Chandra/HETG observations that the observed absorption lines out of the dips are gravitationally red-shifted ($z = 8 \times 10^{-4}$). If this interpretation is correct, it is possible to infer that the hot plasma is located close to $10^8$ cm from the NS while the absorber during the dips is more distant, close to the outer rim of the accretion disk at $\sim 10^{10}$ cm (see Fig. \ref{fig:1916}).
%
Gravitational red-shifted absorption lines, arising in ionized disk atmospheres, may be a common feature in  dippers of the UCXB class, or with short orbital periods, providing detectable gravitational redshift. 
In the next years, if red-shifted absorption lines will be observed in other sources, these will form a new class of absorbers in which distances from the compact object will be measured with high accuracy, providing a powerful diagnostic tool for the plasma in disk atmospheres.

\begin{figure}
\centering
\includegraphics[width=.49\textwidth, angle=0]{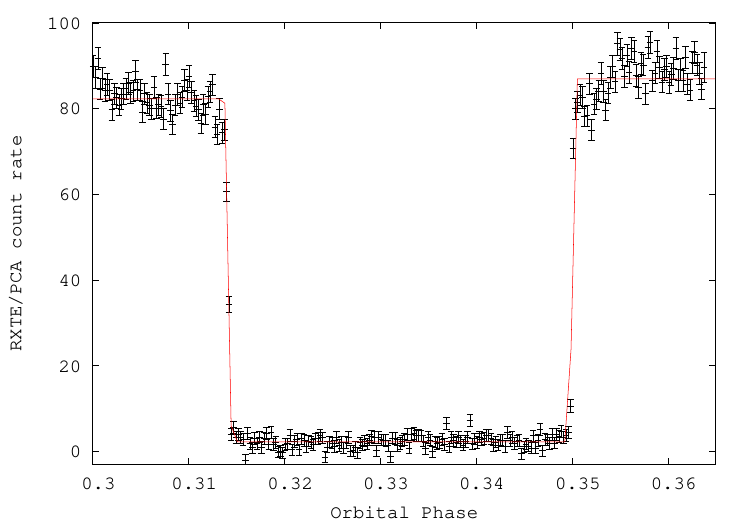}
\includegraphics[width=.47\textwidth, angle=0]{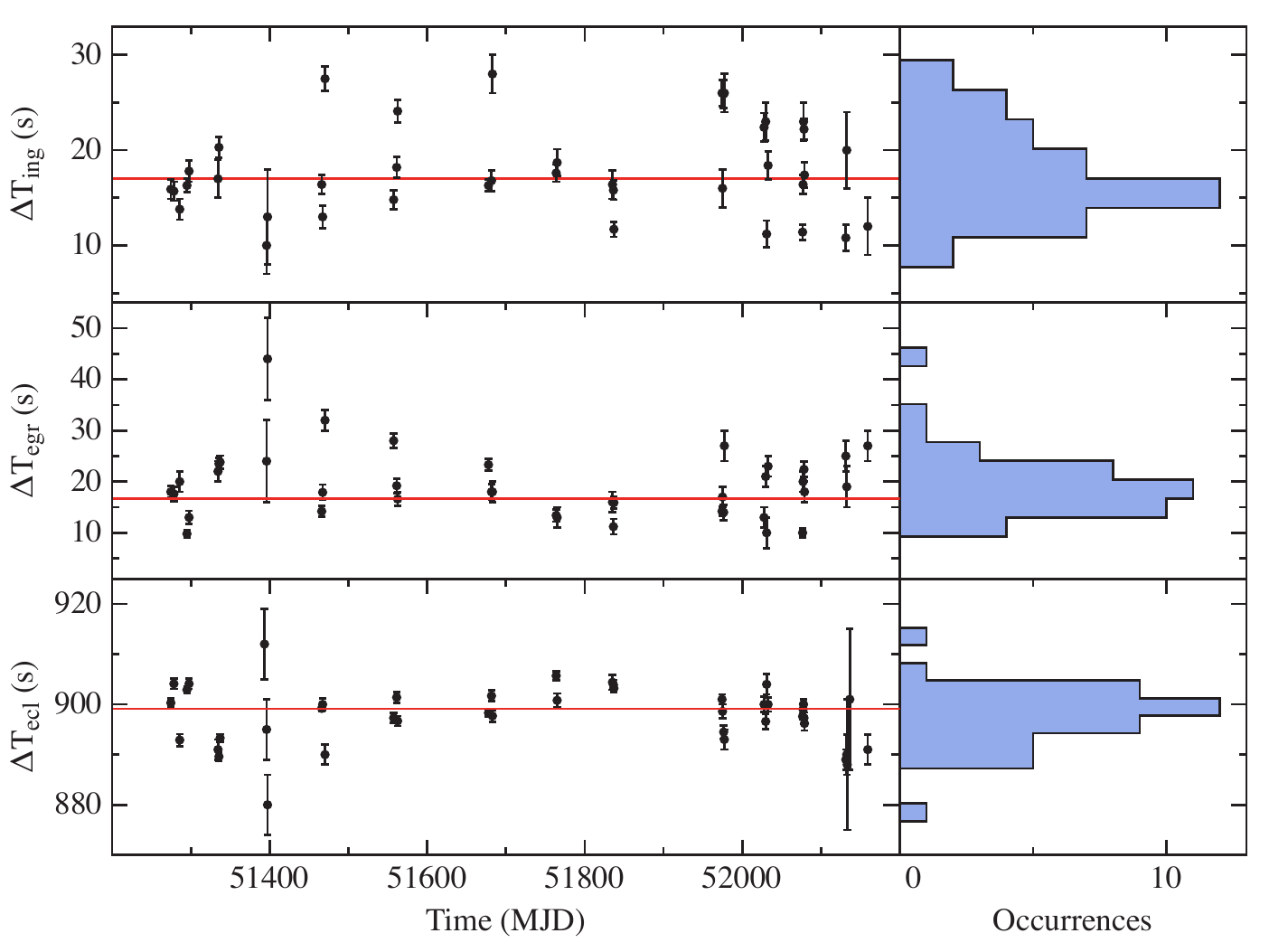}
\caption{Left: Total eclipse of X1659-628 observed by the RXTE/PCA instrument. The superimposed red function is the step-and-ramp function adopted to estimate the eclipse arrival time (see \cite{iaria_18} and references therein).
Right: The ingress, egress and eclipse duration of the eclipse of  X~1659--628 as function of time. The values are obtained from the RXTE/PCA eclipses analyzed by \cite{iaria_18}.   The red lines indicate the average values for each duration. From the top-right to the bottom-right   the histograms of the occurrence of the ingress, egress and eclipse duration, respectively, are shown. }
\label{fig:eclipse}
\end{figure}

Eclipsing sources 
(inclination angle $i$ between 75$^\circ$ and 80$^\circ$) also show   total eclipses in addition to dips in their light-curves. The eclipse is caused by the transit of the companion star between the observer and the central region of the system emitting the X-ray radiation (see left panel in Fig. \ref{fig:eclipse} for a typical total eclipse).  As for dippers,  these systems also show absorption lines associated with highly ionized elements in their spectra confirming the presence of  a ionized atmosphere and/or outflowing winds  above the disk
(see e.g.\ \cite{iaria_19} and reference therein). 

NS-LMXBs sources observed with inclination angles larger than 80$^\circ$ show only partial eclipses in their light-curves, and sometimes a modulation of the X-ray flux with the orbital phase caused by the variable thickness of the outer disk. The high inclination angle, in fact, implies that the direct emission from the innermost region is shaded by the swelling in the external region of the accretion disk, caused by the inflowing matter transferred by the secondary star that hits the outer disk.
Since the direct X-ray emission from the central region is shielded, we observe only the radiation scattered along the line of sight by a large diffuse optically thin corona above the disk, namely the ADC, that extends up to the outer rim of the disk. 
The prototype of the ADC   sources is X~1822-371 and, to date, a total of five ADC sources are known (including also 2S 0921-630, XTE 2123-056, 4U 2129+12 and 4U 2129+47).  The X-ray spectra of X~1822-371 
show emission lines  instead of absorption lines. 
Indeed, emission lines can still be produced in the inner part of the system but we see the fraction of this emission that is diffuse along our line of sight (see e.g. \cite{anitra_21}).
The observed  luminosity of the source is $\sim 10^{36}$ erg s$^{-1}$ and is probably only a small fraction of the intrinsic X-ray luminosity, as indicated by the ratio of the X-ray to optical luminosity of the source, that is a factor $\sim 50$ lower than typical values for LMXBs. Furthermore, timing studies of the source (see below) suggests a super-Eddington mass transfer and hence an intrinsic luminosity at the Eddington limit, implying that the optical depth of the ADC should be $\tau \sim 0.01$ so that only $\sim 1\%$ of the total X-ray luminosity is scattered along the line of sight \cite{iaria_13}.
 
Often, periodic features such as eclipses and dips present in the light-curve of high-inclination LMXBs allow us to  estimate their orbital period $P$   and  orbital period derivative $\dot{P}$. The more accurate method to obtain or refine these two quantities is  the O-C (observed minus calculated) method.  A detailed description of the method is given by \cite{chou_14}. 
To probe the orbital evolution of dippers, the dip arrival time is selected as the time at which the minimum dip intensity occurs. However, the dip is caused by X-ray absorption at  the outer edge of the accretion disk; this implies that dips are  not completely phase locked with the orbital cycle, but sometimes show phase jitter (see \cite{hu_08} and references therein). Furthermore,  the dip  arrival time   is sometimes difficult to uniquely define because  the dip shape can change from an orbital cycle to another.   We show the values of $P$ and $\dot{P}$ of the pure dippers (and other high-inclination sources) for which an estimation has been obtained  with the O-C method in Tab. \ref{tab:tab1ro}.  
To date, the O-C method has been  successfully applied to three dippers (XB 1916-053, X 1254-690, XB 1323-619); an accurate estimation of $\dot{P}$ was obtained only for XB 1916-053, since its short orbital period allows to analyze a large amount of  dip arrival times.  The study of the dip arrival times of XB 1916-053 suggests that the system could be a hierarchical triple system in which the third body 
has a mass of 45 M$_J$ and an orbital revolution period of $24.9\pm 0.8$ yr \cite{iaria_1916_21}.  

The study of the orbital period evolution in eclipsing sources, adopting the O-C method, is done by determining the eclipse arrival times that, contrarily to the dip arrival times, are phase locked with the orbital cycle. However, also for eclipses, small jitters are observed. They can be ascribed to the companion-star atmosphere that influences the measurement  of the ingress and egress times of the eclipse \cite{wolf_09}.  The jitters observed for the eclipsing X~1659-628 are shown in the right panel of Fig. \ref{fig:eclipse}. 

\begin{figure}
\centering
\includegraphics[width=.33\textwidth, angle=-90]{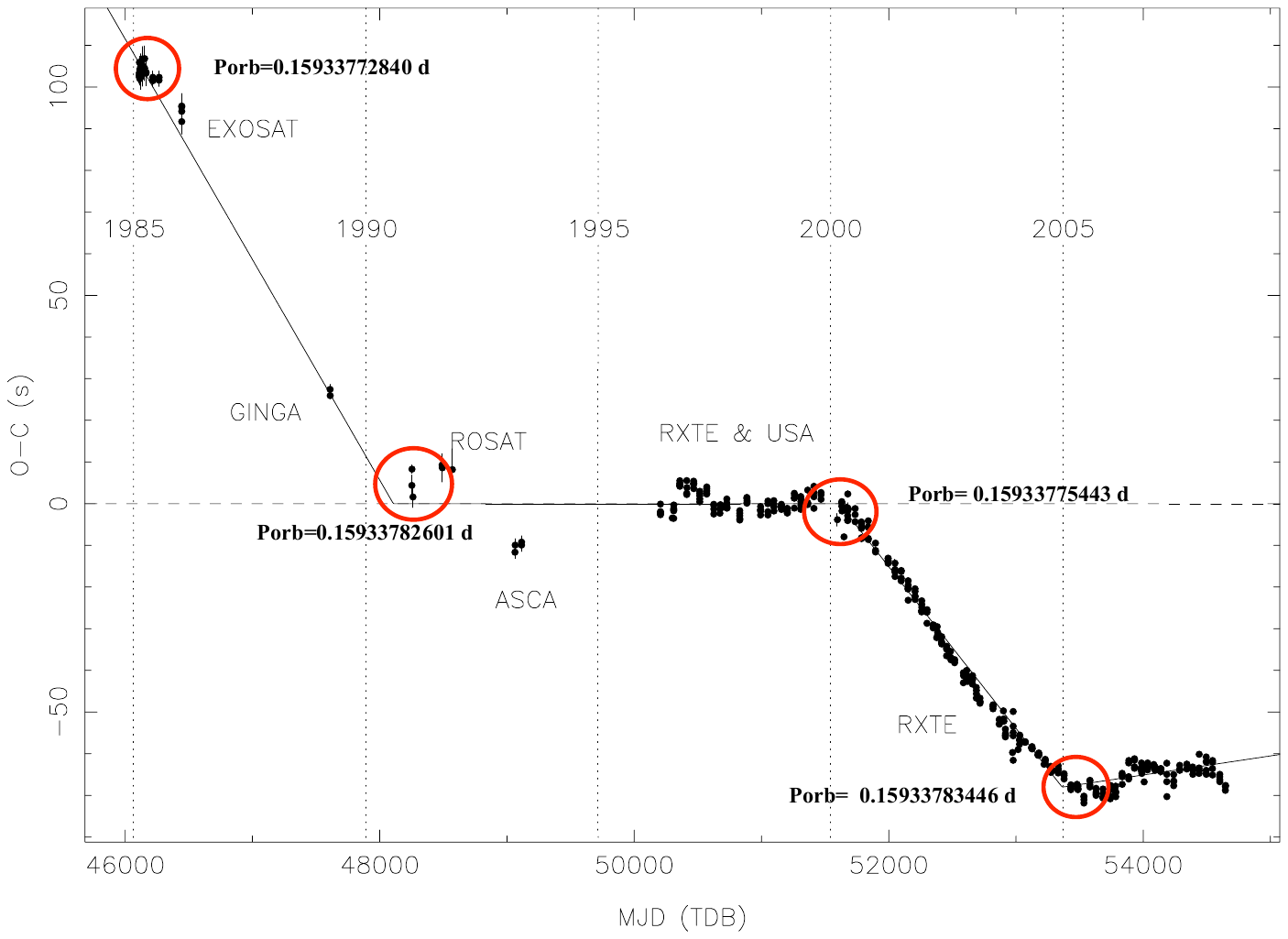}
\includegraphics[width=.37\textwidth, angle=-90]{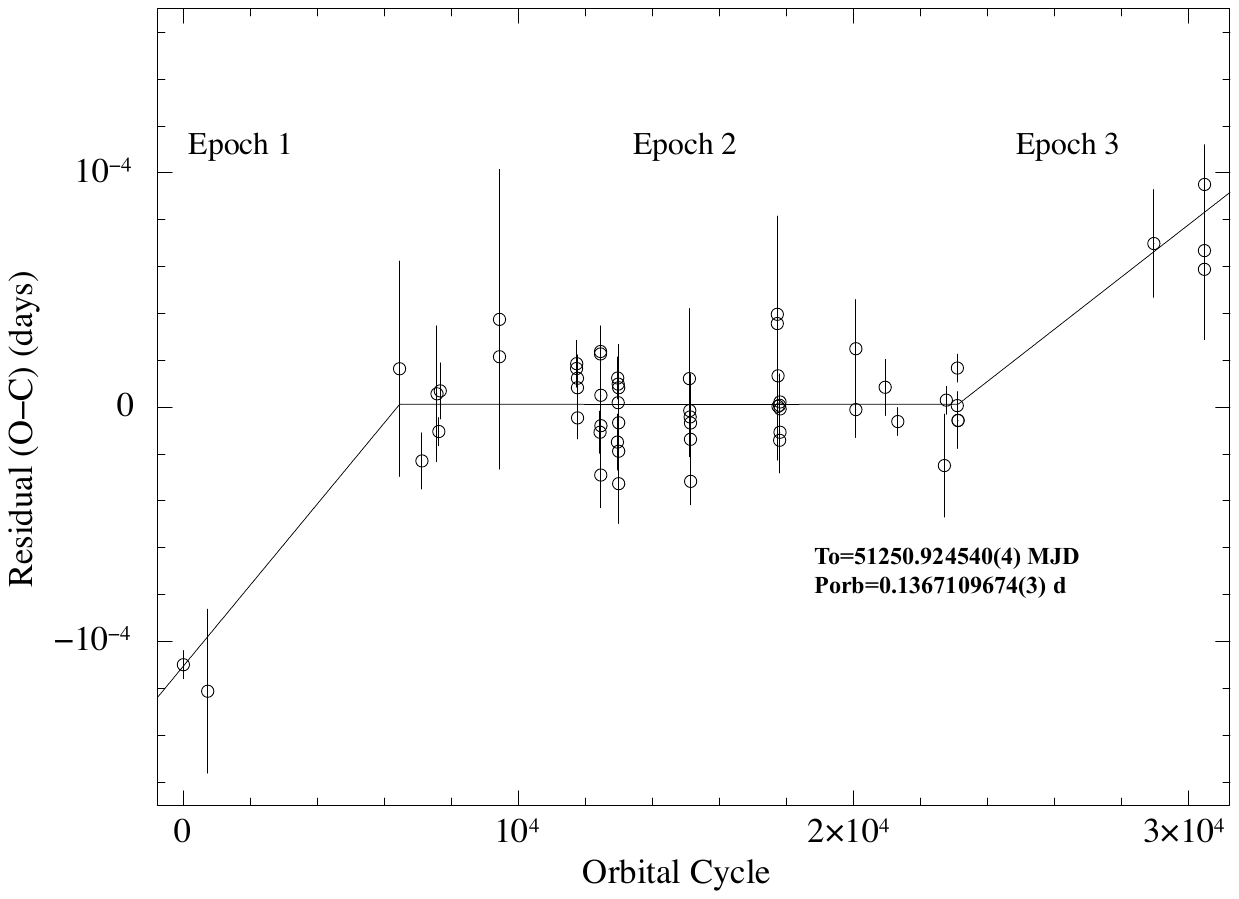}
\caption{Left: O-C residuals of EXO 0748-676. We show the estimated orbital period $P$ at the epoch in the red circle.  
P increases by 8 ms between 46111 and 48107 MJD, decreases by 6 ms between 48107 and  51610 MJD and, finally, increases again between  51610 and 51610 MJD.  Adapted from \cite{wolf_09}.
Right: O-C residuals of XTE J1710-28. The lower limits on orbital period changes are $ \Delta P = 1.4 $ ms  between epoch-1 and epoch-2  and $ \Delta P = 0.9 $ ms  between epoch-2 and epoch-3.  Adapted from \cite{chetana_11}}
\label{fig:wolff09}
\end{figure}

Among the eclipsing sources EXO 0748-676 (in X-ray quiescence since 2008, after 23 years of persistent X-ray activity) is one of the most studied. The orbital period of the source has been tracked since 1985 and shows puzzling residuals with respect to a constant orbital period solution resembling a cyclical variation (see Fig. \ref{fig:wolff09}, left panel). Since the companion star mass is close to 0.4 M$_{\odot}$ the source could have a convective envelope and, hence, a stellar magnetic field. It has been suggested that cyclic changes of the companion star magnetic field may induce cyclic changes in the secondary’s structure and its quadrupole moment which may explain the cyclic variation of the orbital period (\citep[][]{wolf_09} and references therein). Indeed, a changing gravitational quadrupole moment of the companion star can result into changes in the orbital period of the binary system because of the spin-orbit coupling in Roche-lobe filling stars. A similar behavior has been observed in the orbital period for the eclipsing XTE J1710-281 \citep{chetana_11}, as shown in Fig. \ref{fig:wolff09} (right panel), for which, however, the companion star type is not known yet. 

\begin{figure}
\centering
\includegraphics[width=.80\textwidth, angle=0]{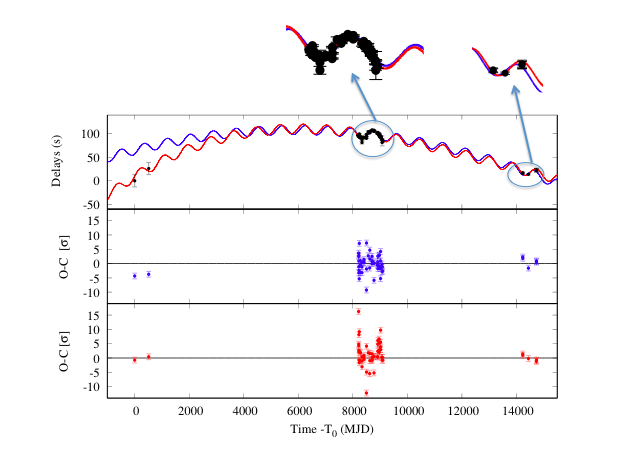}
\caption{Delays and O-C residuals of X~1658-298. Top panel: Estimated delays and best-fit curves taking into account a $\dot{P}_{orb}$  (blue curve) and both $\dot{P}_{orb}$ and $\ddot{P}_{orb}$ (red curve) in addition to a sinusoidal modulation with a period of 2.3 years; the insets show a magnification of the sinusoidal modulation possibly caused by the presence of a third body orbiting around the binary system. Central and bottom panels: residuals in units of $\sigma$ with respect to the blue and red model, respectively. In both the cases jitters are present.   Adapted from \cite{iaria_18}.}
\label{fig:1659}
\end{figure}

A unique  long-term orbital solution was obtained  for the transient eclipsing X~1658-298 spanning 40 yr,  
although the orbital solution is sampled at three epochs due to the transient nature of the source. 
Superimposed to the orbital solution of X~1658-298 a sinusoidal modulation is present (see Fig. \ref{fig:1659}); it  has been interpreted assuming the presence of a third body with a mass of $22\pm 3$ M$_J$ orbiting around the binary system, the estimated  revolution period is  $2.31\pm0.02$ yr (see \cite{iaria_18} and references therein).

\begin{figure}
\centering
\includegraphics[width=1.0\textwidth, angle=0]{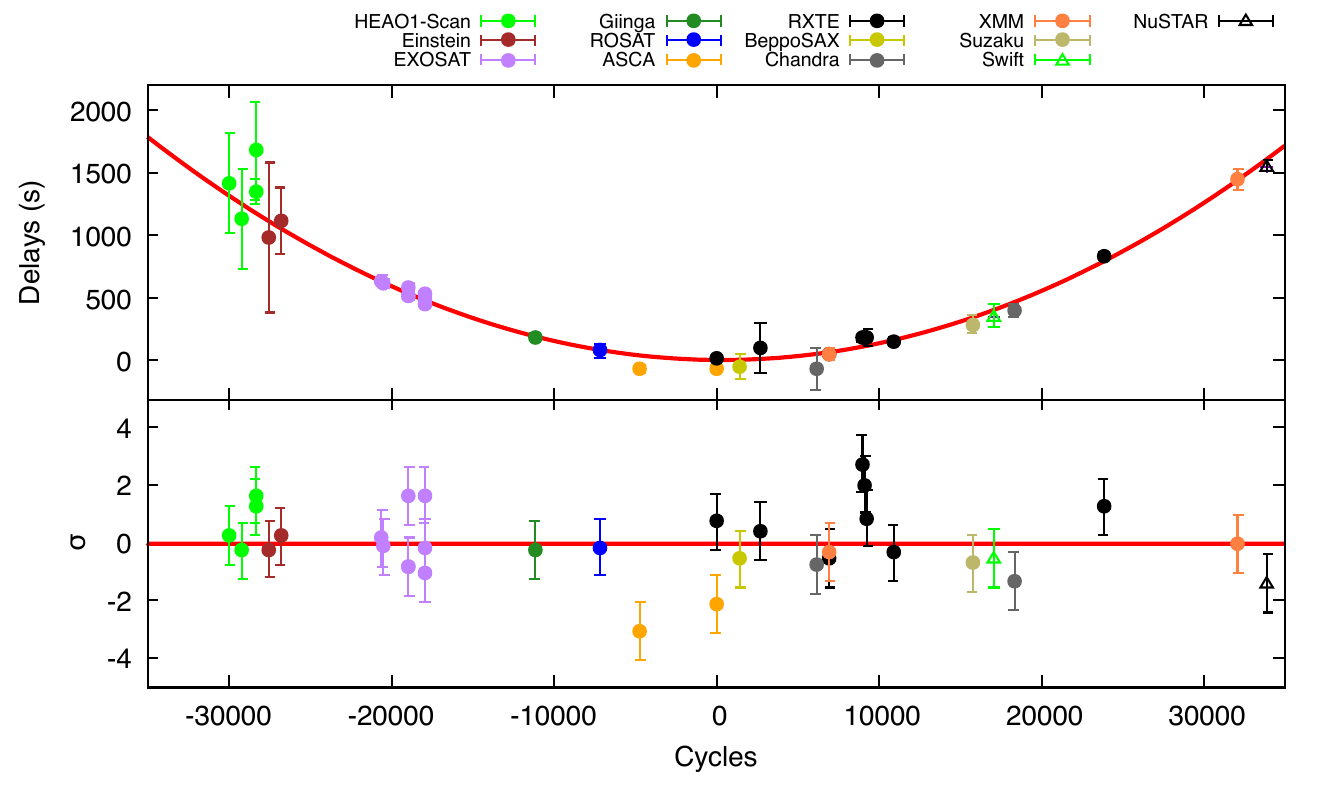}
\caption{Delays and O-C residuals of X 1822-371 over a timespan of 40 years. On the x-axis the orbital cycle is reported. Top panel:   Delays and best-fit curve taking into account a constant $\dot{P}_{orb}$. 
Bottom panel: residuals in units of $\sigma$ with respect to the model above. Courtesy of A. Anitra e R. Iaria 
(see \cite{anitra_21}, and references therein, for further details).}
\label{fig:1822}
\end{figure}

Among the ADC sources X~1822-371 is the prototype. It is a slowly rotating (spin period of $\sim 0.59$ s) X-ray pulsar with an orbital period of $\sim 5.57$ h, measured from the periodic eclipse of the X-ray source and confirmed through the coherent timing analysis of the X-ray pulsations. 
The delays of the eclipse arrival times with respect to a constant orbital period over 40 years show a clear parabolic trend (see Fig. \ref{fig:1822}), which implies a constant orbital period derivative, that is more than three orders of magnitude what is expected from conservative mass transfer driven by MB and GR (e.g.\ \cite{anitra_21} and references therein).

To explain the fast orbital expansion of X~1822-371 mechanisms based on the gravitational quadrupole coupling of the companion star with the orbit have been investigated; however, they resulted not suitable since the $\sim 0.3\, M_\odot$ companion star lacks enough internal (nuclear) power to produce such a large orbital period variation. A possible explanation is given by a highly not conservative mass transfer, in which the companion star transfers mass at a super-Eddington rate. Most of the transferred mass is then expelled from the system by the strong radiation pressure of the central source emitting at the Eddington limit. In fact, it has been proposed that 4U 1822-37 is accreting at the Eddington limit (and just $\sim 1\%$ of the total X-ray luminosity is visible due to the high inclination angle, $\sim 80^\circ - 85^\circ$), whilst the companion star is transferring at a higher rate (of the order of seven times the Eddington limit), and most of the transferred mass is expelled from the system by the radiation pressure producing strong outflows and winds (e.g.\ \cite{burderi_10}). The fact that the observed X-ray luminosity of the source may be an underestimation of the intrinsic source luminosity is also indicated by the timing solution for the NS spin which gives a long-term (over 20 yr) spin-up rate of about $-2.5 \times 10^{-12}$ s/s. For a source luminosity of $\sim 10^{36}$ erg/s, this spin-up rate implies an unphysical magnetic field of $8 \times 10^{16}$ G, whereas for a luminosity at the Eddington limit ($\sim 10^{38}$ erg/s) the the magnetic field assumes a more reasonable value of $8 \times 10^{10}$ Gauss.

\begin{table}
\begin{center}
\caption{Known values of $P$ and $\dot{P}$ of the NS-LMXBs at high inclination angles for which the O-C method has been applied}
\begin{tabular}{ c| c |c|c }
\hline
source& type &  $P$ (s) & $\dot{P}$ (s s$^{-1}$) \\  
\hline
XB 1916-053 &  dipper & $3000.65129(8)$ & $(1.46\pm 0.03)\times 10^{-11}$ \\  
X 1254-690 & dipper & $ 14160.004(6)$  & $(0.0\pm1.4)\times 10^{-10}$ \\ 
XB 1323-619 & dipper& $ 10590.896(2)$ &$ (0.8 \pm 1.3) \times  10^{-11}$ \\ 
XTE J1710-281 & transient, eclipse & $11811.82758(8)$  & $-1.6 \times 10^{-12} \leq \dot{P}  \leq 0.2 \times 10^{-12}$  \\   
AX J1745.6-2901 & transient, eclipse & $30063.6292(6)$  & $(-4.03 \pm 0.27) \times 
10^{-11}$  \\  
EXO 0748-676 & transient, eclipse & $\sim 13766.78$ &   \\ 
X 1658-298& transient, eclipse & $25617.9956(11)$ &$(-8.5\pm 1.2) \times 10^{-12}$ \\  
4U 2129+47 & ADC & $18857.594(7)$ &$ (1.03 \pm 0.13) \times 10^{-10}$ \\  
4U 1822-371  & ADC & $20054.2664(6)$& $(1.41 \pm 0.04) \times 10^{-10}$  \\ 
\hline
\hline

\end{tabular}
\label{tab:tab1ro}
\end{center}
\end{table}


\subsection{Accreting Millisecond Pulsars}
 \label{ss:AMXPs}
 
AMSPs are transient LMXBs that show a coherent modulation in their X-ray light curves at the NS spin period of a few milliseconds (see, e.g. the recent reviews by \citet{campana_18}, \citet{patruno_21} and Di Salvo \& Sanna \cite{DiSalvo_2021}). Their existence demonstrates that mass accretion in a LMXB can spin up a NS to such fast velocities. Already during the seventies, it was clear that accreting matter from a geometrically thin accretion disk would have yielded the high specific angular momentum of the inflowing plasma to the compact object \citep{bisnovatyi_76}. When radio astronomers uncovered a 1.6 ms millisecond pulsar \cite{backer_82}, scientists swiftly argued that a previous X-ray bright evolutionary phase had to have occurred. The weakly magnetized ($B\approx 10^8-10^9$~G) NS had to be spun up by the accretion of the mass transferred by a sub-solar companion star through a disk \cite{alpar_82,radhakrishnan_82}. Indeed, spotting an AMSP in RXTE data of the X-ray transient SAX J1808.4-3658 ended in 1998 a decadal search for a similar system \cite{wijnands_98}. Twenty years later, the list of AMSPs amounts to two dozen systems discovered at a rate of roughly one new source per year (see Table \ref{tab:amsp} for their properties). 

All the AMSPs are X-ray transients with outbursts typically lasting a few weeks, even if at least three sources have been active for several years (see below). In an outburst, they attain a peak luminosity of $10^{36}-10^{37}$ erg/s, i.e. not larger than 10\% of the Eddington rate. The X-ray pulse profiles are typically sinusoidal and described by a couple of harmonics. A fractional amplitude ranging from a few to 20-30 per cent characterizes the component at the  fundamental frequency. The motion in the binary system modulates the observed spin frequencies; modelling the Doppler shifts so induced allows setting a lower limit on the mass of the donor. So far, either main-sequence stars ($M_d\simeq 0.1- 0.5\, M_{\odot}$ , $P_{orb}\approx \mbox{a few hours}$), brown dwarfs ($M_d\approx 0.05\, M_{\odot}$ , $P_{orb}\simeq 1-2 {\rm hr}$) and white dwarfs ($M_d\approx 0.01\, M_{\odot}$ , $P_{orb}\simeq 40 {\rm min}$) have been identified as AMSPs companions. 

The long time it took to discover an AMSP did not owe only to the large effective area that an X-ray telescope requires to spot such relatively faint pulsators. The population of AMSPs is a small fraction of the $\sim 200$ LMXBs currently known. What causes such a rarity is still an open question. Channelling the in-flowing plasma to hotspots close to the magnetic poles of AMSPs yields the observed X-ray pulsations. For this to happen, the rotating magnetosphere of the NS has to truncate the disk flow before the matter reaches the NS surface. On the other hand, termination of the disk much further than the co-rotation radius (i.e., where the field lines of the NS rotate faster than disk plasma in Keplerian rotation) would quench accretion altogether through the propeller effect \cite{illarionov_75}. The co-rotation radius of an AMSP lies at a few tens of km from the NS, i.e. $R_{\rm co}=(GM_* P^2/4\pi^2)^{1/3}=30.9 (P/2.5 {\rm ms})^{2/3} (M/1.4 \, M_{\odot})$~km, and the range of disk truncation radii compatible with the observation of X-ray pulsations is narrow ($\simeq 10-30$~km for a 2.5~ms pulsar). Although the disk plasma is highly conductive, the magnetospheric field lines can thread it and get bent by the differential rotation with the Keplerian rotation of the disk material. When the magnetic stress so generated overcomes the viscous stress in the disk, the plasma starts to follow the magnetic field lines and fall towards the NS surface.
The size of the disk truncation radius depends on the physics of the disk/field interaction and is still a matter of debate. 
If the interaction occurs over a large radial extent, the truncation radius (often also known as magnetospheric radius) is a fraction $\xi$ of the Alfven radius $R_A$ \cite{ghosh_79}. The distance from the NS where the highest possible value of the magnetic stress ($\approx \mu^2 / 8 \pi r^6$) balances the density of gravitational stress of the in-falling plasma ($\approx (2G M)^{1/2} \dot{M}/4\pi r^{5/2} $) defines this radius. This yields:
\begin{equation}
    R_{in} = 15.4 \left(\frac{\xi}{0.5}\right) \left(\frac{\mu}{10^{26}\, {\rm G cm^3}}\right)^{4/7} \left(\frac{\dot{M}}{10^{16}\, \rm{g/s}}\right)^{-2/7} \left(\frac{M_*}{1.4\, M_{\odot}}\right)^{-1/7}\, km.
    \label{eq:rin}
\end{equation}

Alternatively, in the low diffusivity case, the lines cannot thread the disk fast enough and inflate and open as a consequence. The resulting inner disk radius is not far from the values obtained considering $\xi=0.5$ in Eq.(1) and has a flatter dependence on the magnetic moment and the accretion rate, $\propto \mu^{2/5} \dot{M}^{-1/5}$ \cite{spruit_93,dangelo_10}. Observing X-ray pulsations from LMXBs accreting at a mass accretion rate in the range of 0.1-10 times the Eddington rate allowed to estimate through Eq. (1) an NS magnetic field of $10^8-10^9$ G \cite[see, e.g.,][]{psaltis_99}.

\begin{table}[t!]
\begin{center}
\caption{Main properties of the AMSPs discovered so far. Information taken from \citet{patruno_21}, Di Salvo \& Sanna \cite{DiSalvo_2021} and Marino et al. \cite{marino_19}. The median mass of the companion star ${\tilde M}_d$ is evaluated for $i=60^{\circ}$ and a NS mass of 1.4 M$_\odot$. Companion type can be either a main sequence (MS), a white dwarf (WD) or a brown dwarf (BD). $N$ is the number of outbursts observed since the  year $Y$ in which pulsations have been first detected until December 2021. An asterisk marks long outbursts ($>$ year).  } 
\begin{tabular}{ l | c c c c c l}
source& $\nu$ (Hz) & $P_{orb}$ (hr)& ${\tilde M}_d $ ($M_{\odot}$) & Donor Type & $Y$& N \\
\hline
SAX J1808.4--3658	& 401	 & 2.01	& 0.049	& BD       & 1998  &  8  \\
XTE J1751--305 		& 435 	 & 0.71 	& 0.015 & WD	   & 2002  &  4	 \\
XTE J0929--314 		& 185 	 & 0.73 	& 0.009 & WD	   & 2002  &  1	 \\
XTE J1807--294 		& 190 	 & 0.67 	& 0.008 & WD	   & 2003  &  1	 \\
XTE J1814--338 		& 314 	 & 4.27 	& 0.194	& MS 	   & 2003  &  1	 \\
IGR J00291+5934 	& 599 	 & 2.46 	& 0.044 & BD	   & 2005  &  4	 \\
HETE J1900.1--2455 	& 377 	 & 1.39 	& 0.018 & BD	   & 2006  &  1* \\
Swift J1756.9--2508 	& 182 	 & 0.91 	& 0.008 & WD	   & 2007  &  4	 \\
Aql X--1		& 550 	 & 18.95 	& 	& MS 	   & 2008  &  	 \\
SAX J1748.9--2021 	& 442 	 & 8.77 	& 0.117	& MS/SubG? & 2008  &  7	 \\
IGR J17511--3057 	& 245 	 & 3.47 	& 0.157	& MS	   & 2010  &  2	 \\
Swift J1749.4--2807 	& 518 	 & 8.82 	& 0.711	& MS	   & 2010  &  1	 \\
NGC 6440 X--2 		& 206 	 & 0.95 	& 0.008 & WD	   & 2010  &  10 \\
IGR J17498--2921 	& 401 	 & 3.84 	& 0.194	& MS	   & 2011  &  1	 \\
IGR J18245--2452 	& 254 	 & 11.03 	& 0.204 & MS       & 2013  &  1	 \\
MAXI J0911--655		& 340	 & 0.74	& 0.026 & WD	   & 2016  &  1* \\
IGR J16597--3704	& 105	 & 0.77	& 0.007 & WD	   & 2017  &  1	 \\
IGR J17062--6143 	& 163	 & 0.63	& 0.007	& WD	   & 2017  &  1* \\
IGR J17591--2342	& 527	 & 8.80	& 0.416	& MS	   & 2018  &  1	 \\
IGR J17379--3747	& 468	 & 1.88	& 0.063 & BD	   & 2018  &  3  \\
IGR J17494--3030 	& 370    & 1.25	& 0.016	& WD       & 2020  &  1	 \\   
MAXI J1816--195 & 529 & 4.83 & 0.275 & MS & 2022 & 1 \\
MAXI J1957+032 & 314 & 1.01 & 0.017 & BD/WD & 2022 & 1 \\

\hline
\hline

\end{tabular}
\label{tab:amsp}
\end{center}
\end{table}

Thorough searches have not detected a coherent signal from at least 27 persistent and transient LMXBs, with typical upper limits of ~1\% on the pulse amplitude \cite[see][and references therein]{patruno_18}. Likely, most LMXBs  have no magnetosphere, or at least a very weak one. In brighter LMXBs, accretion of unmagnetized material would screen the magnetic field below the NS surface fast enough to impede its re-emergence through Ohmic diffusion \cite[see][and references therein]{cumming_01}. This model would explain why AMSPs are all relatively faint LMXBs. However, even a few comparably faint LMXBs do not show pulsations down to very stringent upper limits on the pulse amplitude \cite[see the discussion in][]{patruno_21}.

Pulsations of AMSPs have been observed throughout an outburst down to a luminosity of $\approx 10^{34}$~erg/s. This value is of the order of the sensitivity threshold of the available instruments at the distance of the closer AMSP known, SAX J1808.4--3658 (d=3.5 kpc). Remarkably, a few AMSPs showed pulsations only intermittently, providing a valuable benchmark to test why most LMXBs do not show X-ray pulsations. The X-ray pulsations of HETE J1900.1--2455 disappeared a few weeks after the onset of its discovery outburst and only reappeared intermittently during its ten-year-long period of X-ray activity \cite{galloway_07}. A decrease of the torque applied on the NS suggested a weakening of the magnetic field, compatible with the disappearance of pulsations below an amplitude of 0.1\% \cite{patruno_12}.  SAX J1748.9–2021 was observed either as a non-pulsing atoll LMXB, as an intermittent pulsar, or as a persistent AMSP \cite{altamirano_08}. MAXI J0911--655 in the globular cluster NGC 2808 has been active for five years, and its pulsations also disappeared after a while \cite{sanna_17A,bult_19}. It is suggestive that these three intermittent pulsars have a long-term X-ray luminosity ($>10^{34}$~erg/s) at least ten time larger than values typically found for AMSPs \cite{marino_19}, since this fits the magnetic screening scenario described above. However, quasi-persistent accretion is perhaps not enough to explain alone the pulse disappearance. X-ray emission from IGR J17062–6143 has persisted during the last $\approx 15$ years, and pulsations have never disappeared so far \cite{bult_21}. On the other hand, the prolific transient Aql X-1 was an even more extreme case as it showed coherent X-ray pulsations for just 120 s over a couple of Ms of data collected over the years \cite{casella_08}. In this case, the time scale of the magnetic field re-emergence would be way faster than expected according to the magnetic screening scenario. Therefore, it seems clear that the problem of understanding what determines the lack of pulsations from most LMXBs remains open.

\subsubsection{Accretion torques}

The possibility of observing a pulsar actually spinning-up due to accretion makes observations of AMSPs most appealing. Matter in the accretion disk rotates at a Keplerian rate $\omega_K(r)=\sqrt{GM_*/r^3}$ and yields its angular momentum to the NS when the magnetosphere drives its motion at $r=R_{in}$. The torque acting on the pulsar is expressed as $N_{\rm mat}\simeq\dot{M}\omega_K(R_{in})R_{in}^2=\dot{M}\sqrt{GM_*R_{in}}$. Accreting mass at a rate of $10^{17}$ g/s (i.e. roughly ten per cent of the Eddington rate) would spin-up a 1.4~$M_{\odot}$ NS with a moment of inertia of $I=10^{45}$~g~cm$^2$ at a rate $\dot{\nu}_{\rm mat}=N_{\rm mat}/(2\pi I)\simeq 3\times10^{-13} (\dot{M}/10^{17} {\rm g/s}) (R_{in}/20 {\rm km})^{1/2}$~Hz/s. Making the dependence of the truncation radius on the mass accretion rate explicit using Eq.~\ref{eq:rin} shows that the dependence of the material torque on the mass accretion rate is almost linear ($\dot{\nu}_{\rm mat}\propto \dot{M}^{6/7}$). In principle, measuring the variation of the spin-up torque during an outburst gives a dynamical estimate of the mass accretion rate onto the NS, and once the efficiency of the process $L_X/\dot{M}c^2$ is assumed, of the accretion luminosity. The proximity of the co-rotation radius of AMSPs to the NS radius also means that spin-down torques due to the field lines which threads the disk beyond the co-rotation surface are likely important \cite{ghosh_79}. The degree of diffusivity of magnetic field lines through the disk and the radial extent of the field/disk interaction set the magnitude of the magnetic spin-down torque. \citet{kluzniak_07} considered the interaction over a large radial extent and expressed such a torque as $N_{mag}=-\mu^2/9R_{0}[3-2(R_{co}/R_{0})^{3/2}]$, where $R_0$ is the radius at which the total torque vanished. This expression reduces to $\simeq \mu^2/9 R_{co}^3$ for a disk truncated near the co-rotation radius. D'Angelo et al. \cite{dangelo_12} considered that a lower field diffusivity made the interaction region much narrower $\Delta r/r<0.1$. A disk truncated near the co-rotation takes the angular momentum yielded by the NS and increases its density to redistribute it outwards. The truncation radius of the disk stays close to co-rotation even if the accretion rate further decreases (a so-called {\it trapped} state), and the NS spins down at a rate of the same order than the expression given above, $N_{mag}\simeq (\mu^2 / R_{in}^3) (\Delta r / R_{in}) $. The spin-down rate due to the field/disk interaction is of the order of $\dot{\nu}_{mag}\simeq - (1/2\pi I) \times \mu^2/9 R_{co}^3\simeq 6\times10^{-15} (B/10^8 {\rm G})^{2} (P/2.5 {\rm ms})^2$~Hz/s. This expression can become comparable to the spin-up term as soon as the magnetic field approaches $10^9$~G and the pulsar is not an extremely fast one. 

To measure such tiny rates of changes of the spin frequency one generally needs to apply timing techniques based on the coherent phase connection of the pulse observed at different times during an outburst. {\it RXTE} and later {\it NICER} performed high-cadence monitoring of the outbursts of AMSPs to measure their timing solution. However, finding accurate solutions was often troublesome. Only a few AMSPs showed a regular evolution of the phase of their X-ray pulse profiles during an outburst. The pulse phases measured by {\it RXTE} during the discovery outburst of IGR J00291+5934 followed a parabolic evolution compatible with a constant spin-up rate of $(5.1\pm0.3)\times10^{-13}$ Hz/s \cite[see, e.g.,][]{burderi_07}. Considering the expected dependence of the accretion torque on the mass accretion rate indicated a spin-up rate roughly ten times larger than expected from the observed flux (at a distance of ~4~kpc).
Unfortunately, {\it RXTE} ended its operations in 2012 and could not monitor the subsequent outburst shown by the source in 2015. A similar spin-up rate characterized data observed from other three AMSPs \citep[XTE J1751-305, XTE J1807-294 and IGR J17511-3057; see][and references therein]{DiSalvo_2021}.
even if irregular variations of the phase complicated the interpretation 
and required considering the pulse phase computed on higher-order harmonics to recover a {\it cleaner} pattern. The opposite curvature of the roughly parabolic trend of the pulse phases of four other AMSPs 
\citep[XTE J0929-314, XTE J1814-338, IGR J17498–2921, IGR J17591-2342, see][and references therein]{DiSalvo_2021} indicated a spin-down compatible with a slightly larger intensity of the magnetic field ($\simeq$ a few $\times 10^8-10^9$~G) compared to other AMSPs. In all cases, the limitations set by outbursts lasting a few weeks and the intrinsic tiny spin frequency derivatives reduced the statistical significance of the measurements. 

Most importantly, a correlation between the phase and the X-ray flux variations was present in a few cases 
\citet{patruno_09} attributed much of the phase variations to a correlation with the X-ray flux, likely due to corresponding movements of the hotspots on the NS surface. Under this hypothesis, the significance of many measurements of the spin frequency derivative would decrease to the point that even determining the sign would be out of reach. However, why AMSPs would show spin frequency derivatives lower by at least one order of magnitude than those predicted by accretion theories, with torques balancing out to a high degree of precision, would remain unexplained. However, Bult et al. \citep{bult_21} have recently managed to measure a spin-up rate of $(3.77\pm0.09)\times10^{-15}$~Hz/s for the quasi persistent AMSP IGR J17062-6143 over 12 years of observations. The spin-up was directly visible in the frequency space and did not suffer from possible biases related to the pulse phase timing noise. Indeed, the low spin-up rate supports a scenario in which the magnetic torque exerted by a $B\simeq 5\times10^8$~G on a disk truncated near the co-rotation radius balances almost perfectly the accretion torque. 

In this complicated picture, the pulse phase timing noise of SAX J1808.4--3658 even defies an easy correlation with the X-ray flux. Irregular movements of the phase computed on the first harmonic of the pulse prevented any meaningful measurement of the torques acting on the NS. However, \citet{burderi_06} noted that the second harmonic phases behaved more orderly and did not show the abrupt jumps seen for the first harmonic. A smoother evolution of even higher-order harmonics than odd ones could result from slight variations of the ratio between the fluxes received from the two almost antipodal hot spots on the surface of AMSPs \citep{riggio_11}. De-occultation of the antipodal spot as the disk recedes during an outburst is an intriguing alternative explanation of the observed behaviour \citep{ibragimov_09}.

\subsubsection{Spin frequency distribution}

Accretion of mass in an LMXB is very efficient in spinning up an NS. A major question is understanding how far it can go in accelerating the rotation of an NS. Spotting an NS spinning at a period close or below a millisecond would directly set constraints on the EoS of NS matter. Mass shedding at the equator for the maximum mass allowed by an EoS sets the minimum spin period accessible, $P_{min}\simeq 0.96 (M/M_{\odot})^{-1/2} (R/10 {\rm km})^{3/2} ms$  \citep[see, e.g.,][and references therein]{lattimer_12}.
Given a certain mass, a stiffer EoS describes large stars that could not spin faster than a thousand times per second. Accretion of mass increases the angular momentum of an NS as far as the Keplerian velocity of the in-flowing plasma is faster than the rotation of the NS field lines at the disk truncation radius $R_{in}$. The accretion torque is assumed to vanish as soon as $R_{in}$ equals the co-rotation radius. A spinning-up 1.4~$M_{\odot}$ NS subject only to accretion torques approaches an equilibrium period defined by this condition:
\begin{equation}
P_{eq}=\frac{2\pi R_{in}^{3/2}}{(G M_*)^{1/2}}\simeq 0.87 \left(\frac{\xi}{0.5}\right)^{3/2} \left(\frac{\mu}{10^{26}\, \rm{G\; cm^3}}\right)^{6/7} \left( \frac{\dot{M}}{10^{16}\, \rm{g/s}}
\right)^{-3/7} 
ms,
\end{equation}
obtained by making use of Eq.~\ref{eq:rin}. Regardless of the details on the expression of $R_{in}$, it is clear that a combination of a higher accretion rate and a lower magnetic field strength brings the truncation radius down to the NS surface. The equilibrium period obtained in that case is as low as $\simeq 0.5$~ms for a 1.4 $M_{\odot}$ NS. 

However, we have not yet found a sub-millisecond pulsar despite decades of searches and improvements in the detectors and the computing power. The fastest accreting millisecond pulsar known has a period of 1.6~ms \citep[IGR J00291+5934;][]{galloway_05}, just slightly slower than the quickest radio pulsar PSR J1748-2446ad \citep[1.4 ms;][]{hessels_06}. Chakrabarty \cite{chakrabarty_08} estimated the distribution of spin frequencies of AMSPs to be cut-off at $\simeq 730$~Hz. Burderi et al. \cite{burderi_01} attributed the absence of sub-millisecond pulsars to the switch on of a rotation-powered pulsar as soon as the mass transfer rate decreased below a given value. The pulsar wind would then efficiently eject the mass transferred by the donor star from the system (the so-called \textit{radio ejection} phase) and complicate any further accretion. 

Several works subsequently analyzed the distribution of the spin frequency of AMSPs and suggested they were on average faster than their assumed descendency of rotation-powered radio MSPs. Spin-down torques act on the NS as the companion star detaches from its Roche lobe and the mass transfer rate declines, possibly explaining this difference. Tauris \citep{tauris_12} used evolutionary calculations to show that the magnetic spin-down torque applied on the NS as $R_{in}$ equals and exceeds the co-rotation radius at the end of the secular mass transfer phase can dissipate more than half of the rotational energy of the pulsar. Papitto et al. \cite{papitto_14} argued that the number of AMSPs was still too low to single out a significant difference compared to radio MSPs but showed that LMXBs that show coherent oscillations only during type-I X-ray bursts were indeed faster than radio MSPs. Patruno et al. \citep{patruno_17} considered the spin distribution of all LMXBs and found statistical evidence for two sub-populations with distributions peaking at $\approx 300$ Hz (i.e. 3.3 ms, similar to radio MSPs) and $\approx 575$ Hz (i.e. 1.7 ms). The very narrow ($\sigma \approx 30$~Hz) distribution of the faster population of LMXBs suggested the existence of a very efficient mechanism to cut off the accretion driven spin-up. A spin-down torque related to the emission of gravitational waves from a rotating object with mass quadrupole $Q$ becomes very effective at high spin frequencies ($N_{\rm gw}\propto Q^2\nu^5$) compared to torques related to the rotation of a NS dipolar magnetic field $\mu$ ($N_{sd} \propto \mu^2 \nu^3$) and represents an intriguing option to cut-off the NS spin-up \cite{bildsten_98}. Either unstable modes of oscillation, crustal or magnetic strains can support the formation of asymmetries in the NS mass distribution \cite{lasky_15}. Although Patruno et al. \cite{patruno_12} showed that the spin equilibrium set by the disk/field interaction was alone enough to explain the observed properties of AMSPs, Bhattacharyya \& Chakrabarty \cite{bhattacharyya_17}
have recently highlighted the effect of transient accretion. They found that at a given average mass accretion rate, the spin equilibrium period of transients is much smaller than for persistent LMXBs. These calculations highlighted the requirement of an additional effect  to explain the observed distribution of spin frequencies. Gravitational radiation stands out as the most intriguing option.

\begin{figure}[t!]
\centering
\includegraphics[width=1.0\textwidth, angle=0]{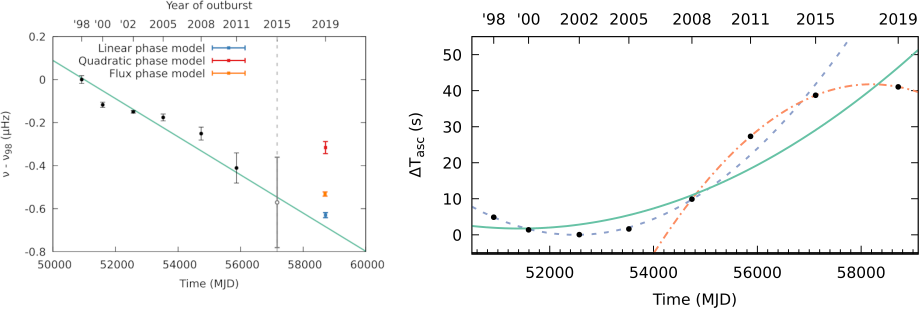}
\caption{\textit{Left:} Secular evolution of the spin frequency of SAX J1808.4--3658 calculated relatively to the value observed in 1998. Black points represent
measurements obtained with RXTE while colored squared represent the NICER measurements obtained for the 2019 outburst of the source for three different models. The solid line indicates the spin evolution best-fit model. \textit{Right:} Orbital evolution of SAXJ1808.4--3658. The dashed (blue), dashed-dot (red)
and solid (green) curves are the best fit parabola between the orbital phases measured in the intervals 1998-2008, 2008-2019,
and 1998-2019, respectively.  Credit: Bult et al., ApJ, 898, 38 (2020), \textcopyright AAS. Reproduced with permission.}
\label{fig:J1808}
\end{figure}

The observed long-term spin behaviour of AMSPs is a valuable testbed for theories of spin evolution. A direct comparison of the spin frequency measured in different consecutive outbursts allowed for a measurement of the spin evolution of AMSPs almost not impacted by the pulse phase timing noise affecting outbursts \citep[see][and references therein for a list of the available measurements]{DiSalvo_2021}. Between 1998 and 2019, SAX J1808.4-3658 has spun down at an average rate of $\dot{\nu}=-1.01(7)\times10^{-15}$~Hz/s (see the left panel of Fig. \ref{fig:J1808}, \cite{bult_21} and references therein).
The spin-down rate of IGR J00291+5934 was harder to estimate since the accretion-driven spin-up taking place during outbursts introduced a sawtooth-like behaviour of the spin frequency. The spin-down between the first two outbursts recorded \citep[see, e.g.][]{papitto_11}
was of the same order as in SAX J1808.4-3658 ($=(-4.1\pm1.2)\times10^{-15}$~Hz/s), while the 2015 accretion episode only allowed to set a coarser upper limit ($|\dot{\nu}|<6\times10^{-15}$~Hz/s).
XTE J175--305, SWIFT J1756.9--2508 and IGR J17494--3030 also showed a long-term spin-down ranging from $-5\times10^{-16}$ to $-2\times10^{-14}$~Hz/s.

The observed spin-down rates of AMSPs are compatible with magnetic fields in the range a few $\times 10^8-10^9$ G at the magnetic poles ($\dot{\nu}_{sd}=-(1/2\pi I) \times \mu^2 / R_{lc}^3$, where $R_{lc}=c P / 2 \pi = 119 (P/2.5\,\mbox{ms})$~km). On the other hand, assuming that the spin-down was due to gravitational radiation led to a relatively small measurement of the average NS mass quadrupole, $Q<2\times 10^{36}$~g~cm$^2$. Under the hypothesis that the emission of gravitational radiation governs the spin-down of AMSPs, the fastest spinning AMSP IGR J00291+5934 would decelerate at a rate $(599/401)^5\simeq 7.5$ times larger than SAX J1808.4-3658. The measurements obtained so far indicate a smaller ratio $\simeq 4\pm1$. This ratio is of the order of the value expected if magneto-dipole rotation is instead the dominant effect ($\simeq (599/401)^3\simeq 3$). Monitoring the future outbursts shown by these AMSPs and widening the sample of AMSPs with a measured long-term $\dot{\nu}$ will allow discriminating between the different models proposed to explain the long-term spin-down of AMSPs.

\subsubsection{X-ray spectra}

The X-ray spectra of AMSPs are typically hard throughout their outbursts. The spectra are dominated by a power-law $dN/dE \propto E^{-\Gamma}$ with a photon index $\Gamma\simeq 2$ which extends up to 100~keV or even beyond \citep{poutanen_06}. The commonly accepted model locates the origin of these hard X-ray photons in the accretion columns that develop above the NS polar caps. The freely-falling plasma moving along the magnetospheric field lines decelerates in a shock just above the NS surface. The electrons energized at the shock up-scatter the soft thermal photons emerging from the underlying surface yielding a spectrum extending up to their equilibrium temperature \cite{gierlinski_02,gierlisnki_05}. Modelling this hard component with a thermal Comptonization model in a slab geometry assumed for accretion columns indicated a moderately optically thick medium ($\tau\approx 1-3$) with electrons characterized by a temperature of $kT_e\simeq 25-50$~keV, up-scattering a seed blackbody spectrum with temperature $kT_{bb}\simeq 0.3-1.0$~keV emitted by a surface with a size $A_{bb}\simeq 20$~km$^2$ compatible with the expectations for a hot-spot \citep[see, e.g.][and references therein]{papitto_20}. 

\begin{figure}
\centering
\includegraphics[width=0.35\textwidth, angle=270]{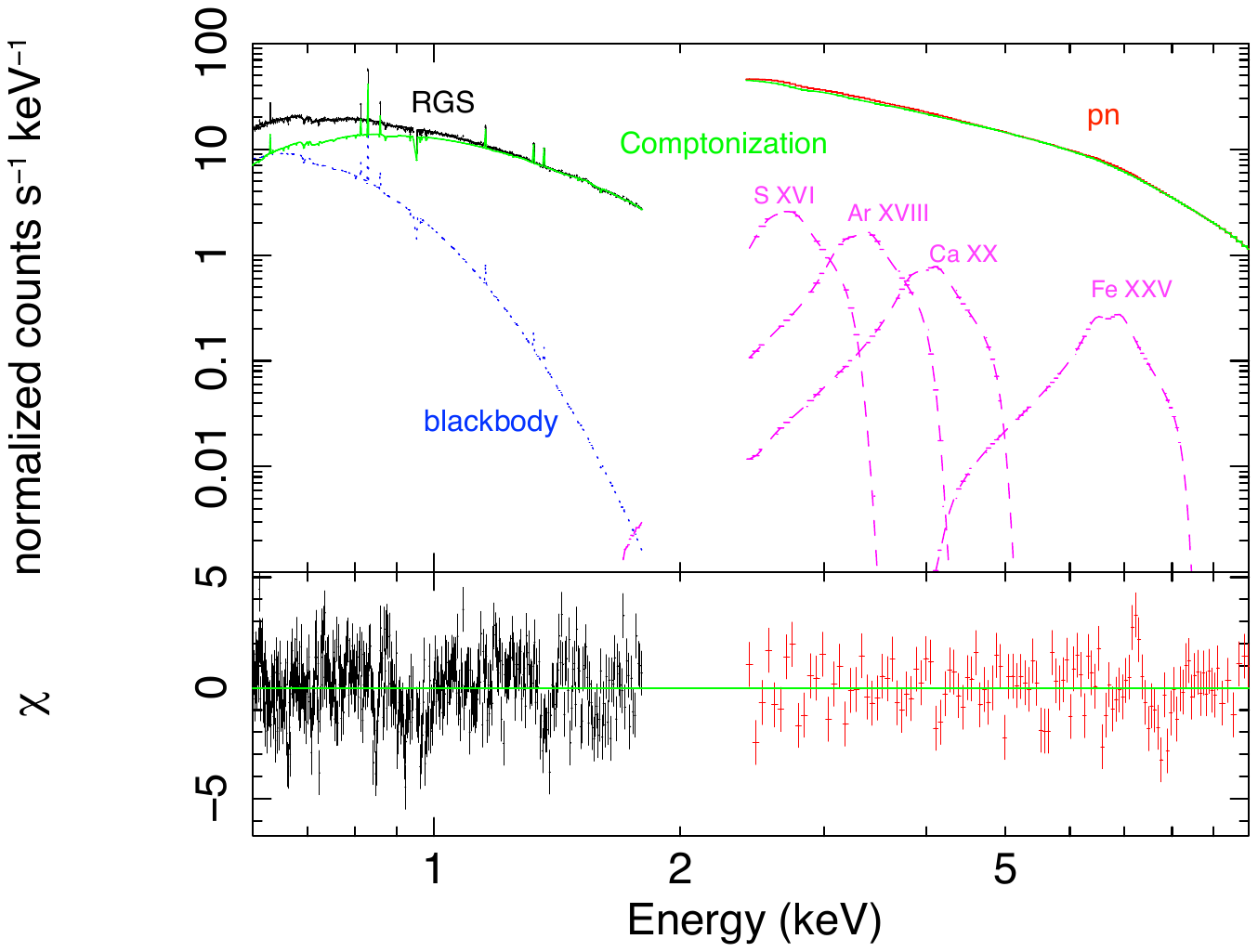}
\includegraphics[width=0.35\textwidth, angle=270]{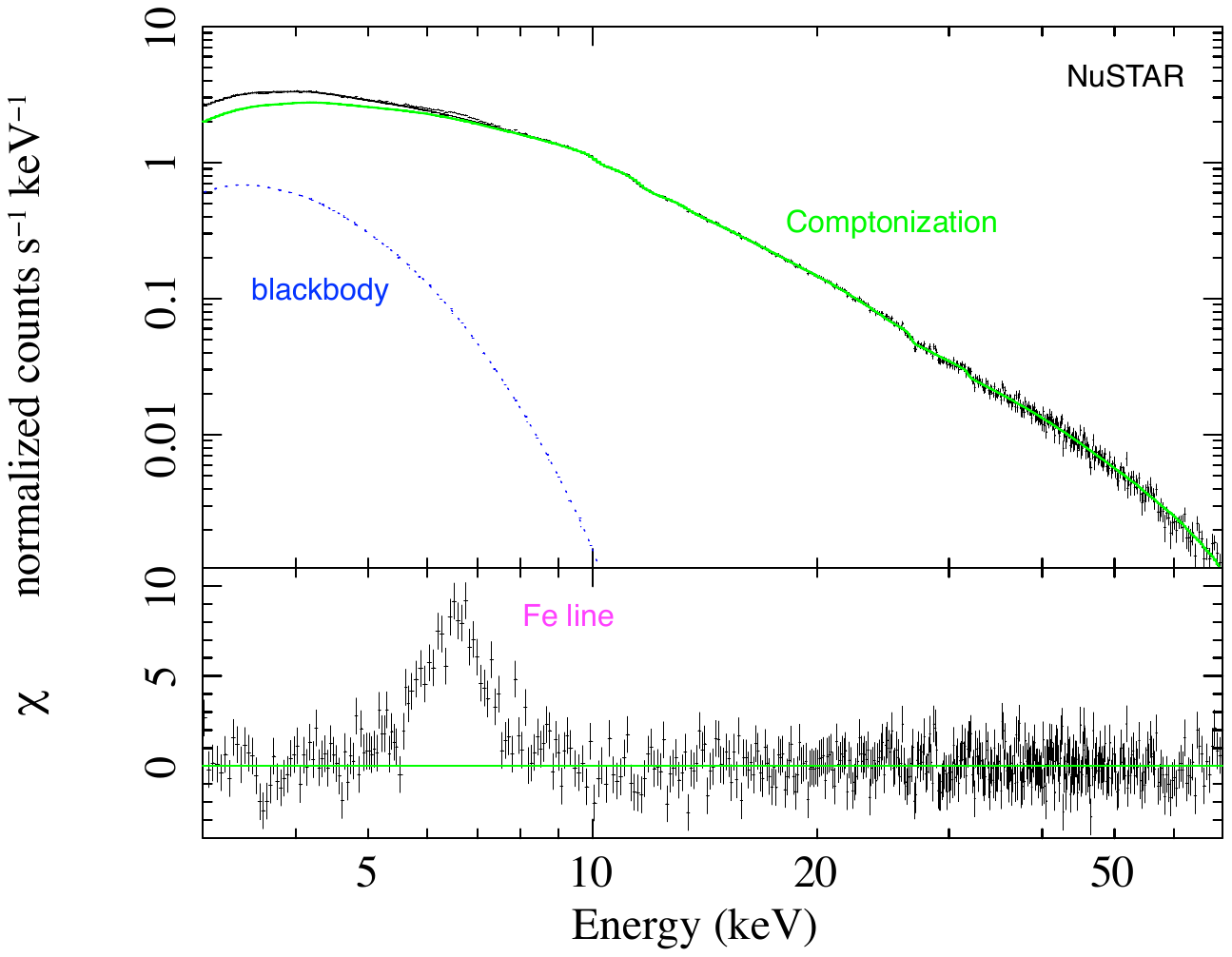}
\caption{\textit{Left:} XMM-Newton RGS and pn spectra of SAX J1808.4–3658 in the energy range $0.6 - 10$ keV (top) and residuals in units of $\sigma$ (bottom) with respect to the best-fit model. The model consists of a blackbody, the Comptonization component \textit{nthComp} and four disklines describing the reflection component. 
\textit{Right:} NuSTAR spectrum in the energy range $3 - 70$ keV (top) and residuals in units of $\sigma$ (bottom) with respect to the continuum model of SAX J1808.4–3658 when the Fe line normalization is set to 0. See \cite{disalvo_19} for further details}
\label{fig:saxj_spect}
\end{figure}

One or two thermal components characterize the softer X-ray energies ($kT\simeq 0.1-1$~keV). The normalization of the colder one ($A>100$~km$^2$) is compatible with emission from the inner rings of the truncated accretion disk, whereas the hotter one comes from the fraction of the NS surface that feeds the accretion columns with soft photons.
In addition, broad ($\sigma \sim$ a few keV) emission lines emerged at the energies of the Iron K-$\alpha$ transition \citep[6.4--7~keV;][]{papitto_09,Cackett_2010}
and sometimes also of ionized species of Sulfur, Argon and Calcium at lower energies \cite{disalvo_19}. 
The estimates of the inner disk radius given by modelling these lines with relativistic disk reflection models were compatible with the tight range expected for the observations of pulsations ($R_*<R_{in}<R_{co}$)
Fitting simultaneously the line profile and the spectral continuum measured at soft and hard X-rays by different telescopes (e.g., XMM-Newton and NuSTAR, see Fig.\ \ref{fig:saxj_spect} for details) further confirmed the presence of a disk reflection component which might play an important role to constrain the geometry of the system \citep[e.g., by allowing a measurement of the system's inclination; see, e.g.,][]{disalvo_19}. 


\subsubsection{Pulse profiles}

Two harmonic components are generally sufficient to model the pulse shapes of AMSPs. The fractional amplitude of the fundamental is typically lower than ten per cent and shows a dependence on the photon energy, even though different trends characterized the various sources. Some showed a marked increase going from a few to 10 keV 
, whereas other AMSPs showed a drop of the amplitude in the same energy range \citep[see][and references therein]{patruno_21}.
The two components contributing to the pulsed emission of AMSPs (thermal photons from the hotspots on the surface and harder photons upscattered in the accretion columns) served to interpret the latter behaviour. Photons upscattered in the accretion columns have a wider angular distribution than the soft thermal photons and yield a weaker modulation at the spin period \citep{poutanen_03}. 
The times of arrival of pulsed photons also showed a dependence on their energy. Photons of a few keV generally arrive later than harder photons by 100-200 µs. 
The broader emission pattern of photons up-scattered in the accretion columns than soft photons emitted from the hotspots possibly explains why an observer would see the former arrive earlier \citep{poutanen_03}.
Phase lags generally saturate around 8-10 keV. On the other hand, IGR J00291+5934 showed a reversal of the lags above ~6 keV. The higher number of scatterings required to reach energies of 50-100 keV could explain the longer time they take to reach the observer \citep[see, e.g.,][and references therein]{papitto_20}.

Modelling the X-ray pulse profiles produced from the surface of MSPs is one of the most powerful techniques to measure simultaneously the mass and the equatorial radius of an NS to constrain its EoS.
The gravitational field of an NS bends the trajectory and energy of the photons emitted by hot spots on its surface in a way proportional to the ratio $M_*/R_{eq}$. This effect causes a drop in the pulse amplitude compared to the non-relativistic case and introduces a characteristic energy dependence. The relativistic motion of the hot spots on the NS surface skews the pulse shape depending on the size of its equator $R_{eq}$. The X-ray pulse shape of MSPs thus encodes the desired information on the mass and radius of the NS \citep{poutanen_03}.
Recently, this technique led to the first measurements of the mass and radius of two isolated rotation-powered MSPs and highlighted the complex multipolar structure of their magnetic field \cite{riley_19, miller_19}. AMSPs are much brighter than isolated rotation-powered MSPs and represent a viable alternative. However, the pulse phase timing noise observed from a few AMSPs and the degeneracy between the geometrical parameters M and $R_{eq}$ and the spot geometrical parameters, such as the binary inclination $i$, the spot co-latitude $\theta$, prevented getting meaningful constraints from RXTE data \citep[see, e.g.,][]{poutanen_03}.
The re-opening of the polarimetric X-ray window with the Imaging X-ray Polarimetry Explorer (IXPE), and later with the planned enhanced X-ray Timing and Polarimetry mission (eXTP) might be a game-changer. 
The expected degree of polarization of the hard X-ray emission produced in the hot shocked regions above the NS surface is $\sim 10-20\%$ \cite{poutanen_20}. {\it IXPE} observations could be already enough to constrain $i$ and $\theta$ within a few degrees \citep{salmi_21}. 
Such estimates would allow to measure the mass and the radius of AMSPs with a relative uncertainty of a few per cent, comparable or slightly smaller than rotation-powered MSPs. 
X-ray polarimetry could then revive the prospects of using AMSPs to measure the EoS of NS.

\subsubsection{X-ray quiescence}
\label{amsp:quiescence}

From the peak of an outburst to the quiescence, the X-ray luminosity of AMSPs drops by more than four orders of magnitude (e.g. from a few $\times 10^{36}$ to $\simeq10^{32}$~erg/s). Following Eq.~\ref{eq:rin}, the truncation radius will expand from a few tens of km to a few hundred. Firstly, it will exceed the co-rotation radius (31 km for a 1.4~$M_{\odot}$ NS spinning at 2.5~ms, such as SAX J1808.4-3658). When this happens, the disk matter cannot provide angular momentum anymore to the NS, and rather it is the rotating magnetosphere that speeds up the rotating disk plasma. Ilarionov \& Sunyaev \cite{illarionov_75} first considered this situation and argued that the magnetospheric centrifugal barrier would quench accretion on the NS altogether. Actually, the disk matter would be flung out of the system only if the angular momentum deposited by the magnetosphere in the disk exceeds the escape velocity ($R_{in}>2^{1/3} R_{co}$, \citep[][]{spruit_93}). For $R_{in} \simeq R_{co}$ the disk readjust itself to redistribute the excess angular momentum outwards, allowing accretion to proceed albeit at a reduced rate \cite{dangelo_10,dangelo_12}. 3D magneto-hydrodynamic simulations confirmed that accretion and ejection of plasma coexist in the propeller state \citep[see][and references therein]{romanova_18}. It is a non-stationary cycle made of matter building up beyond the magnetospheric radius, pushing the inner radius of the disk inwards, opening and inflating the field lines of the magnetosphere (with both accretion and ejection taking place), which expands and re-start the cycle. The efficiency of mass ejection is minimal in the {\it weak} propeller conditions ($R_{in}\geq R_{co}$) and is maximal in the {\it strong} propeller state ($R_{in}> 5 R_{co}$), in which conical jets and a strong wind are launched. The persistence of accretion during the propeller state would explain why pulsations of SAX J1808.4--3658 are detected down to a luminosity of at least a few $10^{34}$~erg/s (the sensitivity limit of {\it RXTE}), roughly a hundred times weaker than the peak luminosity. 
Radial oscillations of the magnetospheric boundary located close to or just beyond the co-rotation boundary are the most likely explanation of the  $\sim 1$~Hz quasi-periodic oscillation appeared in the later stages of some of the outbursts of SAX J1808.4--3658 \cite{Patruno_2009c}. Transitions between a {\it weak} and a {\it strong} propeller occurring on a time scale shorter than a second also provided a satisfactory explanation of the bi-modal magnitude and spectral variability of the transitional millisecond pulsar IGR J18245--2452 in outburst \cite{ferrigno_14}.

In quiescence, the X-ray luminosity of AMSPs is much lower ($L_X\simeq 5 \times 10^{31}$~erg/s for SAX J1808.4--3658). A $\Gamma\simeq 1.4$ power-law described the quiescent X-ray spectrum of SAX J1808.4--3658 \citep[e.g.][]{campana_02}. 
The NS atmosphere had a temperature of at most $\approx 30$~eV already a couple of years after the last X-ray outburst, requiring a fast atmospheric cooling likely related to URCA neutrino emission processes involving protons, hyperons or deconfined quarks \citep{heinke_07}. IGR J00291+5934 is the other AMSP with well defined quiescent properties \citep[e.g.,][]{campana_08}.
A thermal component with a temperature of 50--100~eV (depending on the assumed model) coexisted with the power-law component, indicating a slower cooling which could be related to a lower NS mass. 

Even assuming that residual accretion powers the emission of AMSPs in quiescence, the low X-ray luminosity indicates that the inner disk radius would also exceed the light cylinder radius ($R_{LC}=c P/2\pi = 120\, (P/2.5 {\rm ms})$~km). According to conventional wisdom, the presence of high-density plasma inside the light cylinder prevents the rotation-powered pulsar, or at least the radio-emission mechanism, from working.
Emptying the light cylinder would instead allow the switch-on of a rotation-powered pulsar whose relativistic wind would then eject the plasma transferred by the companion from the system \citep{burderi_01}. Expectations were that a radio pulsar would be active during the quiescent states of transient LMXBs \citep{stella_94}.
Some pieces of evidence indirectly support this scenario. The X-ray luminosity observed in quiescence from SAX J1808.4--3658 was not high enough to irradiate the companion star up to the magnitude observed in the optical band; the spin-down luminosity of a rotation-powered pulsar with a magnetic field of a few $\times 10^8$~G represented an intriguing solution to the problem \citep{burderi_03}.
The orbit of SAX J1808.4--3658 expanded at a rate much faster than conservative mass transfer can explain, whereas ejection of the mass transferred by the companion from the vicinity of the inner Lagrangian point that connects the binary Roche lobes would remove enough angular momentum \citep{disalvo_08}.
Lastly, the emission of a rotation-powered pulsar would explain the detection of a gamma-ray counterpart to SAX J1808.4--3658 \citep{deona_16}. 
However, deep searches for pulsations in the radio band did not succeed \cite[see][and references therein]{patruno_17}. Absorption or scattering of radio waves by matter enshrouding the system could explain such a non-detection. 

Discovering an AMSP that behaves as a radio pulsar during the X-ray quiescence changed the picture \citep{papitto_13}. IGR J18245-2452, with a relatively long orbital period of $\sim 11$ h, is the only transitional millisecond pulsar (see Sec.\ \ref{ss:tMSP}) observed in a bright X-ray outburst. Long eclipses of the radio pulsations observed in quiescence proved the presence of matter ejected by the pulsar wind and partly enshrouding the binary system. Undetected during the accretion outburst, it took less than a couple of weeks for radio pulsations to reappear after it had ended. Observations of IGR J18245-2452 eventually proved that variations of the mass accretion rate make a pulsar swing from a rotation to an accretion powered regime and vice versa. 

The optical and ultraviolet pulsations detected from SAX J1808.4-3658 gave further indication of emission powered by the rotation of the NS magnetic field \citep{ambrosino_21}. Strikingly, the detection took place during the rising and the declining phase of an accretion outburst. Self-absorbed cyclotron emission in accretion columns entrained by a $\approx 10^8$~G magnetic field is a viable mechanism to explain the pulsed emission observed from cataclysmic variables at those wavelengths. However, the optical/UV pulsed luminosity of SAX J1808.4--3658 exceeded by two orders of magnitudes what this process can produce. Similar reasoning rules out reprocessed emission from the irradiated disk. In addition, the sinusoidal shape of the pulse profiles argues against a substantial beaming of the radiation. A mechanism related to the acceleration of electrons and positrons to relativistic energies by the pulsar rotating electromagnetic field seems in order. Enveloping a significant fraction of the pulsar wind by the disk surrounding the pulsar would help attain the high efficiency in converting the spin-down power into radiation required by the observed pulsed optical/UV luminosity. Similar considerations hold for the case of the transitional millisecond pulsar PSR J1023+0038 (see Sec.\ \ref{ss:tMSP}). Strikingly, magnetospheric particle acceleration in SAX J1808.4--3658 would coexist with accretion onto the polar caps, a possibility generally excluded for pulsars.

\subsubsection{Binary evolution}

The transitional millisecond pulsar IGR J18245-2452 bridged the population of AMSPs and irregularly eclipsed rotation-powered MSPs. 
The phenomenology of the latter class of pulsars descends from the interaction between the relativistic pulsar wind and the matter lost by the companion star and ejected from the system. These eclipsed pulsars belong to close binary systems ($P_{orb}<{\rm day}$) and are called \textit{redbacks} or \textit{black widows} depending on the nature of the companion star. A non-degenerate hydrogen-rich star with $M_d\simeq 0.1-1\,M_{\odot}$ is the companion of redback pulsars, whereas a low mass ($M_d<0.06\,M_{\odot}$) degenerate star is the companion of black widows. The existence of a link between AMSPs, redbacks and black widows is not surprising. They all belong to binary systems with a donor that is losing mass and close enough that the energy densities of the accretion flow and the pulsar wind are comparable over the typical binary scale. Slight variations of the accretion rate might then push a system into either the accretion or the rotation-powered regime. 

Observing the variations in the orbital period of AMSPs is a powerful tool to investigate the binary evolution. The fast expansion of the orbit of SAX J1808.4--3658 ($\dot{P}_{orb}\simeq$ a few $\times 10^{-12}$; see the right panel of Fig. \ref{fig:J1808}) is reminiscent of the evolution observed from black widow pulsars, and strongly suggests a non-conservative evolution in which a large fraction of the matter transferred by the donor is ejected by the system \cite{disalvo_08}. In addition, the expected values of the average X-ray luminosity of AMSPs obtained assuming a conservative binary evolution governed by magnetic braking and emission of gravitational waves are much larger than the values obtained by averaging the X-ray emission observed during the oubursts with the much longer quiescence intervals \cite{marino_19}. This finding further strengthened the hypothesis that non-conservative mass transfer characterizes the class of AMSPs as a whole. However, the similarity between SAX J1808.4--3658 and black widows also extended to the quasi-cyclic variations of the orbital phase observed during its last few outbursts \citep[see, e.g.,][]{bult_20, illiano_23b} .
An intense short-term angular momentum exchange between the mass donor and the orbit caused by the gravitational quadrupole coupling with the variable oblateness of the companion might explain such quasi-cyclic variations of the orbital period \citep{applegate_94}. Monitoring the future outbursts of SAX J1808.4--3658 will help understand if the observed variations of the orbital phase are a good tracer of the secular expansion of the orbit or they rather reflect shorter-term variations. 

Interestingly, the orbital evolution of IGR J00291+5934 is much slower ($\dot{P}_{orb}< 5 \times 10^{-13}$) and is compatible with the evolution driven by the emission of gravitational radiation 
\citep[see, e.g.,][]{sanna_17}. Tailo et al. \cite{tailo_18} performed binary evolution calculations and showed the significant role that irradiation of the companion star by the pulsar emission has to play to reproduce the current characteristics of these AMSPs. Irradiation induces cycles of expansion of the donor star followed by mass ejection and contraction on a thermal time scale. SAX J1808.4--3658 and IGR J00291+5934 might be at different stages of this evolution; the former would evolve rapidly by losing mass at high rates, the latter would experience mass transfer mass at much lower rates and evolve much more slowly. 

\subsection{Transitional millisecond pulsars}
\label{ss:tMSP}

Transitional millisecond pulsars (TMSPs) are a small subset of LMXBs,  observed at different times either in an accretion disk state or a rotation-powered radio pulsar regime \citep[see the recent review by Papitto \& de Martino][]{papitto_21}. Variations in the mass accretion rate are the driver of the state transitions observed so far from three systems (see Table~\ref{tab:tmsp}). TMSPs allowed us to witness the possible outcomes of the interaction between the electromagnetic field of an MSP and the matter lost by the companion star. 

When rotation-powered, TMSPs behave as redback pulsars \citep[see, e.g., ][]{strader_19}. Matter ejected by the pulsar wind from the inner Lagrangian Point irregularly eclipses radio pulsations. The pulsar wind/plasma interaction creates an intrabinary shock where particles are accelerated to relativistic energies and radiate synchrotron photons \citep{bogdanov_11}.  Viewing the intrabinary shock at different angles along the orbit yields a distinctive orbital modulation of the X-ray emission so produced. Irradiation of the companion star by the high-energy emission of the pulsar also determines an orbital sinusoidal modulation of the optical emission \citep[see, e.g.,][]{demartino_15}. 

The behaviour of TMSPs surrounded by an accretion disk is much more peculiar. Only IGR J18245--2452 showed a relatively bright outburst with properties typical of AMSPs (see Sec.~\ref{amsp:quiescence}). When accretion powered, the other two TMSPs PSR J1023+0038 \citep{archibald_09} and XSS J12270-4859 \cite{bassa_14} stayed for years in a very faint state ($L_X\simeq \mbox{a few} \times 10^{33}$~erg/s) with peculiar multiwavelength properties. Note that IGR J18245-2452 also featured this {\it sub-luminous} disk state, although for a shorter interval. What determines whether a source shows a bright outburst or a much fainter state remains unexplained. 

\begin{figure}
\centering
\includegraphics[width=1.0\textwidth, angle=0]{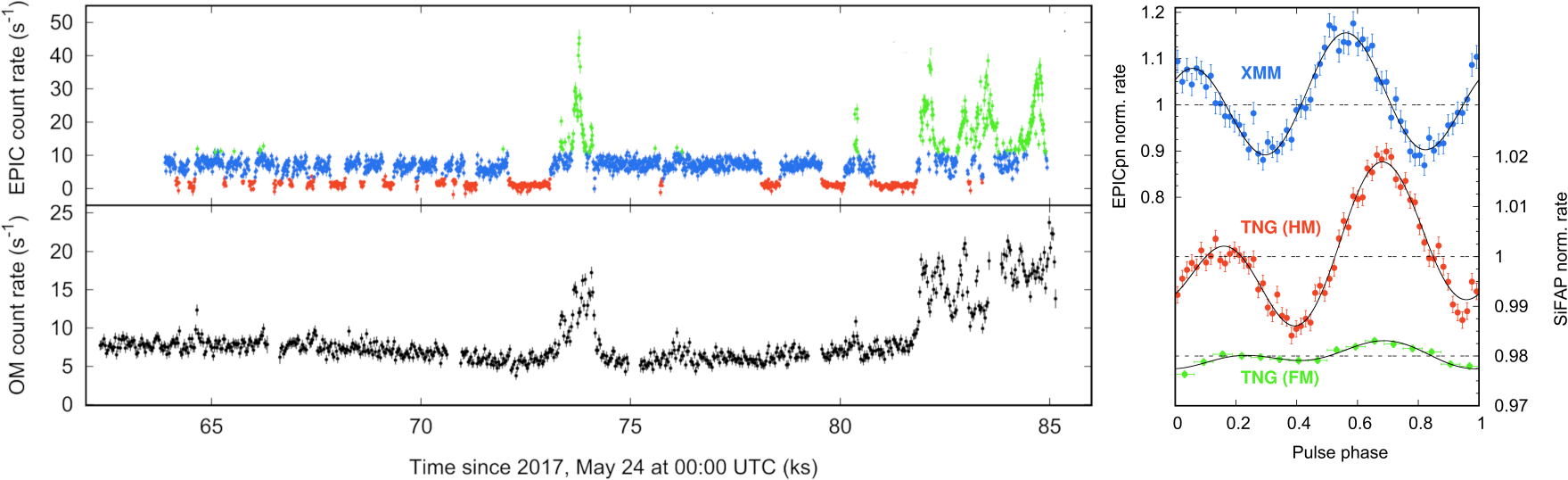}
\caption{\textit{Left:} XMM-Newton $0.3–10$ keV EPIC light curve of PSR J1023+0038 in the {\it sub-luminous} disk state, binned every 10 s (top) and optical monitor light curve, binned every 30 s (bottom) . Blue, red and green points indicate high-, low-, and flaring mode intervals. 
\textit{Right:} X-ray pulse profile observed during the high mode (blue points) and the optical pulse profiles in the high (red points) and flares (green points) detected by SiFAP2 at the INAF TNG Galileo Telescope. Credit: Papitto et al., ApJ, 882, 104 (2019), \textcopyright AAS. Reproduced with permission.}
\label{fig:J1023_modes}
\end{figure}

PSR J1023+0038 is the best-studied case of a TMSP in the {\it sub-luminous} disk state, although the others show similar phenomenology. The X-ray light curve of TMSPs in this state shows rapid ($\sim 10$~s) switching between two roughly constant flux levels usually dubbed as {\it high} and {\it low} modes \cite{bogdanov_15} (an example is shown in the left panel of Fig.\ \ref{fig:J1023_modes}). The ratio between the flux in the {\it high} and {\it low} modes is $\approx 5-10$. Simultaneous transitions show up in the ultraviolet, optical and near-infrared bands. However, for the latter two, the sign of the correlation with the X-ray/UV band is uncertain. X-ray \cite{archibald_15} and optical (in the case of PSR J1023+0038, \cite{ambrosino_17}) pulsations appear during the {\it high} mode and disappear in the {\it low modes}. Optical pulsations of PSR J1023+0038 lag X-ray ones by $\sim 100-200$~µs and maintain such a delay over the years (\cite{papitto_19, illiano_23a}, see the right panel of  Fig.\ \ref{fig:J1023_modes}). The pulsar spins down at almost the same rate shown during the radio pulsar state \cite{jaodand_16}. In addition, the gamma-ray emission of TMSPs increases by a factor of a few when an accretion disk appears in the system \citep[see, e.g.,][and references therein]{torres_17}. The flat-spectrum radio emission is compatible with the presence of a jet, while the brightening observed simultaneously to X-ray {\it low} modes suggests the ejection of plasmoids \citep{bogdanov_18}. The properties of TMSPs in the {\it sub-luminous} disk state (e.g., a gamma-ray counterpart with a 0.1--100~GeV flux a few times larger than the 0.1--10~keV X-ray flux) are so peculiar that they have been used to identify at least five strong candidates TMSPs \citep[see][and references therein]{papitto_20}.

The observation of characteristics typical of either accreting LMXBs (a disk, enhanced X-ray emission although fainter than bright LMXBs, X-ray pulsations) or rotation-powered pulsars (enhanced gamma-ray emission, spin-down rate, optical pulses) complicate the identification of the nature of the {\it sub-luminous} state. This duplicity is understood considering that the accretion luminosity of TMSPs in the {\it sub-luminous} disk state is of the same order of the pulsar spin-down power ($\approx \mbox{a few} \times 10^{34}$~erg/s), and both processes must be relevant. Earlier models argued that TMSPs were rotation-powered pulsars enshrouded by plasma of the intrabinary shock located far ($\sim10^{9}-10^{10}$~cm) from the pulsar \citep[see, e.g.,][]{cotizelati_14}. The observation of X-ray pulsations and fast switching between X-ray modes hardly fitted this scenario, suggesting that the accretion flow reached the NS surface, albeit at a low rate. Locking of the disk in a trapped low-$\dot{M}$ state \citep{dangelo_12} or the ejection of most of the disk matter through the propeller effect could explain the low X-ray luminosity observed \citep{papitto_14}. More recently, Papitto et al. \cite{papitto_19} and Veledina et al. \cite{veledina_19} reconsidered the hypothesis of an enshrouded rotation-powered pulsar after the discovery of optical pulsations from PSR J1023+0038. Similar problems arise in explaining such a phenomenon in terms of accretion to those mentioned for SAX J1808.4--3658 (see Sect.~\ref{amsp:quiescence}), only exacerbated by the phase lags of optical pulse compared to X-ray ones. Emission of pulsed X-ray and optical synchrotron emission from the termination shock of the pulsar wind by the disk just outside the light cylinder would provide a satisfactory explanation in which the rotation of the magnetic field and mass accretion coexist. In all the mentioned scenarios, the transitions between {\it high} and {\it low} modes could reflect the movement of the disk truncation radius related to changes of the physical mechanism powering the emission (e.g. from the propeller to the rotation-powered state, \cite{campana_16}, or vice versa, \cite{veledina_19}, or from the light-cylinder radius to outwards, \cite{baglio_23, papitto_19}). Reconnection of the magnetic field lines threading the disk (or the donor star) could instead explain the flares observed from the X-rays to the near-infrared band \citep{campana_19}. 

\begin{table}[t!]
\begin{center}
\caption{Main properties of the TMSPs discovered so far. Information taken from Papitto \& de Martino \cite{papitto_21}. The median mass of the companion star ${\tilde M}_d$ is evaluated for $i=60^{\circ}$ and a NS mass of 1.4 M$_\odot$. Companion type can be either a main sequence (MS), a white dwarf (WD) or a brown dwarf (BD). Observed states are labelled as radio-pulsar (RP), {\it sub-luminous} disk state (SLD) and accretion outburst (OUT). Candidates identified on the basis of the similarities of the multiwavelength emission with that observed from TMSPs in the {\it sub-luminous} disk state are listed in the lower part of the table.} 
\begin{tabular}{ l c c c c c l}
TMSPs& $\nu$ (Hz) & $P_{orb}$ (hr)& ${\tilde M}_d $ ($M_{\odot}$) & Donor Type & Obs. states \\
\hline
PSR J1023+0038  & 592 & 4.75 & 0.16 & MS & RP,SLD & \\
XSS J12270--4859 & 592 & 6.91 & 0.25 & MS & RP,SLD \\
IGR J18245--2452 & 254 & 11.0 & 0.21 & MS & RP,SLD,OUT \\
\\

Candidate TMSPs            & $\nu$ (Hz) & $P_{orb}$ (hr)& Modes &  Gamma-rays   & Obs. states\\
\hline
RXS J154439.4--112820   & - &   5.8 &   yes &     yes     & SLD     \\
CXOU J110926.4--650224  & - &       &   yes &     yes     & SLD        \\
4FGL J0407.7--5702       & - &  -    &   ?   &   yes    & SLD         \\
3FGL J0427.9--6704      & - &  8.8   &   flares   &  yes   & SLD          \\
4FGL J0540.0-7552   & - & -  &   flares   &     yes  & RP(?),SLD  \\
\hline
\hline

\end{tabular}
\label{tab:tmsp}
\end{center}
\end{table}


\subsection{Faint and very faint sources}
\label{ss:VFXT}
As already discussed in Subsection \ref{ss:transients}, transient X-ray binaries spend most of their lifetime in quiescence (at very low X-ray luminosities, $<10^{34}$ erg/s, and accretion almost switched off) and sporadically undergo "bright" outbursts, in a range of X-ray luminosities comprised between 10$^{36}$-10$^{39}$ erg/s. Most of the phenomenology presented in this chapter has been observed during these bright outbursts, during which we see accretion at its full power. However, X-ray transients can also undergo fainter outbursts, with X-ray luminosities at the peak of the outburst about 10$^{36}$ erg/s, or even lower. X-ray transients exhibiting this sub-luminous accretion regime are classified as Very Faint X-ray Transients (VFXTs) and they may outnumber the standard "bright" X-ray binaries in the Galactic Center \citep{Muno_2005}. A further sub-classification can be sometimes found between "Faint" and "Very Faint" X-ray Transients depending on the achieved luminosity, i.e. between $10^{35}$-$10^{36}$ erg/s, and $10^{34}$-$10^{35}$ erg/s, respectively. Their generally low fluxes make these objects particularly elusive and indeed they were mainly discovered serendipitously during surveys of the Galactic Center with {\it Swift}, {\it Chandra}, or {\it XMM-Newton}, \citep[see, e.g. ][]{Muno_2005, Bahramian_2021}. These faint outbursts are indeed too dim to be caught by X-ray All-Sky Monitors of the past, such as ASM onboard {\it RXTE}, or of the present, such as BAT onboard {\it Swift} or {\it MAXI}, making dedicated pointings our only chance to investigate this peculiar behaviour. Being so challenging to observe, the number of observations of these objects has been so far rather limited, and, as a consequence, the nature of these objects is often not yet established. 

"VFXTs" is indeed an "umbrella"-term enclosing different species of X-ray emitting sources; a large number of these objects has been identified as NS LMXBs (which justifies their presence in this chapter) mostly thanks to the detection of type-I X-ray bursts (Sec. \ref{ss:bursts}), but among the unidentified objects there are sources suggested to be BH X-ray binaries and (at least in one case) NS High Mass X-ray Binaries (see \cite[][]{Bahramian_2021} for a review). Furthermore,  bright Cataclismic Variables (CVs),  magnetars and symbiotic stars (compact objects accreting matter transferred from a red giant companion through winds)  are in principle able to achieve the range of luminosity associated to VFXTs behaviour as well. It is also noteworthy that an unambiguous inference of the distance of X-ray sources is often not available, making estimates of the X-ray luminosity - and subsequent VFXT classification - sometimes unreliable. It can not be excluded that some of the identified VFXTs may simply be standard "bright" LMXBs in outbursts located at high distances. In the following, we refer to the systems already identified as accreting NSs with low magnetic fields, for coherence with the topic of the chapter. 

Analogously to bright outbursts from X-ray binaries, "faint" outbursts range from being short, from days to weeks, to last for years, as in the case of the quasi-persistent VFXT XMMU J174716.1-281048 \citep{DelSanto_2007}. Some X-ray binaries have displayed both "faint" and "bright" outbursts, the so-called "hybrid" VFXTs \citep{Marino_2019b}. An even more peculiar case is represented by "burst-only" system, NS LMXBs detected only during type-I X-ray bursts
but with a "faint" persistent luminosity.  
VFXTs spectra show two components, one hard, power-law like, presumably connected to the accretion flow, and a soft thermal component at low temperatures, below 0.5 keV (see e.g.\ \citep[][]{Wijnands_2015}, and references therein), presumably radiated from the NS surface. In systems fading from 10$^{35}$ erg/s to 10$^{34}$ erg/s, a progressive spectral softening has been observed, due to a decrease in both the blackbody temperature and the contribution from the hard component. The typically low fluxes detected from these systems make the obtained spectra rather poor, so that a more detailed spectral modelling is often out of reach (but see \citep[][]{Degenaar_2017}, for the possible detection of a reflection component and an absorption line in a VFXT). 

Several explanations have been put forward for the usually long-lived, weak accretion regimes exhibited by VFXTs (see \citep[][]{Heinke_2015} for a review). As several VFXTs are UCXBs, i.e. with orbital periods of tens of minutes, the faint luminosity of these objects has been ascribed to their small-sized accretion disks which could fit in such compact orbits. However, this theory can not be universal, as witnessed by several UCXBs reaching bright luminosities, e.g. 4U 1820-30, and several VFXTs with measured orbital periods higher than one hour, e.g. Swift J1749.4-2807. Furthermore, high mass-transfer rates induced by gravitational radiation are expected for such short-period systems, making the observed faint luminosity odd, unless strongly non-conservative mass-transfer scenarios are invoked (e.g.\ \citep[][]{DiSalvo_2008}). Alternatively, the accretion flow in these objects can be truncated at relatively large distances from the NS due to the magnetic field, making the X-ray emission rather faint. Indeed, even in the case of weakly magnetized NSs, for sufficiently low mass-accretion rates the magnetospheric radius 
can be located at tens of gravitational radii. This is the case of e.g. IGR J17062-6143, where the inner edge of the accretion disk was found at $>$100 Rg 
\citep{Degenaar_2017}. Magnetic inhibition may trigger the onset of a propeller phase, with following outflows as an observational signature \citep[][]{Degenaar_2017}, as in 
the sub-luminous X-ray state of the TMSPs (see Sec.\ \ref{ss:tMSP}). Indeed, it was proposed that the majority of the VFXTs may be TMSPs undergoing their "active" X-ray state \citep{Heinke_2015}. 
Other possibilities invoke: (i) a brown dwarf or even a planet as donors in VFXTs \citep{King_2006}, already present as companion stars at the formation of the binary system, (ii) low \citep{King_2006} or (iii) high (e.g.\ \citep[][]{Muno_2005}) inclination angles; iv) the existence of a "period gap", as in Cataclismic Variables (CVs), where accretion via Roche lobes overflow is inactive but a low-level accretion via stellar wind is present \citep{Maccarone_2013}. Due to the heterogeneous phenomenology displayed by these puzzling sources, a single explanation for all the systems is unlikely to exist and these enlisted mechanisms may be at play in different objects.

\section{Multiwavelength observations of NS LMXBs}

So far, we have been focusing on the rich and variegated phenomenology of mainly the X-ray emission from NS LMXBs across different classes. However, these objects are bright all over the electromagnetic spectrum, from radio up to $\gamma$-ray frequencies. First of all, in the X-ray window we only see part of the accretion disk multi-color blackbody spectrum, the one corresponding to the innermost regions of the disk. The outer regions of the disk are bright especially in the UV and optical realm and usually dominate these realms, especially if the contribution from the donor star is negligible, i.e. for UCXBs. 
At longer wavelengths, the disk emission becomes negligible and the spectral contribution from the jet becomes dominant. Jets are collimated outflows of ionized, relativistic particles which are almost ubiquitous in accretion-powered sources (see \citep[][]{Belloni_2010} for a review), including low-magnetized NSs in X-ray binaries. Jets are observed in both Z and atolls sources. Sources in the former class, in addition to be brighter in X-rays, are usually more radio-loud than atoll sources (with the exception of the bright atoll GX~13+1). Radiation from jets extend from radio to mid-IR and it is ascribed to the superposition of self-absorbed synchrotron spectra emitted at different wavelengths by the different portions of the jet.

\begin{figure}[t]
\centering
\includegraphics[width=0.8\textwidth, angle=0]{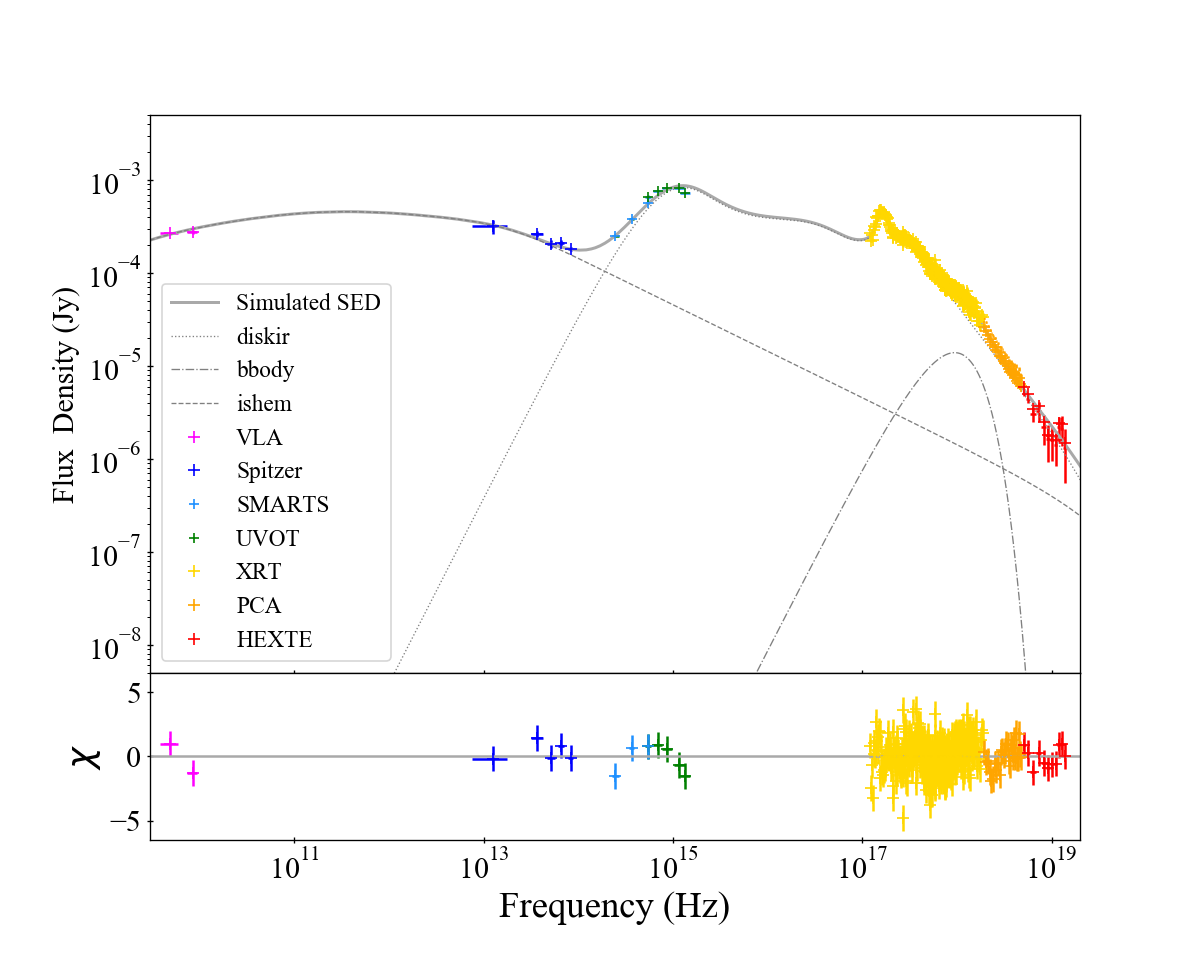}
\caption{Spectral Energy Distribution of the NS LMXB 4U 0614+091 modelled with the Internal Shocks \texttt{ishem} model for the jet, an irradiated disk model \texttt{diskir} and a blackbody model for the emission from the NS surface / boundary layer \citep{Marino_2020}. }
\label{fig:sed_ns}
\end{figure}

With only a few exceptions, such as Sco X-1 \citep{Fomalont_2001}, jets in LMXBs are typically not resolved and are only visible as point-like radio sources, hence they are often called compact jets. Spectra from compact jets, i.e. which can be described as $S_\nu \propto \nu^\alpha$ are typically flat\footnote{The flatness of these spectra is actually not expected, and it witnesses the existence of some internal energy replenishment mechanisms, such as magnetic reconnection or internal shocks (see, e.g., \citep[][]{Malzac_2014} and references therein.} ($\alpha \approx 0$) or inverted ($\alpha >0$) at low energies, i.e. where the jet is opaque, up to a spectral break typically located at IR frequencies, after which the jet becomes optically thin and the spectrum is steep, well described by a power-law. A representative Spectral Energy Distribution is shown in Fig. \ref{fig:sed_ns} ({\it left}), where the separate contributions from jet and accretion flow are highlighted. Two physical processes are universally accepted to produce jet: the Blandford-Znajek (B-Z) mechanism for rotation-powered jets and the Blandford-Payne (B-P) mechanism for disk-driven jets, although the former mechanism is only plausible for jets in BH systems (see, however, \citep[][]{Migliari_2011} for a discussion on rotation-powered jets in NS binaries). On the contrary, the B-P mechanism is applicable to NS LMXBs, but it ignores completely the role, if any, played by the NS own magnetic field in producing the jet, recently included instead in the models by e.g. \cite[][]{Parfrey_2017}.

\subsection{Facts (and peculiarities) of NS LMXBs jets}

Despite being apparently different phenomena, several observational evidences firmly indicate that matter accretion and jets are inextricably intertwined. Decades of radio and X-rays observations of BH X-ray binaries have indeed shown that the physical properties of the jet change drastically depending on whether the system is in hard, intermediate or soft spectral state. 
In hard state, steady compact jets with typical radio-flat spectra are produced. While systems evolve through the intermediate states, radio emission becomes flared and variable and the jet becomes "transient", i.e. composed of discrete, radio bright, knots. Finally, in soft state, radio emission is suppressed and jets are quenched. Furthermore, in hard state, radio and X-rays emission are clearly correlated (see, e.g.\ \citep[][]{Tetarenko_2016}). 
The emerging picture points out the existence of a solid disk-jet interconnection, at least in systems harbouring BHs. 

\begin{figure}[t]
\centering
\includegraphics[width=1.0\textwidth, angle=0]{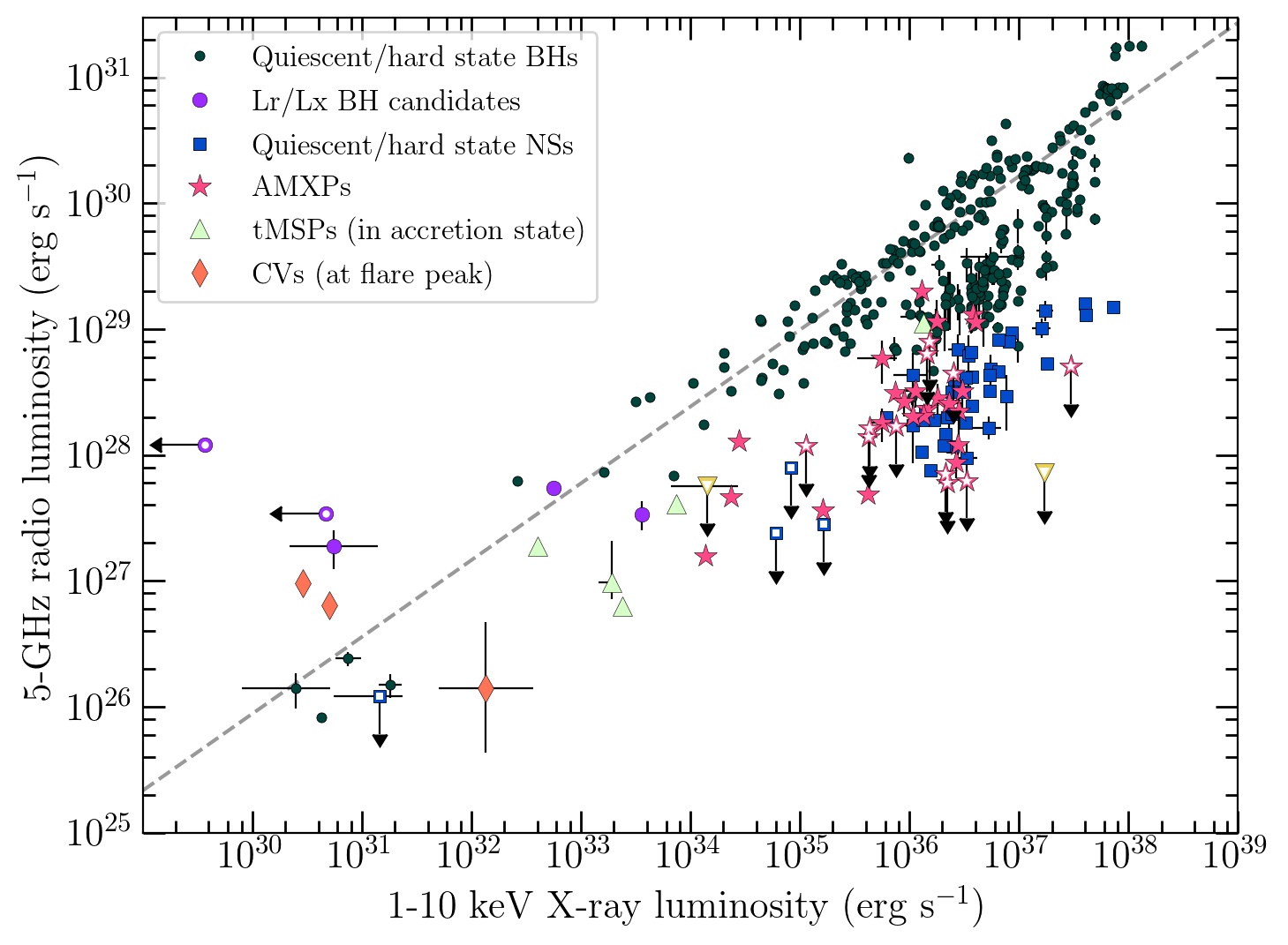}
\caption{Radio/X-ray correlation for the full sample of accreting stellar mass BHs (black circles) and NSs (blue squares), subdivided by classes. Data taken from \url{https://doi.org/10.5281/zenodo.1252036}. \citep{Bahramian_2018}}
\label{fig:jets}
\end{figure}

The pattern of behavior showed by jets in atolls, which have spectral and timing characteristics similar to BH systems, on the other hand, seem to be less clear (see \citep[][]{Migliari_2006, vandenEijnden_2021} for comprehensive studies). When observed, jets are systematically fainter in systems hosting NSs, i.e. typically of a factor $\sim$20, when compared to jets in BH systems \citep[][]{Gallo_2018}. Jet suppression in soft state is observed but it is not the norm in NS LMXBs, with several systems not showing quenching at all \citep[][]{Migliari_2006}. Interestingly, \cite[][]{Russell_2021} showed that jet quenching in the persistent UCXB 4U 1820-303 was triggered not when the system transitioned to the soft state, as in BH systems, but when the flux exceeded a certain threshold, shedding light on the possibility that jets may be more responsive to the accretion rate rather than the spectral state in NS LMXBs. As shown in Fig. \ref{fig:jets}, radio and X-rays luminosities in hard state are correlated in NS LMXBs as well, but their distribution on a radio--X-rays diagram is scattered and harder to interpret when compared to BH systems (see, e.g.\ \citep[][]{Tetarenko_2016}). It has been suggested that jets in NS LMXBs may also have different geometries or couplings with the disk \citep[][]{Marino_2020}. It is still unclear whether the presence of a magnetic field and/or a stellar surface, the mass and spin of the accretor or a different structure of the accretion flow may account for these differences between BH and NS systems.

\section{Future perspectives}

NS with low magnetic fields are important as they provide us with a unique opportunity to study several aspects of the extreme physics of these objects and of the accretion flow around them. As we have discussed in this chapter, these systems are in fact useful for obtaining information on: i) the constraints on the EoS models of matter at supra-nuclear density in the nuclei of NS through the measurement of stellar parameters, such as mass and radius; ii) the physics of the atmosphere of NS as well as the physical characteristics of the crust and nucleus; iii) the geometry and physical properties of the accretion flow in a strong gravitational field regime around these objects; iv) general relativity tests and alternative theories of gravity. In principle, LMXBs may be useful to put constraints on alternative theories of gravity, if their orbital periods can be measured with high accuracy. Since the difference in the orbital period evolution of binaries interpreted with General Relativity (GR) and other theories of Gravity (e.g.\ {\it Brans-Dicke gravity}) is related to the mass difference of the two members of the binary system, these sources provide prime candidates for constraining deviations from GR (see \cite{Psaltis_2008} and references therein).

Despite more than 50 years of research and observations of systems containing NS with low magnetic fields, to date there are still several questions to be clarified regarding these systems, many of which have been briefly discussed in this review. Trying to understand these still unclear issues is, however, of capital importance to obtain information on several aspects of the fundamental physics of these objects and the surrounding environment.
The understanding of the complex long-term orbital residuals in these systems is of fundamental importance not only to constrain the orbital period evolution in these systems, which will provide a way to study the evolutionary path leading to the formation of MSPs, but also the precise determination of the orbital period derivative caused by mass transfer will give in perspective the possibility to constrain alternative theories of Gravity.
Other astrophysical issues that the observation of these systems could clarify concern 
the possibility that these systems can form and contain exoplanets; the study of the long-term accretion process and of the structures formed by the material transferred from the companion star and/or accreting onto the compact object; the finding of an observational way to distinguish systems containing a BH from those containing NS and consequently find the signature of the presence (or absence) of a solid surface for the compact object.

This can be done with future more sensitive instruments/observers with large collecting area, and good spectral and temporal resolution, using simultaneous multiwavelength observations, and with the use of large monitors capable of frequently scanning the entire X-ray sky with sufficient sensitivity to discover and characterize new transients during their X-ray outbursts and to spot peculiar behaviors of these sources.
In this regard, space missions planned for the future, such as Athena, eXTP, XRISM, will certainly be important. The detailed spectral and spectroscopic analysis that the next generation telescopes such as Athena and XRISM will allow, could provide strong constraints on the radius and mass of NS, and hence on the EoS of ultra-dense matter, as well as on the geometry of the accretion flow. Furthermore, the combination of the good timing and polarimetry capabilities of eXTP will enable parallel ways to constrain the accretion geometry, e.g.\ by studying time lags between the different spectral components or the polarization of the incoming radiation. It is also noteworthy that analogous polarimetry studies are now possible with IXPE, 
launched in December 2021. 
Finally, the improved sensitivity of these instruments might also allow the discovery and characterization of similar sources in other Galaxies, allowing a comparison with the behavior of known Galactic sources.


\end{document}